\documentclass[11pt,letterpaper]{article}
\RequirePackage{pdf14}
\usepackage[utf8]{inputenc}
\usepackage{graphicx,tikz}
\usepackage{amsfonts,amsmath,amssymb}
\usepackage{array,setspace,mathrsfs,yfonts,dsfont,bbm,colonequals,amscd,mathtools}
\usepackage{relsize,suffix,cancel,bbm,capt-of,upgreek,verbatim,xcolor}
\usepackage{dsfont}
\usepackage{subcaption}
\usepackage{colortbl}
\usepackage[mode=image|tex]{standalone}
\usepackage{./auxiliary/jheppub}
\usepackage{./auxiliary/genyoungtabtikz}
\usepackage{./auxiliary/arydshln}

\usetikzlibrary{snakes}

\numberwithin{equation}{section}

\usepackage{amsmath}

\numberwithin{equation}{section}

\def\bea{\begin{eqnarray}}
\def\eea{\end{eqnarray}}

\def\ksu2{k_{2d}^{\suf(2)}}

\DeclarePairedDelimiterX\MeijerM[3]{\lparen}{\rparen}%
{\begin{smallmatrix}#1 \\ #2\end{smallmatrix}\delimsize\vert\,#3}

\newcommand\MeijerG[8][]{%
  G^{\,#2,#3}_{#4,#5}\MeijerM[#1]{#6}{#7}{#8}}

\WithSuffix\newcommand\MeijerG*[7]{%
  G^{\,#1,#2}_{#3,#4}\MeijerM*{#5}{#6}{#7}}

\def\tr{\mathop{\mathrm{tr}}\nolimits}
\def\rank{\mathop{\mathrm{rank}}\nolimits}

\def\bZ{\mathbb{Z}}

\def \beg#1{\begin{#1}} 
\def \be{\beg{equation}}
\def \bea{\beg{eqnarray}}
\def \eea{\end{eqnarray}}
\def \ee{\end{equation}}

\def \suf{\mf{su}}

\def \restr#1#2{{\left.\kern-\nulldelimiterspace#1\vphantom{\big|}\right|_{#2}}}

\def \mf{\mathfrak}

\def \nn{\nonumber}

\def \ie{{\it i.e.}}

\def \Tr{{\rm Tr}}

\def \Zb{\mathbb{Z}}

\def \II{\mathcal{I}}

\def \NN{\mathcal{N}}
\def \OO{\mathcal{O}}

\def \SS{\mathcal{S}}
\def \TT{\mathcal{T}}

\def\beq{\begin{equation}}
\def\eeq{\end{equation}}
\def\bea{\begin{eqnarray}}
\def\eea{\end{eqnarray}}

\begin{document}


\title{New $\mathcal N=2$ superconformal field theories from $\mathcal S$-folds}

\author{Simone Giacomelli, Carlo Meneghelli, Wolfger Peelaers}

\affiliation{Mathematical Institute, University of Oxford, Woodstock Road, Oxford, OX2 6GG, United Kingdom}

\abstract{We study the four-dimensional $\mathcal N=2$ superconformal field theories that describe D3-branes probing the recently constructed $\mathcal{N}=2$ $\mathcal{S}$-folds in F-theory. We introduce a novel, infinite class of superconformal field theories related to $\mathcal{S}$-fold theories via partial Higgsing. We determine several properties of both the $\mathcal{S}$-fold models and this new class of theories, including their central charges, Coulomb branch spectrum, and moduli spaces of vacua, by bringing to bear an array of field-theoretical techniques, to wit, torus-compactifications of six-dimensional $\mathcal{N}=(1,0)$ theories, class $\mathcal{S}$ technology, and the SCFT/VOA correspondence.}

\maketitle
\setcounter{page}{1}


\section{Introduction and summary}
\label{s:intro} 
The discovery of the Seiberg-Witten solution a little over a quarter century ago \cite{Seiberg:1994rs, Seiberg:1994aj} ignited an explosion of research in the study of four-dimensional $\NN=2$ supersymmetric quantum field theories. This burst of activity was motivated in large part by the realization that the extended supersymmetry of these models makes feasible a reliable analysis of nonperturbative effects in gauge theories. Indeed, over the years a rich assortment of tools has been developed to compute exactly their observables. What's more, their investigation has led to the discovery of intrinsically strongly-coupled superconformal field theories (SCFTs)\footnote{For foundational work on superconformal field theories see \cite{Sohnius:1981sn, Howe:1983wj}.} such as Argyres-Douglas models \cite{Argyres:1995jj} and theories of class $\mathcal{S}$ \cite{Gaiotto:2009we}. By now a stupendous number of $\NN=2$ theories has been found, throwing into sharp relief the need to define an organizing principle to sort this plethora of models. The constraints following from the rigid special K\"{a}hler structure of the Coulomb branch of vacua constitute one such classifying principle, most successfully implemented for theories with one-complex-dimensional Coulomb branch. See \cite{Argyres:2020nrr} for a recent review. Similarly, the stratified, hyperk\"ahler structure of the Higgs branch of vacua, in combination with the correspondence between $\NN=2$ SCFTs and vertex operator algebras \cite{Beem:2013sza}, provides another one \cite{WIP_BMMPR}. In this paper, we will focus on a class of theories organized according to a third principle, namely their admitting of a uniform geometric realization in string theory.

We will consider four-dimensional superconformal field theories residing on the worldvolume of a stack of D3-branes probing an F-theory singularity. Since the rank of the SCFT equals the number of D3-branes, this setup has the immediate advantage that it facilitates a detailed understanding of the properties of these theories at all ranks. For example, rank-$r$ instanton SCFTs are described as a stack of $r$ D3-branes probing a flat seven-brane with constant axio-dilaton \cite{Sen:1996vd, Banks:1996nj, Dasgupta:1996ij}. At rank one, this set includes in particular the Minahan-Nemeschansky theories with exceptional global symmetry \cite{Minahan:1996fg, Minahan:1996cj} and a number of Argyres-Douglas theories discovered by means of Seiberg-Witten theory in the nineties.\footnote{Their higher-rank analogues were recently revisited from the point of view of their associated vertex operator algebra in \cite{Beem:2019snk}.} Moreover, the first examples of $\mathcal{N}=3$ supersymmetric quantum field theories in four dimensions were constructed using an F-theory setup, with D3-branes probing fourfold terminal singularities called $\mathcal{S}$-folds \cite{Garcia-Etxebarria:2015wns, Aharony:2016kai}. 

Recently, in \cite{Apruzzi:2020pmv}, it was shown that one can combine the $\mathcal{N}=3$ terminal singularities with seven-branes. Probing the resulting $\NN=2$ $\SS$-folds with a single D3-brane provides an F-theoretical realization of all (but one) rank-one SCFTs that are not discrete gaugings. Increasing the number of probe-branes generalizes these models to arbitrary higher rank. We denote these theories $\SS_{G,\ell}^{(r)}$, indicating their rank $r$, the $\bZ_\ell$ $\SS$-fold, and the gauge group $G$ supported on the worldvolume of the seven-brane. In \cite{Apruzzi:2020pmv}, various properties of these models were extracted from the geometric description, including their Coulomb branch spectrum and $a$ and $c$ conformal anomaly coefficients. We extend this F-theoretical analysis of $\NN=2$ $\SS$-folds by calculating the flavor central charges of the simple factors of the global symmetry group. Table \ref{rank-r S-folds} summarizes these properties.

\renewcommand{\arraystretch}{1.5}
\begin{table}[t]
\centering
\begin{small}
\begin{tabular}{c|c||c|c|c|c|c}
$\ell$ & $G$ & $D_i$ & $a$ & $c$ & $h$ & Flavor Symmetry\\
\hline\hline
2& $E_6$& $6,12,\dots,6r$&$\frac{36r^2+42r+4}{24}$&$\frac{36r^2+54r+8}{24}$& $4+r$& $Sp(4)_{6r+1}\times SU(2)_{6r^2+r}$\\
\hdashline[1pt/2pt]
2& $D_4$& $4,8,\dots,4r$&$\frac{24r^2+24r+2}{24}$&$\frac{24r^2+30r+4}{24}$ & $2+r$& $Sp(2)_{4r+1}\times SU(2)_{8r}\times SU(2)_{4r^2+r}$\\
\hdashline[1pt/2pt]
2& $A_2$& $3,6,\dots,3r$&$\frac{18r^2+15r+1}{24}$&$\frac{18r^2+18r+2}{24}$& $1+r$& $Sp(1)_{3r+1}\times U(1)\times SU(2)_{3r^2+r}$\\
\hline
3& $D_4$& $6,12,\dots,6r$&$\frac{36r^2+36r+3}{24}$& $\frac{36r^2+42r+6}{24}$& $3+r$& $SU(3)_{12r+2}\times U(1)$\\
\hdashline[1pt/2pt]
3& $A_1$& $4,8,\dots,4r$ &$\frac{24r^2+20r+1}{24}$&$\frac{24r^2+22r+2}{24}$& $1+r$& $U(1)\times U(1)$\\
\hline
4& $A_2$& $6,12,\dots,6r$&$\frac{36r^2+33r+2}{24}$& $\frac{36r^2+36r+4}{24}$& $2+r$&  $SU(2)_{12r+2}\times U(1)$\\
\end{tabular}
\end{small}%
\caption{\label{rank-r S-folds}Properties of rank-$r$ $\NN=2$ $\SS$-fold theories $\SS_{G,\ell}^{(r)}$ specified by their $\bZ_\ell$ $\mathcal{S}$-fold and a choice of gauge group $G$ supported on the worldvolume of the seven-brane. We list the Coulomb branch spectrum $D_i, i=1,\ldots, r$, the $a$ and $c$ Weyl anomaly coefficients, the quaternionic dimension $h$ of the enhanced Coulomb branch fiber, and the flavor symmetry. We have indicated the flavor central charges as subscripts. For $r=1$ the flavor symmetry enhances.}
\label{tableSfoldIntro}
\end{table}

The geometric description is not omniscient. For example, the F-theory realization indicates that the flavor symmetry of the superconformal field theory on the stack of $r$ probe branes does not depend on the rank $r$ of the theory, but comparison with the classification of rank-one theories shows that an enhancement must occur when $r=1$ \cite{Apruzzi:2020pmv}. The geometric construction is also blind to the full structure of the enhanced Coulomb branch (ECB) of the $\mathcal{N}=2$ $\mathcal{S}$-fold SCFTs.\footnote{\label{footnoteECB}Recall that the existence of an enhanced Coulomb branch indicates that the low-energy effective theory in a generic point of the Coulomb branch includes a collection of free hypermultiplets. An immediate consequence is that the Higgs branch of vacua of the theory contains a rank-preserving singular subvariety, namely the locus where the enhanced Coulomb branch intersects the Higgs branch: if we activate vacuum expectation values for local operators in such a way that we move along this subvariety, the low-energy effective theory has the same rank as the parent UV SCFT.} The F-theory setup only makes visible that part of the ECB that is captured by the position of the D3-branes along the transverse directions. At the origin of the Coulomb branch, the Higgsing associated to this motion triggers a renormalization group flow whose endpoint is easy to predict from the geometric picture: it involves the superconformal field theories described by D3-branes probing flat seven-branes, \ie{}, the instanton-SCFTs, and possibly lower-rank $\NN=2$ $\SS$-fold SCFTs. We will show that, besides this geometrically visible part, the ECB includes another inequivalent locus not captured by the motion of D3-branes and thus inaccessible in F-theory. What's more, we will argue that the enhancement of the flavor symmetry of rank-one $\SS$-folds goes hand in hand with the renormalization group flow triggered by Higgsings along any direction inside the intersection of the ECB and the Higgs branch being equivalent to the flow described above. For higher-rank theories, however, we will see that the infrared fixed point of the renormalization group flow initiated by a partial Higgsing along the inequivalent locus defines a novel, infinite family of SCFTs.  They are in one-to-one correspondence with $\mathcal{N}=2$ $\mathcal{S}$-fold models $\SS_{G,\ell}^{(r)}$ -- we denote them as $\TT_{G,\ell}^{(r)}$ -- and have similar properties. For example, they all have an ECB and their flavor symmetry does not depend on the rank except for an enhancement in the rank-two case. Table \ref{rank-r T-theories} summarizes some of their properties as derived in this work.

\renewcommand{\arraystretch}{1.5}
\begin{table}[t]
\centering
\begin{small}
\begin{tabular}{c|c||c|c|c|c|c}
$\ell$ & $G$ & $D_i$ & $a$ & $c$ & $h$ & Flavor Symmetry\\
\hline\hline
2& $E_6$& $6,12,\dots,6(r-1),3r$&$\frac{6 r^2+r}{4}$&$\frac{6r^2+3r}{4}$& $r$& $(F_4)_{6r}\times SU(2)_{6r^2-5r}$\\
\hdashline[1pt/2pt]
2& $D_4$& $4,8,\dots,4(r-1),2r$&$r^2$&$\frac{4r^2+r}{4}$& $r$& $SO(7)_{4r}\times SU(2)_{4r^2-3r}$\\
\hdashline[1pt/2pt]
2& $A_2$& $3,6,\dots,3(r-1),\tfrac{3}{2}r$&$\frac{6r^2-r}{8}$&$\frac{3r^2}{4}$& $r$& $SU(3)_{3r}\times SU(2)_{3r^2-2r}$\\
\hline
3& $D_4$& $6,12,\dots,6(r-1),2r$&$\frac{ 3r^2-r}{2}$&$\frac{6r^2-r}{4}$&$r$& $(G_2)_{4r}\times U(1) $\\
\hdashline[1pt/2pt]
3& $A_1$& $4,8,\dots,4(r-1),\tfrac{4}{3}r$ &$\frac{2r^2-r}{2}$&$\frac{12r^2-5r}{12}$& $r$& $SU(2)_{\frac{8}{3}r}\times U(1)$\\
\hline
4& $A_2$& $6,12,\dots,6(r-1),\tfrac{3}{2}r$&$\frac{12r^2-7r}{8}$&$\frac{6r^2-3r}{4}$& $r$&  $ SU(2)_{3r}\times U(1)$\\
\end{tabular}
\end{small}%
\caption{\label{rank-r T-theories}Properties of $\TT_{G,\ell}^{(r)}$ specified by the same data as the $\NN=2$ $\SS$-fold theory from which they can be obtained via partial Higgsing. We list the Coulomb branch spectrum $D_i, i=1,\ldots, r$, the $a$ and $c$ Weyl anomaly coefficients, the quaternionic dimension $h$ of the enhanced Coulomb branch fiber, and the flavor symmetry. We have indicated the flavor central charges as subscripts. For $r=2$ the flavor symmetry enhances: the $SU(2)_k$ factors of the global symmetry of the $\ell=2$ theories enhance to $SU(2)_{\frac{k}{2}}\times SU(2)_{\frac{k}{2}}$, the $U(1)$ factor of the flavor symmetry of the $\ell=3$ cases enlarges to $SU(2)_{14}$ and $SU(2)_{10}$ for $G=D_4$ and $A_1$ respectively, and the $U(1)$ factor of the symmetry of the $\ell=4$ theory enhances to $SU(2)_{14}$.}
\label{tableTIntro}
\end{table}

In this paper, we bring to bear an array of purely field-theoretical constructions and techniques to further our understanding of $\mathcal{N}=2$ $\mathcal{S}$-fold SCFTs and their partial Higgsings, and, in particular, to elucidate the above-mentioned features invisible via F-theory. We propose a universally valid formula for the enhanced Coulomb branch of four-dimensional $\NN=2$ superconformal field theories. Specializing to $\NN=2$ $\SS$-fold theories, we make a uniform proposal for their ECB, and leverage that description to derive the properties of the $\TT_{G,\ell}^{(r)}$ theories, which arise as a partial Higgsing along the intersection of the ECB and the Higgs branch, presented in table \ref{rank-r T-theories}. In the process, we find a uniform expression for the enhanced Coulomb branch of the theories $\TT_{G,\ell}^{(r)}$ as well. Using the VOA/SCFT correspondence of \cite{Beem:2013sza}, and more specifically, the technology developed in \cite{Beem:2019tfp,Beem:2019snk}, we analyze and construct the Higgs branches of vacua of both $\SS_{G,\ell}^{(r)}$ and $\TT_{G,\ell}^{(r)}$.

We present a uniform realization of these models as torus-compactifications of six-di\-men\-sion\-al $\NN=(1,0)$ SCFTs \cite{Heckman:2013pva, DelZotto:2014hpa, Heckman:2015bfa} in the presence of almost commuting holonomies for the flavor symmetry along the two nontrivial cycles of the torus. For the $\mathcal{N}=2$ $\mathcal{S}$-fold SCFTs, the relevant six-dimensional theories are realized by placing $r$ M5-branes on an M9-wall which wraps a $\mathbb{C}^2/\Zb_{\ell}$ singularity and turning on a suitable holonomy at infinity for $E_8$ -- see section \ref{6dT2SW} for more details. The effective Lagrangian theory at a generic point of the tensor branch is described by the quiver gauge theory
\be\label{suellquiver}
\begin{tikzpicture}[thick, scale=0.4]
\node[rectangle, draw, minimum width=.6cm,minimum height=.6cm](L1) at (0,0){8};
\node[](L2) at (3,0){$ SU(\ell)$};
\node[](L3) at (7,0){$SU(\ell)$};
\node[](L4) at (10.5,0){$\dots$};
\node[](L5) at (14,0){$SU(\ell)$};
\node[rectangle, draw, minimum width=.6cm,minimum height=.6cm](L6) at (17,0){$\ell$};
\node[](L7) at (8.5,2){$r$};
\node[rectangle, draw, minimum width=.6cm,minimum height=.6cm](L8) at (3,-3){1};

\draw[-] (L1) -- (L2);
\draw[-] (L2) -- (L3);
\draw[-] (L3) -- (L4);
\draw[-] (L4) -- (L5);
\draw[-] (L6) -- (L5);
\draw[snake=brace]  (2,1) -- (15,1);
\draw[snake=zigzag,segment aspect=0]  (3,-.5)  -- (3,-2.3);
\end{tikzpicture}
\ee 
where the leftmost gauge group has eight fundamental and one antisymmetric hypermultiplets. Notice that for $r=1$ the fundamental hypermultiplets on the left and on the right are charged under the same gauge group in perfect harmony with the expected flavor symmetry enhancement. Our torus-compactifications with almost-commuting holonomies generalize the rank-one results of \cite{Ohmori:2018ona} and provide an alternative, purely field-theoretical definition of all $\NN=2$ $\SS$-fold theories. Using this construction, we rederive the theory's Weyl anomaly coefficients, flavor central charges and Coulomb branch spectrum. We find perfect agreement with the geometric computation from F-theory, which can be considered as a highly nontrivial consistency check on the $\mathcal{N}=2$ $\mathcal{S}$-fold construction. Furthermore, since the relevant six-dimensional theories can all be embedded in M-theory as M5-branes wrapping the torus, this construction sheds light on how the M-theory/F-theory duality works in the case of $\mathcal{N}=2$ $\mathcal{S}$-folds.

Similarly, the theories $\TT_{G,\ell}^{(r)}$ have a construction in terms of a torus reduction with almost commuting holonomies turned on. The class of  relevant theories are obtained from the same M-theory setup but with a different choice of $E_8$ holonomy. On a codimension-one submanifold of the tensor branch the effective six-dimensional theory to obtain $\TT_{G,\ell}^{(r)}$ after torus-compactification is
\be\label{su4quiver2sum}
\begin{tikzpicture}[thick, scale=0.4]
\node[rectangle, draw, minimum width=.6cm,minimum height=.6cm](L1) at (-1.5,0){E-string};
\node[](L2) at (3,0){$SU(\ell)$};
\node[](L3) at (7,0){$SU(\ell)$};
\node[](L4) at (10.5,0){$\dots$};
\node[](L5) at (14,0){$SU(\ell)$};
\node[rectangle, draw, minimum width=.6cm,minimum height=.6cm](L6) at (17,0){$\ell$};
\node[](L7) at (8.5,2){$r-1$};
\node[rectangle, draw, minimum width=.6cm,minimum height=.6cm](L8) at (3,-2.5){$\ell$};

\draw[-] (L1) -- (L2);
\draw[-] (L8) -- (L2);
\draw[-] (L2) -- (L3);
\draw[-] (L3) -- (L4);
\draw[-] (L4) -- (L5);
\draw[-] (L6) -- (L5);
\draw[snake=brace]  (2,1) -- (15,1);
\end{tikzpicture}
\ee  
where the $SU(\ell)$ gauge group on the left is coupled to the rank-one E-string theory. We observe that for $r=2$ the $\ell$ flavors on the left and on the right are charged under the same gauge group indicating the above-mentioned flavor symmetry enhancement. These six-dimensional constructions allow us to rederive all data of table \ref{rank-r T-theories}.

We have also performed a systematic scan through theories of class $\SS$, and conclude that, except for rank-two instances, the theories under consideration here do not seem to have a class $\SS$ description. Nevertheless, these sporadic cases are useful as they provide easy access to spectral data of the theories in the form of their superconformal indices.

This paper is organized as follows. In section \ref{Fththeories}, we study the geometric realization of four-dimensional SCFTs in F-theory. In particular, subsection \ref{review} briefly reviews the $\NN=2$ $\SS$-fold construction of \cite{Apruzzi:2020pmv} and the properties of the corresponding superconformal theories. Then in subsection \ref{flavorcech} we compute the flavor central charges of these theories, recovering as a special case the known central charges of rank-one theories. Section \ref{sec_modspace} studies in detail the moduli space of vacua of $\NN=2$ $\SS$-fold SCFTs. In particular, it introduces the infinite class of novel models, which we call $\TT_{G,\ell}^{(r)}$, via a partial Higgsing invisible from the F-theory point of view. In section \ref{6dT2SW}, we describe in detail the construction of $\mathcal{N}=2$ $\mathcal{S}$-fold theories and the related models $\TT_{G,\ell}^{(r)}$ via $T^2$ compactification of six-dimensional $\mathcal{N}=(1,0)$ SCFTs. Finally, in section \ref{sec_classS} we discuss class $\SS$ realizations of $\SS_{G,\ell}^{(r)}$ and $\TT_{G,\ell}^{(r)}$. 

\section{Four-dimensional SCFTs from D3-branes probing \texorpdfstring{$\mathcal{N}=2$ $\SS$}{N=2 S}-folds}
\label{Fththeories}
In this section, we start by briefly reviewing the main properties of the $\mathcal{N}=2$ $\SS$-fold SCFTs constructed in F-theory in \cite{Apruzzi:2020pmv}. Then we extend the geometric analysis of these models by computing the flavor central charges of the simple factors of their global symmetry group.

\subsection{Brief review of \texorpdfstring{$\SS$}{S}-fold SCFTs from F-theory}
\label{review} 

The $\NN=2$ $\SS$-fold SCFTs $\SS^{(r)}_{G,\ell}$ arise as the worldvolume theory of a stack of $r$ D3-branes probing a generalized $\SS$-fold, i.e., an F-theory singularity that combines $\SS$-folds, labeled by $\mathbb{Z}_{\ell}$ with $\ell=2,3,4$, or $6$, and seven-branes with constant axio-dilaton, which we specify by the gauge symmetry $G$ they carry: $G=A_1, A_2, D_4, E_6, E_7$, or $E_8$. The seven-brane's most relevant property for us is its deficit angle, usually denoted as $\Delta_7$. This quantity can be written as $\Delta_7 = \frac{h^{\vee}_{G} + 6}{6}$, in terms of the dual Coxeter number of $G$. Table \ref{7brane} tabulates $\Delta_7$.   

\begin{table}[t]
\centering
\begin{tabular}{c||c|c|c|c|c|c}
$G$ & $A_1$ & $A_2$ & $D_4$ & $E_6$ & $E_7$ & $E_8$ \\
\hline\hline
$\Delta_7$ & $\frac{4}{3}$ & $\frac{3}{2}$ & 2 & 3 & 4 &  6 \\
\end{tabular}
\caption{\label{7brane}Deficit angle $\Delta_7 = \frac{h^{\vee}_{G} + 6}{6}$ of seven-branes labeled by the gauge group $G$ supported on their worldvolume.}
\end{table}

The rank-one $\NN=2$ $\SS$-fold SCFTs are well-known models. Indeed, when we probe a seven-brane by a single D3-brane in the absence of $\SS$-folds, the four-dimensional worldvolume theory is the rank-one $G$-instanton SCFT whose single Coulomb branch chiral ring generator has scaling dimension equal to the deficit angle, while in the presence of $\SS$-folds one finds the following identifications
\begin{align}
&\SS^{(1)}_{E_6,2} ~\longleftrightarrow~[II^* ,C_5]\;, \quad  &&\SS^{(1)}_{D_4,2}~\longleftrightarrow~[III^*,C_3 C_1]\;, \quad && \SS^{(1)}_{A_2,2}~\longleftrightarrow~[IV^*,C_2U_1]\;,\nn \\
&\SS^{(1)}_{D_4,3}~\longleftrightarrow~[II^*,A_3{\!\rtimes}Z_2]\;, \quad &&\SS^{(1)}_{A_1,3} ~\longleftrightarrow~[III^*,A_1U_1{\!\rtimes}Z_2]\;, \quad && \nn\\ 
&\SS^{(1)}_{A_2,4} ~\longleftrightarrow~[II^*,A_2{\!\rtimes}Z_2]\;. &&&& \label{dictionary} 
\end{align}
Here we used the notations of \cite{Argyres:2020nrr}: the first entry within the square brackets indicates the Kodaira type of the Coulomb branch singularity, while the second one denotes the flavor symmetry of the theory. The scaling dimension of the unique Coulomb branch chiral ring generator equals $\ell \Delta_7$. All in all, we recover all (but one) rank-one models that are not discrete gaugings.

Theories with $r>1$ constitute a higher rank generalization of these models. Their $r$ Coulomb branch chiral ring generators have scaling dimensions 
\be\label{CBspectrum} D_i= \ell\Delta_7,\; 2\ell\Delta_7,\dots,\; r\ell\Delta_7.\ee 
When there is no seven-brane (i.e., $\Delta_7=1$) the theory has enhanced $\mathcal{N}\geq 3$ supersymmetry. The central charges were determined in \cite{Apruzzi:2020pmv}. The theories  $\SS_{G,\ell}^{(r)}$ satisfy the Shapere-Tachikawa relation $8a-4c=\sum_i(2D_i-1)$ \cite{Shapere:2008zf}, hence from (\ref{CBspectrum}) one confirms 
\be\label{2a-c}4(2a-c)=r(r+1)\ell\Delta_7-r\;.\ee 
The combination $c-a$ can be computed to be
\be\label{c-a}24(c-a)=(6r+\ell)(\Delta_7-1)\;,\ee 
and we should notice that (\ref{c-a}) holds only for $\ell>1$. In the absence of $\SS$-folds, namely in the case of instanton theories (which we denote as $\mathcal{I}^{(r)}_G$), the formula actually reads $6r(\Delta_7-1)$. We would also like to remark that (\ref{c-a}) is the quaternionic dimension of the Higgs branch.\footnote{This follows from 't Hooft anomaly matching provided the theory can be Higgsed to free hypers. We know this is the case for $\mathcal{S}$-fold theories since they can be Higgsed to the corresponding instanton theories (D3-branes probing a flat seven-brane)  \cite{Apruzzi:2020pmv} and these models in turn can be Higgsed to free hypermultiplets.}. 
Finally, the quaternionic dimension of the enhanced Coulomb Branch (ECB) fiber can be determined to be \cite{Apruzzi:2020pmv}\footnote{See footnote \ref{footnoteECB} for the definition of an enhanced Coulomb branch.}
\be\label{ECB} h=\ell(\Delta_7-1)+r\;.\ee 
This expression can be understood as capturing two distinct contributions. Indeed, by moving all the D3-branes away from the seven-brane and the $\SS$-fold singularity (i.e., going to a generic point on the Coulomb branch), each D3-brane carries a single hypermultiplet (so we have $r$ of them) and the generalized $\SS$-fold contributes $\ell(\Delta_7-1)$ extra hypermultiplets. In the case of instanton theories instead, the number of extra hypermultiplets is of course zero. 

The F-theory description of the $\NN=2$ $\SS$-fold SCFTs makes manifest two distinct contributions to their global symmetry: one from isometries of the background and another one from the gauge symmetry supported on the seven-brane. Let's start by discussing the former. The obvious $U(1)$ isometry rotating the complex plane transverse to the seven-brane worldvolume is identified with the $U(1)_{\mathsf r}$ symmetry of the superconformal algebra (we will denote by $\mathsf r$ the corresponding generator). The directions of the seven-brane worldvolume transverse to the D3-branes is $\mathbb{C}^2/\mathbb{Z}_{\ell}$, which has $U(2)$ symmetry for $\ell>2$ and $SO(4)$ for $\ell=2$. An $SU(2)$ subgroup is identified with the $SU(2)_R$ symmetry of the superconformal algebra (we will denote the Cartan generator by $R$) and the commutant becomes a global symmetry of the SCFT. We therefore find a $U(1)$ global symmetry for $\ell>2$ and $SU(2)$ for $\ell=2$. As for the contribution from the seven-brane gauge symmetry $G$, it was found in \cite{Apruzzi:2020pmv} that the $\mathbb{Z}_{\ell}$ acts nontrivially on $G$ and the resulting flavor symmetry of the four-dimensional theory is the $\mathbb{Z}_{\ell}$-invariant subgroup $H$. Table \ref{global7} presents the relevant invariant subgroups and summarizes the total flavor symmetry of the models $\SS^{(r)}_{G,\ell}$ made manifest in their geometric realization. We claim that the flavor symmetry groups thus obtained represent the full global symmetry of the SCFT for $r>1$, but the symmetry enhances for $r=1$ as in (\ref{dictionary}).

\begin{table}
\centering
\begin{tabular}{c|c||c|c||c}
$G$ & $\Zb_\ell$ & $H$ & $I_{H\hookrightarrow G}$ & Flavor symmetry of $\SS_{G,\ell}^{(r)}$ \\
\hline\hline
$E_6$ & $\Zb_2$ & $Sp(4)$ & 1 & $Sp(4)\times SU(2)$\\
$D_4$ & $\Zb_2$ &$Sp(2)\times SU(2)$& $I_{Sp(2)}=1;\; I_{SU(2)}=2$& $Sp(2)\times SU(2)\times SU(2)$\\
$D_4$ & $\Zb_3$ &$SU(3)$& 3 &$SU(3)\times U(1)$\\
$A_2$ & $\Zb_2$ &$SU(2)\times U(1)$& $I_{SU(2)}=1$ & $SU(2)\times U(1)\times SU(2)$\\
$A_2$ & $\Zb_4$ &$SU(2)$& 4 & $SU(2)\times U(1)$\\
$A_1$ & $\Zb_3$ &$U(1)$& & $U(1)\times U(1)$\\
\end{tabular}
\caption{\label{global7}For each seven-brane of type $G$, the table lists the relevant $\Zb_\ell$-invariant subgroup $H$ and its embedding index  into the gauge group $G$, denoted as $I_{H\hookrightarrow G}$. Including also the contribution from the isometries of the background, the geometric description of the model $\SS_{G,\ell}^{(r)}$ makes manifest the flavor symmetry indicated in the last column. For $r=1$ an enhancement occurs.}
\end{table}

\subsection{Geometric derivation of flavor central charges}
\label{flavorcech}

To compute the flavor central charges of the simple factors of the global symmetry $G_F$ of the theories $\SS_{G,\ell}^{(r)}$, see table \ref{global7}, it will be helpful to first review the derivation of the $a$ and $c$ central charges of \cite{Apruzzi:2020pmv}. That analysis combined techniques developed in \cite{Aharony:2007dj} and \cite{Aharony:2016kai} and, as we will show, can be generalized to the computation of flavor central charges as well. 

We make use of the well-known formulae
\be\label{thooft} \Tr \; \mathsf r\, R^2=2(2a-c)\;,\quad \Tr\; \mathsf r^3=\Tr\; \mathsf r = 48(a-c)\;, \quad k_{G_F}=-2\Tr\; \mathsf r\, G_F^2\;,\ee 
and of the D3-brane charge $\epsilon$ of the generalized $\SS$-fold \cite{Apruzzi:2020pmv}
\be\label{d3charge}\epsilon=\frac{\ell-1}{2\ell}\;.\ee 
The total D3-brane charge is thus $r+\epsilon$. The $a$ and $c$ central charges can be parametrized as follows: 
\begin{equation}\label{param}
2(2a-c)=\alpha r^2 +\beta r + \gamma\;, \qquad 48(c-a)=\delta r + \mu\;.
\end{equation}
Our task is to determine the coefficients $\alpha, \beta, \gamma, \delta$, and $\mu$. One can do so from the F-theory description via a holographic computation. The holographic description of the theories $\SS_{G,\ell}^{(r)}$ was determined in \cite{Apruzzi:2020pmv}: it is given by Type IIB string theory on $AdS_5\times M_5$, where $M_5=\widetilde{S}_5/\mathbb{Z}_{\ell}$ and $\widetilde{S}_5$ is a five-sphere in which the angular coordinate around the seven-brane has periodicity $2\pi/\Delta_7$. Locally $M_5$ is the same as $S^5/\mathbb{Z}_{\ell}$. Also note that in (\ref{param}) we have a priori discarded a contribution of order $r^2$ to $c-a$ because in the large $r$ limit $a=c$ at leading order. 

The leading contribution is proportional to the square of the total D3-brane charge, i.e., $(r+\epsilon)^2$, divided by the volume of $M_5$ appearing in the holographic description. More precisely, if we normalize the radius of curvature of $M_5$ to one, we have the formula \cite{Aharony:2007dj}
\be\label{genlead} a\big|_{\text{leading}} = c\big|_{\text{leading}} =\frac{(r+\epsilon)^2\pi^3}{4\text{Vol}(M_5)}\;.\ee
In the case at hand, since $\text{Vol}(M_5)$ is the volume of the round five-sphere divided by $\ell\Delta_7$, we find
\be\label{leadingcc} a\big|_{\text{leading}} = c\big|_{\text{leading}} = \ell\Delta_7\frac{(r+\epsilon)^2}{4}\;.\ee 
Notice that this contribution fixes the value of $\alpha$ but also contributes to the other coefficients. 

Next, we take into account a subleading contribution of the seven-branes. Because the seven-brane action is linear in $r$, seven-branes do not contribute at order $r^2$. The contribution to the 't Hooft anomalies for $\mathsf r$ and $R$ turns out to be proportional to the D3-brane charge times the volume of the three-manifold (inside $M_5$) wrapped by the seven-brane and divided by the volume of $M_5$. This quantity was computed for $\ell=1$ in \cite{Aharony:2007dj}. Since in our case the volume of both the three- and five-manifolds is divided by $\ell$ with respect to the $\ell=1$ case, we can simply use the formula found in \cite{Aharony:2007dj} with the replacement $N\rightarrow r+\epsilon$. We thus have 
\be\label{subleading}2(2a-c)\big|_{\text{subleading}} = \frac{(r+\epsilon)(\Delta_7-1)}{2}\;, \qquad 48(a-c)\big|_{\text{subleading}}= -12(r+\epsilon)(\Delta_7-1)\;.\ee
Combining (\ref{d3charge}), (\ref{leadingcc}) and (\ref{subleading}) we can determine all the coefficients in (\ref{param}) except $\gamma$ and $\mu$ which are harder to evaluate holographically. However, we can bypass this difficulty by exploiting the fact that $\gamma$ and $\mu$ are the only surviving contributions for $r=0$ and we know that when there are no D3-branes the four-dimensional theory is a collection of $\ell(\Delta_7-1)$ hypermultiplets. These do not contribute to $2a-c$ (therefore $\gamma=0$) and contribute  $2\ell(\Delta_7-1)$ to $48(c-a)$. In this way we reproduce (\ref{2a-c}) and (\ref{c-a}). 

We perform a similar analysis to determine holographically the flavor central charges. We will focus on nonabelian factors of $G_F$: the $SU(2)$ factor that is always present for models with $\ell=2$ -- we call the corresponding central charge $k_{2}$ -- and the nonabelian factors of the invariant subgroup $H$ of the seven-brane gauge symmetry $G$, whose central charge will be denoted as $k_H$.  In \cite{Henningson:1998gx,Aharony:2007dj} it was argued that in the holographic setup 't Hooft anomalies depend (at most) quadratically on the D3-brane charge, therefore we can parametrize  flavor central charges similarly to (\ref{param}): 
\begin{equation}\label{param2}
k_2=\alpha' r^2 +\beta' r + \gamma'\;, \qquad k_H=\delta' r + \mu'\;.
\end{equation}
In the above formula we did not include an order $r^2$ term for $k_H$ because this quantity comes from seven-branes and, as mentioned before, the seven-brane action is linear in $r$.
Let's start with the order $r^0$ contribution. To determine it, we use that the $\ell(\Delta_7-1)$ hypermultiplets transform as
\begin{align}
\text{for }\ell=2:\qquad & (\mathbf 1,\mathbf {4\Delta_7-4}) \text{ of } Sp(1)\times Sp(2\Delta_7-2)  \\
\text{for }G=D_4, \ell=3:\qquad & \mathbf 3_{+} \text{ of } SU(3)\times U(1) \\
\text{for }G=A_2, \ell=4:\qquad & \mathbf 2_{+} \text{ of } SU(2)\times U(1) 
\end{align}
These transformation rules will be derived independently in sections \ref{sec_modspace} and \ref{6dT2SW}. Note that they are obviously compatible with the enhanced flavor symmetry of the rank-one models. Also note that we cannot make any statements for $\SS_{A_1,3}^{(r)}$ theories since the global symmetry is $U(1)^2$ and does not include nonabelian factors. Overall, for $\ell=2$ we conclude that $\gamma'=0$ since the bulk hypermultiplets are not charged under $SU(2)$. The values of $\mu'$ are given in table \ref{muprime}. 
\begin{table}[t]
\centering
\begin{tabular}{c|c|c|c|c|c}
Theory & $\SS_{E_6,2}^{(r)}$ & $\SS_{D_4,2}^{(r)}$ & $\SS_{A_2,2}^{(r)}$ & $\SS_{D_4,3}^{(r)}$ & $\SS_{A_2,4}^{(r)}$\\ 
\hline\hline
Value of $\mu'$ & 1 & (1,0) & 1 & 2 & 2 \\
\end{tabular}
\caption{\label{muprime}The values of $\mu'$ which appear in the flavor central charge formula \eqref{param2} of the flavor symmetry $H$. For $\SS_{D_4,2}^{(r)}$ the notation $(1,0)$ means that $\mu'$ is one for the $Sp(2)$-factor and zero for the $Sp(1)$-factor of $H$.}
\end{table}

Let us now consider the other coefficients for $k_2$. The leading term is not affected by the presence of the seven-brane and therefore we can exploit the fact that for $\ell=2$ and $\Delta_7=1$ the theory has enhanced $\mathcal{N}=4$ supersymmetry. The relevant 't Hooft anomaly can therefore be determined, as in the $\ell=1$ case discussed in \cite{Aharony:2007dj}, by decomposing the $SU(4)_R^3$ Chern-Simons interactions of Type IIB on $AdS_5\times S^5/\mathbb{Z}_2$ since, as was mentioned before, $M_5$ is locally the same as $S^5/\mathbb{Z}_{\ell}$. The decomposition includes the  $\Tr (\mathsf r\, SU(2)^2)$ anomaly, which has a fixed ratio relative to the strength of the R-symmetry Chern-Simons term. We therefore conclude that the relation between $ k_2\big|_{\text{leading}}$ and $ a\big|_{\text{leading}}$ is the same as for $\mathcal{N}=4$ theories: $ k_2\big|_{\text{leading}}=4 a\big|_{\text{leading}}$. From (\ref{leadingcc}) we get 
\be\label{k2lead}k_2\big|_{\text{leading}}=2\Delta_7(r+\epsilon)^2.\ee
The subleading $\OO(r)$ contribution to $k_2$ comes entirely from the Chern-Simons terms on the seven-brane, since there are no bulk contributions at this order. This term can be determined by noticing that the $SU(2)$ flavor symmetry factor arises from an $SO(4)$ isometry, with the other $SU(2)$ being part of the R-symmetry of the theory. Since the two $SU(2)$ groups are on equal footing we conclude that $\Tr (\mathsf r\, R^2)=(4a-2c))\big|_{\text{subleading}}$, which we have already computed, is equal to $\Tr (\mathsf r\, SU(2)^2)=-k_2\big|_{\text{subleading}}/2$. From (\ref{k2lead}) and (\ref{subleading}) we therefore conclude that
\be\label{su2cc} k_2=2\Delta_7 r^2+r\;,\ee 
where we have used $\gamma'=0$.

The computation of $k_H$ can be done as in \cite{Aharony:2007dj}: The contribution originates entirely from the Chern-Simons interaction $C_4\wedge \Tr(F_G\wedge F_G)$ on the seven-brane, where $F_G$ is the field-strength of the gauge symmetry supported on the seven-brane. This is due to the decomposition $C_4\simeq A_{\mathsf r}\wedge\omega$, where $A_{\mathsf r}$ is the $U(1)_{\mathsf r}$ gauge field and $\omega$ is the volume form of the three-manifold $M_3$ wrapped by the seven-brane so, upon integrating over $M_3$, we get the desired Chern-Simons interaction in the bulk. As explained above, the overall coefficient is proportional to the ratio between the volume of $M_3$ and the volume of the compact five-manifold $M_5$. Furthermore, as was noticed before, this quantity does not depend on the parameter $\ell$ and therefore the result is the same as in \cite{Aharony:2007dj} (namely $k_G=2N\Delta_7$), with the usual replacement $N\rightarrow r+\epsilon$. The only difference is that now only the subgroup $H$ is present and therefore we should multiply the expression reported in \cite{Aharony:2007dj} by the embedding index of $H$ in $G$. These embedding indices were reported in table \ref{global7}. They effectively reintroduce a dependence on $\mathbb{Z}_{\ell}$. We therefore find 
\be\label{Hcc} k_H=2\Delta_7rI_{H\hookrightarrow G}+\mu'\;.\ee 
Table \ref{flavorcc} presents the result of the flavor symmetry central charge computations \eqref{su2cc} and \eqref{Hcc}.\footnote{The flavor central charges of rank-two models with $\ell=2$ have recently also been computed with a different method in \cite{Argyres:2020wmq}.}

\begin{table}[t]
\centering
\begin{tabular}{c|c}
Theory & Flavor symmetry\\
\hline\hline
$\SS_{E_6,2}^{(r)}$ & $Sp(4)_{6r+1}\times SU(2)_{6r^2+r}$ \\ 
\hdashline[1pt/2pt]
$\SS_{D_4,2}^{(r)}$  &$Sp(2)_{4r+1}\times SU(2)_{8r}\times SU(2)_{4r^2+r}$ \\
\hdashline[1pt/2pt]
$\SS_{A_2,2}^{(r)}$ & $Sp(1)_{3r+1}\times U(1)\times SU(2)_{3r^2+r}$ \\
\hline 
$\SS_{D_4,3}^{(r)}$  & $SU(3)_{12r+2}\times U(1)$ \\ 
\hline 
$\SS_{A_2,4}^{(r)}$  & $SU(2)_{12r+2}\times U(1)$\\ 
\end{tabular}
\caption{\label{flavorcc} Flavor symmetry groups and their flavor central charges of $\SS_{G,\ell}^{(r)}$ theories.}
\label{tableSfoldmainbody}
\end{table}

Let us conclude by remarking that for $r=1$ and $\ell=2$, we reproduce the known flavor central charges of the enhanced flavor symmetry. Indeed, the nonabelian factors in the flavor symmetry group have embedding index one into the enhanced global symmetry group and they are equal. 

\section{Moduli space of vacua of \texorpdfstring{$\NN=2$ $\SS$}{N=2 S}-fold SCFTs and partial Higgsings}\label{sec_modspace}

In this section, we set out to analyze in detail the moduli space of vacua of $\NN=2$ $\SS$-fold theories $\SS_{G,\ell}^{(r)}$ and to scrutinize their partial Higgsings. We start by discussing general aspects of enhanced Coulomb branches of four-dimensional $\NN=2$ superconformal field theories, including a universal proposal for their structure and an examination of rank-preserving partial Higgsings triggered by vacuum expectation values along the intersection of the Higgs branch and the enhanced Coulomb branch. Upon specializing to $\SS_{G,\ell}^{(r)}$, we present their enhanced Coulomb branch and discern a partial Higgsing to a novel, infinite class of theories, which we denote $\TT_{G,\ell}^{(r)}$. These partial Higgsings are invisible in F-theory: they roughly correspond to turning on a vacuum expectation value for the $\ell(\Delta_7-1)$ hypermultiplets localized at the $\SS$-fold singularity.\footnote{Naturally, the rank-preserving partial Higgsings corresponding to the motion of D3-branes in transverse directions are also faithfully captured by the structure of the ECB.} The construction of $\TT_{G,\ell}^{(r)}$ as partial Higgsings of $\SS_{G,\ell}^{(r)}$ allow us to derive all their properties summarized in table \ref{rank-r T-theories}. Next, we turn attention to the Higgs branches of the models $\SS_{G,\ell}^{(r)}$ and $\TT_{G,\ell}^{(r)}$. Exploiting the ideas put forward in \cite{Beem:2019tfp,Beem:2019snk}, we construct these varieties via a chain of ``unHiggsings.''

\subsection{Structure of enhanced Coulomb branches of four-dimensional \texorpdfstring{$\NN=2$}{N=2} SCFTs}
A four-dimensional $\mathcal{N}=2$ superconformal field theory possesses an enhanced Coulomb branch ($\mathsf{ECB}$) if its low-energy effective field theory in a Coulomb branch vacuum includes neutral, massless hypermultiplets.\footnote{The terminology enhanced Coulomb branch was introduced in \cite{Argyres:2016xmc}}. This condition implies that the moduli space of vacua of the SCFT contains a branch, the $\mathsf{ECB}$, that is locally a product of the Coulomb branch ($\mathsf{CB}$) and a number of copies of $\mathbb C^2$ capturing the moduli of the hypermultiplets. In other words, locally the enhanced Coulomb branch takes the form $\mathsf{CB}\times \mathbb{C}^{2h}$ for some integer $h$. Globally, we propose that
\beq
\label{ECBingeneral}
\mathsf{ECB}=\frac{\mathbb{C}^{2h}\times \widetilde{\mathsf{CB}}}{\Gamma}\,,
\eeq
where $\Gamma$ is a crystallographic complex reflection group and $\widetilde{\mathsf{CB}}\simeq \mathbb{C}^{r}$.\footnote{See, e.g., \cite{Aharony:2016kai, Bonetti:2018fqz, Tachikawa:2019dvq} and references therein for instances in which complex reflection groups have appeared in the description of the moduli space of vacua of theories with extended supersymmetry.} We will further specify the action of $\Gamma$ on each factor momentarily. Note that an expression of the form \eqref{ECBingeneral} captures the $\mathsf{ECB}$s of all Lagrangian theories, all currently known  $\mathcal{N}\geq 3$ theories,\footnote{See, e.g.,~\cite{Aharony:2016kai,Bonetti:2018fqz,Argyres:2019ngz}.} many, and quite possibly all, class $\SS$ theories, and all rank-one theories (see \cite{WIP_BMMPR}), instanton SCFTs, and more generally theories that arise from probe branes in F-theory (as our $\SS$-folds).

Both the $U(1)_{\mathsf r}$ and $SU(2)_R$ R-symmetry groups act on the $\mathsf{ECB}$. The loci of the  $\mathsf{ECB}$ where respectively $SU(2)_R$ and   $U(1)_{\mathsf r}$ remain unbroken are
\beq
\label{fromECBtoWandCB}
\mathsf{ECB} \supset \,\, \mathsf{CB}= \frac{\widetilde{\mathsf{CB}}}{\Gamma}\;,
\qquad
\,\,\,
\mathsf{ECB}\supset \,\, W:=\frac{\mathbb{C}^{2h}}{\Gamma} \subset \mathcal{M}_{\text{Higgs}}\;.
\eeq
The action of $\Gamma$ on $\widetilde{\mathsf{CB}}\simeq \mathbb{C}^{r}$ is constrained by the requirement that it commute with  $U(1)_{\mathsf r}$. If the scaling weights of $\widetilde{\mathsf{CB}}$ are $\widetilde{\Delta}_k$ with multiplicity $r_k$, such that $\sum_k r_k=r$, then we deduce that $\Gamma\subset\prod_k GL(r_k,\mathbb C)$. If we further assume that the Coulomb branch chiral ring is freely generated,\footnote{This fact was first explicitly conjectured in \cite{Tachikawa:2013kta}. Some potential counterexamples have been discussed recently in, e.g., \cite{Bourget:2018ond,Argyres:2018wxu,Argyres:2019ngz}.}
 we conclude, via a famous theorem of Chevalley, Shephard, and Todd that $\Gamma$
acts  on $\widetilde{\mathsf{CB}}$ as a complex reflection group. More explicitly, $\Gamma=\prod_{k}\,\Gamma_k$ is a product of irreducible complex reflection groups, compatible with the  $U(1)_{\mathsf r}$ condition. Irreducible complex reflection groups were classified by  Shephard and  Todd, see e.g.,~\cite{reflectiongroups}. What's more, we anticipate, but have not worked out in detail, that the the crystallographic condition follows from electric-magnetic duality of the low-energy effective theory on the Coulomb branch.

The action of $\Gamma$ on $\mathbb{C}^{2h}$ is also constrained by $\mathcal{N}=2$ superconformal symmetry, because $W$ should have an $SU(2)_R$  group of (non-holomorphic) isometries. This requirement translates into the condition that $\Gamma$ is an element of the $Sp(h)$ factor of $Sp(h)\times SU(2)_R\subset SO(4 h)$ which act on flat $\mathbb{C}^{2h}\simeq \mathbb{R}^{4h}$. It then immediately follows that the space $W=\mathbb{C}^{2h}/\Gamma$ is a so-called symplectic singularity.\footnote{For more details on symplectic singularities, see \cite{Beauville2000SymplecticS}. Our case of interest is discussed in his example (2.5).}

The inclusion of $W$ in the Higgs branch of vacua in \eqref{fromECBtoWandCB} requires a comment. In general $W$ is not a subvariety of $\mathcal{M}_{\text{Higgs}}$, but rather the normalization of such (singular) subvariety, see \cite{WIP_BMMPR}. In particular, this implies that some of the element of the chiral ring $\mathbb{C}[W]$ do not descend from elements of the Higgs branch chiral ring.

Theories whose moduli space of vacua contains a nontrivial $\mathsf{ECB}$ possess a distinguished set of partial Higgsings triggered by vacuum expectation values corresponding to points in $W$. These Higgsings are special because they preserve the rank of the theory. More precisely, let us take $w\in W$ and denote by $\Gamma_w \subset \Gamma$ the subgroup of $\Gamma$ that fixes $w$. The Coulomb branch of the theory one obtains after a Higgsing corresponding to the point $w$ is given by
\beq
\mathsf{CB}_{\text{IR}}= \frac{\widetilde{\mathsf{CB}}}{\Gamma_w}\;.
\eeq
When $w$ is a generic point of $W$, one has $\Gamma_w=\text{id}$.\footnote{If this were not the case, we would have $W=\mathbb{C}^{2 h_0}\times W'$, indicating the presence of $h_0$ decoupled free hypers.}
 Thus, $\widetilde{\mathsf{CB}}$ is identified with $\mathsf{CB}_{\text{IR}}$ for these generic partial Higgsings. While the relation between Coulomb branches of the UV and IR theories is simple, the relation between the UV and IR Higgs branches is  more involved and discussed in the examples below.

\subsection{The enhanced Coulomb branch of \texorpdfstring{$\mathcal{S}^{(r)}_{G,\ell}$}{S} theories}
We propose that the enhanced Coulomb branch of the $\NN=2$ $\SS$-fold theories is given by
\beq
\label{ECBSfolds}
\mathsf{ECB}[\mathcal{S}^{(r)}_{G,\ell}]
\,=\,
\frac{\mathbb{C}^{M} \otimes (v^{}_{\Gamma}\oplus v^*_{\Gamma})\otimes \left(V^{}_{\Gamma}\oplus V^{*}_{\Gamma} \right)\otimes
 \left( V^{}_{\Gamma}\otimes \mathbb{C}_{u_{IR}}
\right) }{\Gamma}\,,
\qquad
\Gamma\,=\,
G(\ell,1,r)\,,
\eeq
Let us describe the ingredients in this formula.
\begin{itemize}
\item The action of the complex reflection group $G(\ell,1,r)$ on its fundamental representation $V_{\Gamma}\,\simeq\, \mathbb{C}^r$, which we parametrize by $(z_1,\dots,z_r)$, is generated by the symmetric group $S_r$ together with $(\mathbb{Z}_{\ell})^r$ transformations generated by $z_k\mapsto \omega^{\delta_{jk}}\,z_k$, for $j=1,\ldots,r$ and with $\omega=e^{2\pi i/\ell}$.\footnote{In fact, $G(\ell,1,r)$  coincides with the so-called  wreath product of $\mathbb{Z}_{\ell}$  with $S_r$.} In particular, any  $g \in G(\ell,1,r)\subset GL(V_{\Gamma})$ can be parametrized as $g=\Omega\,P$ where $\Omega\in (\mathbb{Z}_{\ell})^r$ and $P\in S_r$.
\item The symbol $v_{\Gamma}\simeq \mathbb{C}$ denotes the nontrivial one-dimensional representation of $G(\ell,1,r)$, which transforms as
 $v_{\Gamma}\mapsto \left(\det \Omega \right) \, v_{\Gamma}$ where $\Omega\in (\mathbb{Z}_{\ell})^r$ was specified in the previous bullet.
\item The superscript $*$ denotes complex conjugation.
\item $\mathbb{C}_{u_{IR}}$ is a copy of the complex plane with complex coordinate $u_{IR}$ of scaling dimension $\Delta_{u_{IR}}=\Delta_7$. Recall that $\Delta_7 = \frac{h^{\vee}_G + 6}{6}$, and thus depends on $G$. See table \ref{7brane}.
\end{itemize}

By comparing \eqref{ECBSfolds} with the general expression \eqref{ECBingeneral} one identifies 
\begin{equation}\label{WSwithoutGamma}
\mathbb{C}^{2h} \simeq  \mathbb{C}^{M} \otimes  (v^{}_{\Gamma}\oplus v^*_{\Gamma}) \otimes \left(V^{}_{\Gamma}\oplus V^{*}_{\Gamma} \right)\;.
\end{equation}
Clearly, $M = h-r$, and $h$ was given in table \ref{rank-r S-folds} -- concisely, $M=\ell h^{\vee}_{G} /6 $, where $h^{\vee}_{G}$ is the dual Coxeter number of $G$. The Coulomb branch of $\SS_{G,\ell}^{(r)}$ is
 \beq
 \label{CBSfolds}
 \mathsf{CB}[\mathcal{S}^{(r)}_{G,\ell}]
\,=\,
\frac{
V^{}_{\Gamma}\otimes \mathbb{C}_{u_{IR}} }{\Gamma}\,,
\qquad
\Gamma\,=\,
G(\ell,1,r)\,.
  \eeq
The scaling dimensions of the Coulomb branch generators of the $\SS$-fold theories can be found from \eqref{CBSfolds}
after recalling that the degrees of the invariants of $\Gamma=G(\ell,1,r)$ are $(\ell,2\ell,\dots,r\ell)$. We then produce the list
 \beq
 \label{Scalingweightsv2}
\{
\ell,2\ell,\dots,r\ell
\}\times \Delta_{u_{IR}}\,,
\qquad
\Delta_{u_{IR}}=\Delta_7\,.
 \eeq
The spectrum \eqref{Scalingweightsv2} reproduces \eqref{CBspectrum} providing a nice consistency check of \eqref{ECBSfolds}.
According to the general discussion above, the Higgsing associated to a generic point of $W$ results in a theory whose Coulomb branch is described by $\widetilde{\mathsf{CB}} = V^{}_{\Gamma}\otimes \mathbb{C}_{u_{IR}}$. We thus find a theory whose Coulomb branch spectrum is $\{\Delta_7,\Delta_7,\dots,\Delta_7\}$. We identify this theory as the product of $r$ copies of the one-instanton theory $\mathcal{I}^{(1)}_{G}$.

In the Higgs branch of $\SS_{G,\ell}^{(r)}$, from each generic point of $W$ thus sprout $r$ copies of the Higgs branch of the one-instanton SCFT $\mathcal{I}^{(1)}_{G}$. Adding the dimensions, we easily find the dimension of $\mathcal{M}_{\text{Higgs}}$ to be
\beq
\label{dimMhiggsSfold}
\text{dim}(\mathcal{M}_{\text{Higgs}})=
\text{dim}(W)+r \times \text{dim}(\mathcal{M}_{\text{Higgs}}[\mathcal{I}^{(1)}_{G}]) \,.
\eeq
Recall that the quaternionic dimension of $W$ is $h$ and of $\mathcal{M}_{\text{Higgs}}[\mathcal{I}^{(1)}_{G}]$ is $h^{\vee}_{G}-1$. This information, together with the Coulomb branch spectrum \eqref{Scalingweightsv2}, the Shapere-Tachikawa relation, and anomaly matching on the Higgs branch, i.e., $24(c-a)=\text{dim}_{\mathbb{H}}\mathcal{M}_{\text{Higgs}}$,\footnote{This expression is valid when the theory on a generic point of the Higgs branch contains only hypermultiplets.} provides a shortcut to the determination of $a$ and $c$. We easily reproduce \eqref{2a-c}-\eqref{c-a}, or, equivalently, the data in table \ref{rank-r S-folds}, by recalling that $M = h - r = \ell h^{\vee}_{G} /6 $.

It is useful to spell out the isometries of the  space $W$, defined in \eqref{fromECBtoWandCB}, for the example of $\SS$-fold theories. Indeed, these isometries appear as flavor symmetry subgroups of the SCFT. The manifest symmetry of $W$ is $U(M) \times U(1)$. However, various enhancements take place. First, for $\ell=2$, $G(\ell,1,r)$ coincides with  the Weyl group $B_r$. The representations $V^{}_{\Gamma}$ and $v^{}_{\Gamma}$ are then real so that $V^{}_{\Gamma}\oplus V^{*}_{\Gamma}=V^{}_{\Gamma}\otimes  \mathbb{C}^2$ and $v^{}_{\Gamma}\oplus v^*_{\Gamma}=v^{}_{\Gamma}\otimes  \mathbb{C}^2$. This implies that $U(M) \times U(1)$ enhances to $SP(M) \times SU(2)$ when $\ell=2$. Second, when $r=1$, $V^{}_{\Gamma}=v^{}_{\Gamma}$ and there is a further symmetry enhancement to $SP(M+1)$ (for $\ell=2$), while for $\ell\neq2$ one finds $U(M+1)$ .

As pointed out in the previous subsection, theories with an $\mathsf{ECB}$ possess a distinguished set of rank-preserving Higgsings. Inequivalent Higgsing are characterized by complex reflection subgroups of $\Gamma$. It is instructive to illustrate this point in the case of $\ell=2$, $r=2$, so that $\Gamma=B_2$. The inequivalent  $\mathsf{ECB}$  Higgsings are summarized in the following table\footnote{Notice that two different entries have the same Coulomb branch spectrum.}
\begingroup
\renewcommand{\arraystretch}{1.3}
\begin{equation}
 \begin{tabular}{| c | c | c|c|} 
 \hline
$w\,\in\,\{\mathbb{C}^{2M};\mathbb{C}^2,\mathbb{C}^2\}$   &   $\Gamma_{w}$
  & $\mathcal{T}_{\text{IR}}$ & CB spectrum  \\
\hline 
\hline 
$\{0;0,0\}$  &   $B_2$  & $\mathcal{S}^{(2)}_{G,2}$ & $(2,4)\Delta_7$  
\\
\hline
$\{\mathbf{e};0,0\}$  &   $D_2$  &  $\mathcal{T}^{(2)}_{G,2}$ & $(2,2)\Delta_7$  
\\
\hline
$\{0;e,0\}$
or
$\{0;0,e\}$  &   $B_{1}$  & $\mathcal{S}^{(1)}_{G,2}\otimes 
\mathcal{I}^{(1)}_{G}$  & $(1,2)\Delta_7$  
\\
\hline
$\{0;e,e\}$  &   $S_2$  & $\mathcal{I}^{(2)}_{G}$  &  $(1,2)\Delta_7$   
\\
\hline
generic
 &   id  & $\mathcal{I}^{(1)}_{G} 
\otimes  \mathcal{I}^{(1)}_{G}$
 &
 $ (1,1)\Delta_7$  \\
\hline
\end{tabular}
\label{HiggsingsB2}
\end{equation}
\endgroup
Here $\mathbf{e}$ and $e$ are non-vanishing elements of $\mathbb{C}^{2M}$ and $\mathbb{C}^2$ respectively. The pattern for higher $r$ (and still $\ell=2$) is similar, reflecting the fact that the subgroups of $B_r$ are products of $S_{r'}$, $B_{r''}$, and $D_{r'''}$, see, e.g., \cite{ReflectionSUBgroups}. The theory obtained in the infrared is a product of instanton theories, $\SS$-fold theories and a new family of theories $\mathcal{T}^{(r)}_{G,2}$ associated to the $D$-factors. We will describe these theories, and their cousins for other values of $\ell$, in detail in the next subsection. Before doing so, let us close this subsection with a pictorial overview of the various $\mathsf{ECB}$  Higgsings, as well more generic Higgsings of $\SS_{G,2}^{(2)}$:
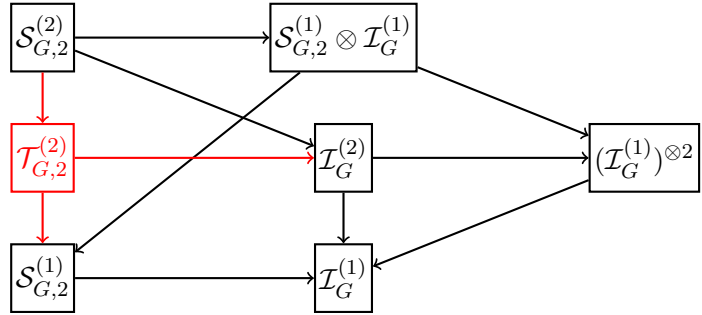
\begin{center}
\begin{tikzpicture}[thick, scale=0.4]
\node[rectangle, draw, inner sep=1.7,minimum height=.9cm](L1) at (0,4){$\SS^{(2)}_{G,2}$};
\node[rectangle, draw, color=red, inner sep=1.7,minimum height=.9cm](L2) at (0,0){$\TT^{(2)}_{G,2}$};
\node[rectangle, draw, inner sep=1.7,minimum height=.9cm](L3) at (10,4){$\SS^{(1)}_{G,2}\otimes \II_{G}^{(1)}$ };
\node[rectangle, draw, inner sep=1.7,minimum height=.9cm](L4) at (10,0){$\II_{G}^{(2)}$};
\node[rectangle, draw, inner sep=1.7,minimum height=.9cm](L5) at (0,-4){$\SS^{(1)}_{G,2}$};
\node[rectangle, draw, inner sep=1.7,minimum height=.9cm](L6) at (20,0){$(\II_{G}^{(1)})^{\otimes 2}$};
\node[rectangle, draw, inner sep=1.7,minimum height=.9cm](L7) at (10,-4){$\II_{G}^{(1)}$};

\draw[->,color=red] (L1) -- (L2);
\draw[->] (L1) -- (L3);
\draw[->] (L1) -- (L4);
\draw[->] (L3) -- (L5);
\draw[->, color=red] (L2) -- (L4);
\draw[->, color=red] (L2) -- (L5);
\draw[->] (L4) -- (L6);
\draw[->] (L5) -- (L7);
\draw[->] (L4) -- (L7);
\draw[->] (L6) -- (L7);
\draw[->] (L3) -- (L6);
\end{tikzpicture}
\end{center}
Here we have omitted a few obvious Higgsings. All renormalization group flows represented in the diagram by a black arrow can be easily understood in the context of the geometric realization in F-theory, but the red entries are invisible in F-theory.

\subsection{The \texorpdfstring{$\mathcal{T}^{(r)}_{G,\ell}$}{T} theories}

We introduce the theories $\mathcal{T}^{(r)}_{G,\ell}$ by performing the partial Higgsing of the  $\mathcal{S}^{(r)}_{G,\ell}$ theories associated to a point $w$ on the factor $\mathbb{C}^{M} \otimes v^{}_{\Gamma}$ of \eqref{ECBSfolds}. (For example, in \eqref{HiggsingsB2}, we are considering the Higgsing in the second row.) When $w$ lies in $\mathbb{C}^{M} \otimes v^{}_{\Gamma}$, then $\Gamma_w=G(\ell,\ell,r)$, which is the subgroup of  $G(\ell,1,r)$  consisting of elements $g=\Omega\,P \in G(\ell,1,r)$,  $\Omega\in (\mathbb{Z}_{\ell})^r$ and $P\in S_r$, such that $\det \Omega=1$. Note that for $\ell=2$ one finds the Weyl group $G(2,2,r)=D_r$, while $G(\ell,\ell,1)=\text{Id}$. The $\mathsf{ECB}$ of the resulting theories is thus
 \beq
\label{ECBTtheories}
\mathsf{ECB}[\mathcal{T}^{(r)}_{G,\ell}]
\,=\,
\frac{\left(V^{}_{\Gamma}\oplus V^{*}_{\Gamma} \right)\otimes
 \left( V^{}_{\Gamma}\otimes \mathbb{C}_{u_{IR}}
\right) }{\Gamma}\,,
\qquad
\Gamma\,=\,
G(\ell,\ell,r)\,.
\eeq
Similarly to the case of $\SS$-fold theories the Coulomb branch has the structure
 \beq
 \label{CBSTolds}
 \mathsf{CB}[\mathcal{T}^{(r)}_{G,\ell}]
\,=\,
\frac{
V^{}_{\Gamma}\otimes \mathbb{C}_{u_{IR}} }{\Gamma}\,,
\qquad
\Gamma\,=\,
G(\ell,\ell,r)\,,
  \eeq 
 so that the CB spectrum is determined by the degrees of the invariants of $G(\ell,\ell,r)$ as\footnote{Note that the theory $\mathcal{T}^{(2)}_{A_1,3}$ has spectrum $\{3;2\}\times \tfrac{4}{3}=\{4,\tfrac{8}{3}\}$. This pair is compatible with the constraints of \cite{Argyres:2018zay} and also with the construction of \cite{Caorsi:2018zsq}, although this pair of numbers does not appear in their table of allowed spectra at rank two. This is just due to a different choice of branch of the logarithm in the dimension formula of \cite{Caorsi:2018zsq}. A similar fact was observed in footnote 16 of \cite{Ohmori:2018ona}.}
  \beq
 \label{Tscalingweightsv2}
\{D_i[\mathcal{T}^{(r)}_{G,\ell}]\} = \{\ell,2\ell,\dots,(r-1)\ell; r \}\times \Delta_{u_{IR}}\,,
\qquad
\Delta_{u_{IR}}=\Delta_7\,.
 \eeq
 As for $\mathcal{S}$-fold theories, the $a$ and $c$ central charges follows immediately from the Coulomb branch spectrum just derived and the dimension of the Higgs branch. As in \eqref{dimMhiggsSfold}, the latter quantitiy can be computed by adding the dimension of $W\simeq \mathbb{C}^{2r}/G(\ell,\ell,r)$ to that of $r$ copies of the one-instanton SCFT. Indeed, also now, the infrared theory at a generic point of $W$ is $r$ copies of the one-instanton theory. We then find 
  \begin{subequations}
\begin{align}
a[\mathcal{T}^{(r)}_{G,\ell}] &=\tfrac{1}{4}\,r
\left(\ell \Delta_7(r-1)+(3 \Delta_7-2)\right)\,, \\
c[\mathcal{T}^{(r)}_{G,\ell}] &= \tfrac{1}{4}\,r
\left(\ell \Delta_7(r-1)+(4 \Delta_7-3)\right)\,, 
 \end{align}
  \end{subequations}
 where we have used once again the relations 
 $\text{dim}_{\mathbb{H}}\mathcal{M}_{\text{Higgs}}[\mathcal{I}_{G}^{(1)}]=
 h^{\vee}_{G}-1$ and $\Delta_7=\tfrac{h^{\vee}_{G}+6}{6}$,
 see e.g.~\cite{Beem:2019snk}.

The isometry-group of $W$ is in general only $U(1)$. However, for $\ell=2$, it enhances to $SU(2)$ for $r>2$ because $V_{\Gamma}$ is real, and further enhances to $SU(2)\times SU(2)$ for $r=2$ due to the fact that $G(2,2,2)=D_2=D_1\times D_1$. In the latter case  $W=(V_{D_2}\times\mathbb{C}^2)/D_2=(\mathbb{C}^2/\mathbb{Z}_2)^2$. For $r=2$ and $\ell=3,4$ the symmetry $U(1)$ is enhanced to $SU(2)$ as follows from the low-rank identification $G(3,3,2)=S_3$, $G(4,4,2)= B_2$, which are real.\footnote{For $r=1$ $W=\mathbb{C}^2$ corresponding to the fact that $\mathcal{T}^{(1)}=\mathcal{I}^{(1)}\times \text{1 hyper}$.} These isometry-groups manifest themselves as flavor symmetries of the theories $\TT_{G,\ell}^{(r)}$. Indeed, they are given by the the second factor of the flavor symmetry as given in table \ref{tableTIntro}.
One can easily determine the first flavor symmetry factor in full by analyzing the relevant Higgsing of $\mathcal{S}^{(r)}$. We present examples of the corresponding ``unHiggsings'' later.
 
Similarly to the $\SS$-fold theories, $\mathcal{T}$-theories admit $\mathsf{ECB}$-type Higgsings. These produce products of lower-rank theories of types $\mathcal{T}^{(r')}_{G,\ell}$ and  $\mathcal{I}^{(r'')}_{G}$, following the pattern of subgroups of $\Gamma$ see \cite{ReflectionSUBgroups,ReflectionSUBgroupsC}. They also admit non-$\mathsf{ECB}$-Higgsings that decrease their rank. We will argue below that the Higgsing associated to giving a minimal nilpotent vacuum expectation value to the moment map of the first factor in the flavor symmetry group in table \ref{tableTIntro}, which is always simple and whose Lie algebra we will denote by  $\mathfrak{f}$, produces the Higgsing  
 $\mathcal{T}^{(r)}_{G,\ell} \mapsto \mathcal{S}^{(r-1)}_{G,\ell}$.\footnote{The relation between Higgs branches of $\mathcal{T}^{(r)}_{G,\ell}$ and $\mathcal{S}^{(r-1)}_{G,\ell}$ is
 \beq
 \label{Slodowyintersection}
 \mathcal{M}_{\text{Higgs}}[\mathcal{T}^{(r)}_{G,\ell}]
 \cap
  \mathscr{S}=
  \mathcal{M}_{\text{Higgs}}[\mathcal{S}^{(r-1)}_{G,\ell}]\times
   \mathbb{C}^{2 (h_{\mathfrak{f}}^{\vee}-2)}
 \eeq
where  $ \mathscr{S}:=e_{\theta}+\mathbb{C}\,f_{\theta}$ denotes the Slodowy slice at the minimal nilpotent element $e_{\theta}$. Note that this is not quite the standard definition of a Slodowy slice which is $ \mathscr{S}_e:=e+\text{ker}_{f}$, with $\text{ker}_{f}=
 \{x\in\,\mathfrak{f}| [f,x]=0\}$, where $(e,f,h)$ is the $\mathfrak{sl}_2$ triple associated to the nilpotent element in question. The definition used in \eqref{Slodowyintersection} differs from the latter only by the  $\mathbb{C}^{2 (h_{\mathfrak{f}}^{\vee}-2)}$ factor. We also point out that the equality
$\text{dim}[ \mathcal{M}_{\text{Higgs}}[\mathcal{T}^{(r)}]]=
\text{dim}[ \mathcal{M}_{\text{Higgs}}[\mathcal{S}^{(r-1)}]]+(h_{\mathfrak{f}}^{\vee}-2)+1$ follows from the identity $h^{\vee}_{G}-h^{\vee}_{\mathfrak{f}}+1-M=0$, upon also using \eqref{dimMhiggsSfold} and $\text{dim}[ \mathcal{M}_{\text{Higgs}}[\mathcal{I}^{(1)}_{G}]]=h_{G}^{\vee}-1$.} All in all, we thus find the chain of Higgsings
 \beq
 \label{chainHiggsings}
\mathcal{S}^{(r)} \rightarrow \mathcal{T}^{(r)}   \rightarrow  \mathcal{S}^{(r-1)} 
 \rightarrow\dots 
 \rightarrow \mathcal{T}^{(2)}
  \rightarrow  \mathcal{S}^{(1)}  \rightarrow  \mathcal{I}^{(1)}\,,
\eeq
which defines a distinguished path in the Hasse diagram. We will put it to use to compute a number of properties of the $\TT$-theories. In particular, we will use it to find the flavor central charges of $\TT_{G,\ell}^{(r)}$.
 
The Higgsing $ \mathcal{T}^{(r)}   \rightarrow  \mathcal{S}^{(r-1)}$ can be used to determine the level of the flavor symmetry subalgebra $\mathfrak f$ of the $\TT_{G,\ell}^{(r)}$ models. (Recall that $\mathfrak f$ is the Lie algebra of the first factor of the flavor symmetry as tabulated in table \ref{rank-r T-theories}.) We do so by matching the $c$ central charge of the UV theory with the central charge of the IR theory together with the contribution of $h_{\mathfrak f}^\vee -2$ free hypermultiplets and an additional hypermultiplet with non-canonical transformation properties.\footnote{This relation was derived in \cite{Beem:2019tfp}. It can also be derived from anomaly matching on the Higgs branch by noticing that the IR $SU(2)_R$ can be identified in the UV, see \cite{WIP_BMMPR} for more details.}
 This gives the relation
   \beq
  \label{cmatchinHiggsing}
  -12\,c[\mathcal{T}^{(r)}_{G,\ell}]=-12\,c[\mathcal{S}^{(r-1)}_{G,\ell}]
   - (h_{\mathfrak{f}}^{\vee}-2) + 2 (-\tfrac{3}{2}k_{\mathfrak{f}}+1)\;.  \eeq
Since we have already determined the values of $c$ of the $\TT$-theories, this formula provides a derivation for the level $k_{\mathfrak{f}}$. 

From the same Higgsing $\mathcal{T}^{(r)}\rightarrow \mathcal{S}^{(r-1)}$ we can also derive the following relation involving the flavor central charge of the $\mathfrak{a}_1$ algebra in the last factor of the flavor symmetry groups in tables \ref{tableSfoldIntro} and  \ref{tableTIntro}, which is present for $\ell=2$
\beq
\label{a1contraintsfromHiggsing}
k_{\mathfrak{a}_1}[\mathcal{T}^{(r)}_{G,2}]=
k_{\mathfrak{a}_1}[\mathcal{S}^{(r-1)}_{G,2}]+
k_{\mathfrak{c}_M}[\mathcal{S}^{(r-1)}_{G,2}]\,.
\eeq
This expression expresses the fact that UV $SU(2)$ flavor symmetry arises as the diagonal of the IR $SU(2)$ flavor symmetry group and an $SU(2)$ subgroup of $Sp(M)$. See below for more details. At this point, we have derived all properties of the $\TT_{G,\ell}^{(r)}$ theories that were presented in table \ref{rank-r T-theories}.

However, further constraints can be derived, which now serve as consistency checks. Let us now consider the Higgsing $\mathcal{S}^{(r)}\rightarrow \mathcal{T}^{(r)}$. Similarly to the rank-one case discussed in \cite{WIP_BMMPR}, when  $\ell=2$ this Higgsing corresponds to a nilpotent vacuum expectation value for a moment map. In fact, in this case it is a minimal nilpotent vacuum expectation value for the first $Sp(M)$ factor of the $\SS$-flavor symmetry in table \ref{tableSfoldIntro}. The same  logic as used to determine \eqref{cmatchinHiggsing} applies, so we have
 \beq
  \label{cmatchinHiggsingchainstep2ellis2}
  -12\,c[\mathcal{S}^{(r)}_{G,2}]=-12\,c[\mathcal{T}^{(r)}_{G,2}]
 +(-1) (M-1) + 2 (-\tfrac{3}{2}k_{\mathfrak{c}_M}+1)\;,
  \eeq
  which can be used to fix $k_{\mathfrak{c}_M}$. 
In the cases $\ell=3,4$, a similar, but slightly more involved, analysis can be performed to determine the level of the $SU(M)$ factors. 

Finally, from the Higgsing $\mathcal{S}^{(r)}\rightarrow \mathcal{T}^{(r)}$ we can derive
the flavor central charges of the $\mathfrak{a}_1$ factors present when $\ell=2$,
\beq
k_{\mathfrak{a}_1}[\mathcal{S}^{(r)}_{G,2}]=
k_{\mathfrak{a}_1}[\mathcal{T}^{(r)}_{G,2}]+
k_{\mathfrak{f}}[\mathcal{T}^{(r)}_{G,2}]\,.
\eeq
At low ranks there are extra symmetries whose levels can be fixed  by similar methods.
All in all, it is remarkable that all the levels can be fixed entirely by this procedure.

  \subsection{Un-Higgsings and free-field realizations}

In this subsection, we will explain how the ideas introduced in \cite{Beem:2019tfp,Beem:2019snk,WIP_BMMPR} can be used, on the one hand, to determine various properties of $\mathcal{S}$ and $\mathcal{T}$ theories, some of which have been described just now, and on the other hand, obtain a rather precise description of their Higgs branches as affine varieties. The idea is to consider a particular partial Higgsing of a UV theory to an IR theory, and to use the IR building blocks to reconstruct the Higgs branch and the associated vertex operator algebra of the UV theory. The real power of this method is to derive many properties of the UV theory from very little input. 

Let us start our endeavor by proposing the Higgs branch generators of $\mathcal{S}^{(r)}_{G,2}$\footnote{As the \dots\, in \eqref{generatorsSpart4} indicate, this is not the full list of generators. The missing generators can
be in principle detected and determined by closure of the Poisson algebra of the listed generators. 
 } 
 \begin{subequations}
 \label{allgeneratorsS}
\begin{align}
\label{Wgenerators}
&
\mathsf{W}^{(1)}_{ ({1},  {3})}\,,
\mathsf{W}^{(2)}_{({1}, {5})}\,,
\dots\,,
\mathsf{W}^{(r)}_{({1},{2r+1})}\,,
\\
\label{higherCMgen}
&
\mathsf{W}^{(1)}_{(\mathfrak{c}_M,{1})}\,,
\mathsf{W}^{(2)}_{(\mathfrak{c}_M,{3})}\,,
\dots\,,
\mathsf{W}^{(r)}_{(\mathfrak{c}_M,{2r-1})}\,,
\\
&
\mathsf{W}^{(1)}_{(\mathfrak{g}_*,{1})}\,,
\mathsf{W}^{(2)}_{(\mathfrak{g}_*,{3})}\,,
\dots\,,
\mathsf{W}^{(r)}_{(\mathfrak{g}_*,{2r-1})}\,,
\\
\label{generatorsSpart3}
&\mathsf{W}^{(3/2)}_{(\mathfrak{g}_1,{2})}\,,
\mathsf{W}^{(5/2)}_{(\mathfrak{g}_1,{4})}\,,
\dots\,,
\mathsf{W}^{(r+\frac{1}{2})}_{(\mathfrak{g}_1,{2 r})}\,,
\\
\label{generatorsSpart4}
& 
\mathsf{W}^{(\frac{r+1}{2})}_{({2M},{r+1})}\,,
\mathsf{W}^{(\frac{r+2}{2})}_{(\mathfrak{R},{r})}\,,
\mathsf{W}^{(\frac{r+3}{2})}_{(\mathfrak{R}',{r-1})}\,,
\mathsf{W}^{(\frac{r+4}{2})}_{(\mathfrak{R}'',{r-2})}\,,
\dots
\end{align}
   \end{subequations}
 where $\mathsf{W}^{(1)}_{({1}, {3})}:=\mathsf{J}_{\mathfrak{a}_1}$, 
 $\mathsf{W}^{(1)}_{(\mathfrak{c}_M,{1})}:=\mathsf{J}_{\mathfrak{c}_M}$ and
 $\mathsf{W}^{(1)}_{(\mathfrak{g}_*,{1})}=\mathsf{J}_{\mathfrak{g}_*}$
  are flavor currents  and 
   $\mathfrak{g}_1$ and $\mathfrak{R},\mathfrak{R}',\mathfrak{R}''\dots$ 
   denote the following  representations of 
   $\mathfrak{c}_M \times \mathfrak{g}_*$
   (where $\mathfrak{g}_*=\varnothing,\mathfrak{a}_1,\mathfrak{u}_1$ for $M=4,2,1$)
\beq
\mathfrak{g}_1=
\begin{cases}
\,{42}\,,\\
({5},{3})\,, \\
\,\,
2_{\pm}
\end{cases}
\quad
\mathfrak{R}=
\begin{cases}
\,{48}\,,\\
({4},{3})\,, \\
\,\,1_{\pm}\,,
\end{cases}
\quad
 \mathfrak{R}'=
\begin{cases}
\,{160}\,,\\
({16},{1})  \,, \\
\varnothing \,,
\end{cases}
\dots
\eeq
The superscript in parenthesis denotes the $SU(2)_R$ weight of the corresponding generator.
At rank one, the generators of dimension one recombine to form $\mathsf{J}_{\mathfrak{c}_{M+1}}$ currents and the ones of dimension $3/2$ recombine in a representation of  $\mathsf{J}_{\mathfrak{c}_{M+1}}\times \mathfrak{g}_{\star}$. For $r=2$, this list agrees with the index results of section \ref{sec_classS}. What's more, for any $r$, it is not hard to convince oneself that the set \eqref{Wgenerators} together with the $\mathfrak{c}_M$ currents and $\mathsf{W}^{(\frac{r+1}{2})}_{{2M},{r+1}}$ coincide with the set of generators of $\mathbb{C}[W]$ where $W=\frac{\mathbb C^{2h}}{\Gamma}$ with $\mathbb C^{2h}$ given in \ref{WSwithoutGamma}.
Similarly, we propose for the generators of the Higgs branch of  $\mathcal{T}^{(r)}_{G,2}$  
 \begin{subequations}
 \label{HBgeneratorsOFTthoeries}
\begin{align}
\label{WgeneratorsT}
&
\mathsf{W}^{(1)}_{(1,3)}\,,
\mathsf{W}^{(2)}_{(1,5)}\,,
\dots\,,
\mathsf{W}^{(r-1)}_{(1,2r-1)}\,,
\\
\label{higherf4generatorsT}
&
\mathsf{W}^{(1)}_{(\mathfrak{f},1)}\,,
\mathsf{W}^{(2)}_{(\mathfrak{f},3)}\,,
\dots\,,
\mathsf{W}^{(r)}_{(\mathfrak{f},2r-1)}\,,
\\
\label{generatorsTpart3}
&
\mathsf{W}^{(3/2)}_{(\mathfrak{r},2)}\,,
\mathsf{W}^{(5/2)}_{(\mathfrak{r},4)}\,,\dots\,,
\mathsf{W}^{(r-\frac{1}{2})}_{(\mathfrak{r},2(r-1))}\,,
\\
\label{generatorsTpart4}
&
\mathsf{W}^{(\frac{r}{2})}_{(1,r+1)}\,,
\mathsf{W}^{(\frac{r+1}{2})}_{(\mathfrak{r},r)}\,,
\mathsf{W}^{(\frac{r}{2}+1)}_{(\mathfrak{r}',r-1)}\,,
\mathsf{W}^{(\frac{r}{2}+1)}_{(\mathfrak{f},r-1)}\,,
\dots
\end{align}
\end{subequations}
   where $\mathfrak{f}=\mathfrak{f}_4,\mathfrak{so}_7,\mathfrak{a}_2$ and 
   $\mathfrak{r}={26},{7},\varnothing$,
   $\mathfrak{r}'={26},\varnothing,\varnothing$ for 
   $G =E_6, D_4, A_2$ respectively.
  As before $\mathsf{W}^{(1)}_{(1,3)}:=\mathsf{J}_{\mathfrak{a}_1}$ and 
   $\mathsf{W}^{(1)}_{(\mathfrak{f},1)}:=\mathsf{J}_{\mathfrak{f}}$ are currents.
   The set of generators \eqref{WgeneratorsT}, together with $\mathsf{W}^{(\frac{r}{2})}_{(1,r+1)}$ in  
   \eqref{generatorsTpart4}, is part of the set of generators of $\mathbb{C}[W]$, see \cite{Bonetti:2018fqz}.
  For $r=1$ the list \eqref{HBgeneratorsOFTthoeries} reduces to $\mathfrak{g}\simeq \mathfrak{f}\oplus\mathfrak{r}$ flavor currents and one free hypermultiplet.
  For $r=2$ there is an extra $\mathfrak{a}_1$ current and the set of generators of dimension $3/2$
  extracted from  \eqref{HBgeneratorsOFTthoeries} is $\mathsf{W}^{(3/2)}_{(\mathfrak{r},2,1)}$, 
 $\mathsf{W}^{(3/2)}_{(\mathfrak{r},1,2)}$. Our proposal agrees with the index results of section \ref{sec_classS} for $r=2$.
  \subsubsection{Inverting the Higgsing $\mathcal{T}^{(r)}\rightarrow \mathcal{S}^{(r-1)}$}
 
We will start with a concrete example and later spell out the general structure.
 
\paragraph{$\mathcal{T}^{(2)}$ from $\mathcal{S}^{(1)}$.}
As the last Higgsing in 
\eqref{chainHiggsings} is discussed in great detail in \cite{WIP_BMMPR},  let us describe the next to last Higgsing of the chain \eqref{chainHiggsings}. To keep the discussion somewhat concrete we will focus on the example of  
 $\mathcal{T}^{(2)}_{E_6,2}\rightarrow \mathcal{S}^{(1)}_{E_6,2}$. 
 As shown in \cite{WIP_BMMPR},\footnote{While we were completing this manuscript, this fact also appeared in 
\cite{Bourget:2020asf}.
}
 the generators of the  Higgs branch of  $\mathcal{S}^{(1)}_{E_6,2}$
 are the  $\mathfrak{c}_5$ flavor symmetry currents, which we will denote as $\mathsf{J}^{\text{IR}}_{\mathfrak{c}_5}$,
 and additional generators of weight $R=3/2$ transforming in the representation ${132}$, that will be denoted as $\mathsf{W}^{\text{IR}}_{{132}}$.
 The theory in the UV is $\mathcal{T}^{(2)}_{E_6,2}$ and has flavor symmetry algebra
  $\mathfrak{f}_4\oplus \mathfrak{c}_1\oplus \mathfrak{c}_1$.
  As advertised above \eqref{chainHiggsings}, the relevant Higgsing corresponds to giving a minimal nilpotent vacuum expectation value to the $\mathfrak{f}_4$ current. This singles out a $\mathfrak{c}_3 \subset \mathfrak{f}_4$ as the commutant of the embedded
   $\mathfrak{sl}_2$ in $\mathfrak{f}_4$. The associated Goldstone bosons transform in the ${14}'$ of $\mathfrak{c}_3$.
Following the ideas presented in  \cite{Beem:2019tfp} we 
introduce a dense open set in the Higgs branch of the UV theory, which in this case is\footnote{The case of $\mathcal{T}^{(r)}_{G,\ell}$ is essentially identical.} 
\beq
\label{OpenpatchTtoSlowerrank}
\mathcal{M}_{H}[\mathcal{T}^{(2)}_{E_6,2}]
\supset \mathcal{U}\,\simeq\,
\frac{\mathcal{M}_{H}[\mathcal{S}^{(1)}_{E_6,2}]\times 
\mathbb{C}^{14}_\xi \times \mathsf{T}^*(\mathbb{C}^*)}{\mathbb{Z}_2}\;,
\eeq
where the factor $\mathbb{C}^{14}_\xi $ corresponds to $14$ half-hypermultiplets and $\mathsf{T}^*(\mathbb{C}^*)$ denotes the cotangent bundle of $\mathbb{C}^*$, which we  coordinatize by $(\mathsf{e}^{1/2},\mathsf{h})$.
The symplectic structure of the UV Higgs branch follows from the symplectic structure of the IR ingredients, which for $\mathsf{T}^*(\mathbb{C}^*)$ is $\{\mathsf{h},\mathsf{e}^{1/2}\}=\mathsf{e}^{1/2}$.
All the chiral ring relations of the UV theory (conjecturally) follow algebraically from the one in the IR theory.
If we restrict the first factor $\mathcal{M}_{H}[\mathcal{S}^{(1)}_{E_6,2}]$ to its origin, \eqref{OpenpatchTtoSlowerrank} reduces to the approximation of the minimal nilpotent orbit of $\mathfrak{f}_4$ discussed in  \cite{Beem:2019tfp}.

The group $\mathbb{Z}_2$ acts by negation on the coordinates $\xi$ and $\mathsf{e}^{1/2}$, its action on the generators of $\mathcal{M}_{H}[\mathcal{S}^{(1)}_{E_6,2}]$ will be specified momentarily.

  It is useful to keep track of the factor of the flavor symmetry which is visible from the UV to the IR.
  For the Higgsings in question this is the commutant of $\mathfrak{sl}(2)_{\theta}$ in the flavor symmetry of the UV theory, which for $\mathcal{T}^{(2)}_{G,2}$ is 
  $\mathfrak{f}^{\natural}\oplus \mathfrak{c}_1 \oplus \mathfrak{c}_1$, see \eqref{matchingfnatural}.
  In the case $G=E_6$ we have $\mathfrak{f}^{\natural}= \mathfrak{c}_3$
  and the Higgs branch generators $\mathsf{J}^{\text{IR}}_{\mathfrak{c}_5}$, $\mathsf{W}^{\text{IR}}_{{132}}$ of the IR theory decompose as follows
\beq
 \mathfrak{c}_5\rightarrow 
 {\color{blue}{\mathfrak{c}_3}}\oplus
  {\color{blue}{\mathfrak{c}_1}} \oplus 
  {\color{blue}{\mathfrak{c}_1}}
  \oplus 
  {\color{magenta}{(6,2,1)}}
  \oplus 
  {\color{magenta}{(6,1,2)}}\oplus
  {\color{cyan}{(1,2,2)}}\,,
 \eeq 
 \beq
 \label{132witholive}
 132\rightarrow 
 {\color{orange}{(14',2,2)}}\oplus
 {\color{red}{ (14',1,1)}}
 \oplus 
 {\color{blue}{(14,2,1)}}
 \oplus 
 {\color{blue}{(14,1,2)}}\oplus 
 {\color{olive}{(6,1,1)}}\,,
 \eeq
where the color code is introduced for later convenience.
After this preparation, we are in a position to describe the action of
$\mathbb{Z}_2$ on the generators of $\mathcal{M}_{H}[\mathcal{S}^{(1)}_{E_6,2}]$:
the components 
$ {\color{blue}{\mathfrak{c}_3}}\oplus
  {\color{blue}{\mathfrak{c}_1}} \oplus 
  {\color{blue}{\mathfrak{c}_1}}
  \oplus 
  {\color{cyan}{(1,2,2)}}
  \simeq \mathfrak{c}_3\oplus \mathfrak{c}_2 $ 
  and     ${\color{blue}{(14,2,1)}}
 \oplus 
 {\color{blue}{(14,1,2)}}\simeq (14,4)$
 are even, the remaining 
  $ 
  {\color{magenta}{(6,2,1)}}
  \oplus 
  {\color{magenta}{(6,1,2)}}
  \simeq (6,4)$ and  $ {\color{orange}{(14',2,2)}}\oplus
 {\color{red}{ (14',1,1)}}
 \oplus 
 {\color{olive}{(6,1,1)}} \simeq (14',5)\oplus (1,6)$ are odd.

We are now ready to present the form of the generators of the Higgs branch of the UV 
theory in terms of the IR building blocks entering \eqref{OpenpatchTtoSlowerrank}.
The UV flavor symmetry takes the form\footnote{Recall that 
$ \mathfrak{f}_4\rightarrow  \mathfrak{c}_1 \oplus(14',2)\oplus \mathfrak{c}_3$.
}   
 \beq
 \mathsf{J}^{\text{UV}}_{\mathfrak{f}_4}=
 \begin{cases}
    \left(\mathsf{e},\mathsf{h},\left( \mathsf{S}^{\natural}-\tfrac{1}{4}\,\mathsf{h}^2\right)\mathsf{e}^{-1}\right)\\
  \left(
  \xi_{14'}\,\mathsf{e}^{1/2},
  \left(   \mathsf{O}_{14'}+  \xi_{14'}\,  \mathsf{h} \right)\mathsf{e}^{-1/2}
  \right)\,,\\
  \, \mathsf{J}^{\text{IR}}_{{\color{blue}{\mathfrak{c}_3}}
    }+\xi \xi \,,
 \end{cases}
 \qquad 
  \mathsf{J}^{\text{UV}}_{\mathfrak{c}_1^{(A)}}= \mathsf{J}^{\text{IR}}_{{\color{blue}{\mathfrak{c}^{(A)}_1}}}
\,,
 \qquad 
  \mathsf{J}^{\text{UV}}_{\mathfrak{c}_1^{(B)}}= \mathsf{J}^{\text{IR}}_{{\color{blue}{\mathfrak{c}^{(B)}_1}}}\,.
 \eeq
 The quantities $\mathsf{S}^{\natural}$ and $\mathsf{O}_{14'}$ are composites of the IR ingredients excluding
 the $\mathsf{T}^*(\mathbb{C}^*)$ coordinates $(\mathsf{e}^{-1/2},\mathsf{h})$ with scaling weight $R=2$ and $R=3/2$ respectively.
 Their explicit form is fixed by starting with an ansatz with the correct $R$ and flavor symmetry quantum numbers and imposing that the algebra
  $\mathfrak{f}_4$ is realized by Poisson brackets.\footnote{The commutation relations of $\mathfrak{f}_4$ in this basis can be found in \cite{Beem:2019tfp}.
 }
  The schematic form of the generators obtained in this way is
  $\mathsf{O}_{14'}=  \#\, \xi^3
  +  \#\,\xi  \,\mathsf{J}^{\text{IR}}_{{\color{blue}{\mathfrak{c}_3}}}+
  \#\,  \mathsf{W}^{\text{IR}}_{{\color{red}{(14',1,1)}}}$
  and 
  $\mathsf{S}^{\natural}\,=\,
 \# \,\xi^4+
  \#\, \xi^2 \,\mathsf{J}^{\text{IR}}_{{\color{blue}{\mathfrak{c}_3}}}
  +
   \#\, \xi\, \mathsf{W}^{\text{IR}}_{{\color{red}{(14',1,1)}}}+
    \#\, \left(  \mathsf{J}^{\text{IR}}_{{\color{blue}{\mathfrak{c}_3}}}\right)^2$.

The Higgs branch of $\mathcal{T}^{(2)}_{E_6,2}$  contains additional generators.
Their form in terms of the IR ingredients is\footnote{Recall that $26\rightarrow (6,2)\oplus (14,1)$.}
 \beq
 \mathsf{W}^{\text{UV}}_{(26,2,1)}=
 \begin{cases}
  \mathsf{J}^{\text{IR}}_{{\color{magenta}{(6,2,1)}}
  } 
  \mathsf{e}^{1/2}+\text{desc}\,,\\
   \mathsf{W}^{\text{IR}}_{{\color{blue}{(14,2,1)}}}+
   \mathsf{J}^{\text{IR}}_{{\color{magenta}{(6,2,1)}}} \xi_{14'}\big{|}_{(14,2,1)}  \,,
 \end{cases}
 \quad
 \mathsf{W}^{\text{UV}}_{(26,1,2)}=
 \begin{cases}
  \mathsf{J}^{\text{IR}}_{{\color{magenta}{(6,1,2)}}
  } 
  \mathsf{e}^{1/2}+\text{desc}\,,\\
   \mathsf{W}^{\text{IR}}_{{\color{blue}{(14,1,2)}}}
   +
   \mathsf{J}^{\text{IR}}_{{\color{magenta}{(6,1,2)}}} \xi_{14'}\big{|}_{(14,1,2)}\,,
 \end{cases}
 \eeq
 \beq
 \mathsf{W}^{\text{UV}}_{(52,2,2)}=
 \begin{cases}
   \mathsf{J}^{\text{IR}}_{{\color{cyan}{(1,2,2)}}}
   \mathsf{e}+\text{desc}\,,\\
\mathsf{W}^{\text{IR}}_{{\color{orange}{(14',2,2)}}}
\mathsf{e}^{1/2}+\text{desc}\,,\\
(\dots)_{(21,2,2)} \,,\\
 \end{cases}
 \quad
  \mathsf{W}^{\text{UV}}_{(26,1,1)}=
 \begin{cases}
\mathsf{W}^{\text{IR}}_{{\color{olive}{(6,1,1)}}}
\mathsf{e}^{1/2}+\text{desc}\,,\\
(\dots)_{(14,1,1)}\,,\\
 \end{cases}
 \eeq
 where $\text{desc}$ indicates $\mathfrak{sl}(2)_{\theta}$ descendants.
 Notice that since we already provided the form of the  flavor symmetry  
 $\mathfrak{f}_4\oplus \mathfrak{c}_1\oplus \mathfrak{c}_1$, the explicit expression of each irreducible 
 $\mathsf{W}$ generator can be obtained from any of its components by acting with the symmetry generators.
 This remark applies in particular to  the expressions $(\dots)_{(21,2,2)}$ and $(\dots)_{(14,1,1)}$ 
 that we have left unspecified. 
 
 One of the most interesting aspects of this construction is that (in favorable circumstances) it admits a straightforward ``affine uplift'' where the geometric IR ingredients in \eqref{OpenpatchTtoSlowerrank} 
 are replaced by VOAs building blocks. In the case of \eqref{OpenpatchTtoSlowerrank}, these VOA ingredients are $\mathbb{V}[\mathcal{S}^{(1)}_{\mathfrak{e}_6,2}]$, the symplectic boson VOA and a pair of chiral bosons associated to the $\mathsf{T}^{*}(\mathbb{C}^*)$ factor, see \cite{Beem:2019tfp}.
 This construction is used to determine the central charge of the UV theory, see \eqref{cmatchinHiggsing},
 and the levels of the flavor symmetry.
 It is currently an open question to establish which functions on the open patch $\mathcal{U}$ extend to the whole Higgs branch. However, it appears that going through the affine uplift helps solving this problem by a mechanism that we illustrate in the following simple example.
We require that the VOA operator corresponding to the generators $\mathsf{W}^{\text{UV}}$ are affine Kac-Moody (AKM) primaries. The function $\mathsf{J}^{\text{IR}}_{{\color{cyan}{(1,2,2)}}}$, being $\mathbb{Z}_2$ even,
is a good function on $\mathcal{U}$, but it is easy to verify that the corresponding VOA generator is not 
AKM primary. The function 
$\mathsf{J}^{\text{IR}}_{{\color{cyan}{(1,2,2)}}} \mathsf{e}$  on the other hand
corresponds to an $\mathfrak{f}_4$ AKM primary, its geometric avatar is thus included in the set of Higgs branch generators. 
%

\paragraph{The cases of $\mathcal{T}^{(2)}_{D_4,2}$ and  $\mathcal{T}^{(2)}_{A_2,2}$} follow the same pattern as  $\mathcal{T}^{(2)}_{E_6,2}$.  In the case of 
$D_4$ we decompose the IR Higgs branch generators as
 \beq
 \mathfrak{c}_3 \rightarrow 
 {\color{blue}{\mathfrak{c}_1}}\oplus
  {\color{blue}{\mathfrak{c}_1}}
   \oplus 
  {\color{blue}{\mathfrak{c}_1}}
  \oplus 
  {\color{magenta}{(2,2,1)}}
  \oplus 
  {\color{magenta}{(2,1,2)}}\oplus
  {\color{cyan}{(1,2,2)}}\,,
  \qquad
  \mathfrak{c}_1 \rightarrow 
 {\color{blue}{\mathfrak{c}_1}}
 \eeq 
 \beq
 (14',3)\rightarrow 
 \left(
 {\color{orange}{(2,2,2)}}\oplus
 {\color{red}{ (2,1,1)}}
 \oplus 
 {\color{blue}{(1,2,1)}}
 \oplus 
 {\color{blue}{(1,1,2)}}
 ,3\right)\,,
 \eeq
 The expressions for $ \mathsf{J}^{\text{UV}}_{\mathfrak{so}_7}$,
 $\mathsf{W}^{\text{UV}}_{(7,2,1)}$,
  $\mathsf{W}^{\text{UV}}_{(7,1,2)}$
  and 
  $\mathsf{W}^{\text{UV}}_{(21,2,2)}$ are obtained by decoding the color coding according to the above template.
  Notice that in this case there is no ${\color{olive}{\text{olive}}}$ part in the decomposition, compared to 
  \eqref{132witholive},
  so the corresponding $\mathsf{W}^{IR}$ generator is absent.
In the 
$\mathfrak{a}_2$ case the relevant decompositions are
  \beq
 \mathfrak{c}_2 \rightarrow 
 {\color{blue}{\mathfrak{c}_1}}\oplus
  {\color{blue}{\mathfrak{c}_1}}\oplus 
  {\color{cyan}{(1,2,2)}}\,,
  \qquad
  \mathfrak{u}_1 \rightarrow 
 {\color{blue}{\mathfrak{u}_1}}
 \eeq 
  \beq
 5_{\pm}\rightarrow 
 {\color{orange}{(2,2)_{\pm}}}\oplus
 {\color{red}{ (1,1)_{\pm}}}\;.
 \eeq
 The colors encode how to reconstructs the UV generators to be
$ \mathsf{J}^{\text{UV}}_{\mathfrak{a}_2}$,
  and 
  $\mathsf{W}^{\text{UV}}_{(8,2,2)}$.

\paragraph{$\mathcal{T}^{(r)}$ from $\mathcal{S}^{(r-1)}$: general rank.}
The case of general $r$ is very similar to the one of $r=2$, but with the important difference that 
the UV theory $\mathcal{T}^{(r>2)}_{G,2}$ has flavor symmetry $\mathfrak{f}\oplus \mathfrak{c}_1$
instead of the $\mathfrak{f}\oplus \mathfrak{c}_1\oplus \mathfrak{c}_1$ symmetry which is present for $r=2$.
Hence, symmetry that remains unbroken from the UV to IR is now  $\mathfrak{f}^{\natural}\oplus \mathfrak{c}_1$.
We decompose the IR generators  with respect to this manifest symmetry using, for example
 \beq
 \mathfrak{c}_4\oplus \mathfrak{a}_1\rightarrow 
  \mathfrak{c}_3\oplus  \mathfrak{c}_1  \oplus (6,2,1)
  \oplus \mathfrak{a}_1
\rightarrow 
 {\color{blue}{\mathfrak{c}_3}}\oplus
{\color{blue}{\widetilde{\mathfrak{a}}_1}}
  \oplus 
  {\color{magenta}{(6,2)}}
  \oplus
  {\color{cyan}{(1,3)}}\,,
 \eeq 
 \beq
(42,2)\rightarrow 
\left((14',2)\oplus (14,1)
,2\right)
\rightarrow 
 {\color{orange}{(14',3)}}\oplus
 {\color{red}{ (14',1)}}
 \oplus 
 {\color{blue}{(14,2)}}\,.
 \eeq
Here $\tilde{\mathfrak a}_1$ is the diagonal of $\mathfrak a_1$ and $\mathfrak c_1$. The UV flavor symmetry takes the form
 \beq
 \label{flavorf4generalr}
 \mathsf{J}^{\text{UV}}_{\mathfrak{f}_4}=
 \begin{cases}
  \,\mathsf{e}+\text{desc}\,,\\
  \left(
  \xi_{14'}\mathsf{e}^{1/2},
  \left(  \mathsf{W}^{\text{IR},(3/2)}_{{\color{red}{(14',1)}}}+\dots \right)\mathsf{e}^{-1/2}
  \right)\,,\\
  \, \mathsf{J}^{\text{IR}}_{{\color{blue}{\mathfrak{c}_3}}  }+\xi \xi \,,
 \end{cases}
 \qquad 
  \mathsf{J}^{\text{UV}}_{\mathfrak{c}_1}= \mathsf{J}^{\text{IR}}_{{\color{blue}{\widetilde{\mathfrak{a}}_1}}}\,,
 \eeq
The generator in 
\eqref{generatorsTpart3}  and 
\eqref{higherf4generatorsT} follow the pattern
 \beq
 \mathsf{W}^{\text{UV},(3/2)}_{(26,2)}=
 \begin{cases}
  \mathsf{J}^{\text{IR}}_{{\color{magenta}{(6,2)}}
    } 
  \mathsf{e}^{1/2}+\text{desc}\,,\\
   \mathsf{W}^{\text{IR},(3/2)}_{{\color{blue}{(14,2)}}}+
    \mathsf{J}^{\text{IR}}_{{\color{magenta}{(6,2)}}}
    \xi_{14'}\big{|}_{(14,2)}\,,
 \end{cases}
 \qquad
  \mathsf{W}^{\text{UV},(2)}_{(52,3)}=
 \begin{cases}
   \mathsf{J}^{\text{IR}}_{{\color{cyan}{(1,3)}}}
   \mathsf{e}+\text{desc}\\
\mathsf{W}^{\text{IR},(3/2)}_{{\color{orange}{(14',3)}}}
\mathsf{e}^{1/2}+\text{desc}\\
\,\,\,
\dots \,\\
 \end{cases}
 \eeq
 and similar expressions for $ \mathsf{W}^{\text{UV},(k-1/2)}_{(26,2k)}$
 and $ \mathsf{W}^{\text{UV},(k)}_{(52,2k-1)}$ for $k=2,\dots,r$.
Concerning the generators in \eqref{generatorsTpart4}, taking as an example the second entry in 
\eqref{generatorsTpart4}, we have
\beq
\mathsf{W}^{\text{UV},(\frac{r+1}{2})}_{(26,r)}=
\begin{cases}
\mathsf{W}^{\text{IR},(\frac{r}{2})}_{(6,r)}\,\mathsf{e}^{1/2}+\text{desc}\,,\\
\dots
\end{cases}\,.
\eeq

\vspace{0.5cm}
\noindent
\emph{Remark:}
It follows from the realization of the current of $\mathfrak{f}^{\natural}$  in \eqref{flavorf4generalr} as 
$\mathsf{J}^{\text{UV}}_{\mathfrak{f}^{\natural}}=\mathsf{J}^{\text{IR}}_{\mathfrak{f}^{\natural}}+\xi \xi$, that
the two-dimensional levels 
satisfy the following relations
 \begin{subequations}
 \label{matchingfnatural}
\begin{align}
& \text{matching $\mathfrak{f}^{\natural}=\mathfrak{c}_3$}
 \qquad \qquad
k^{(2d)}_{\mathfrak{f}_4}[\mathcal{T}^{(r)}_{\mathfrak{e}_6,2}]=
-\tfrac{5}{2}+k^{(2d)}_{\mathfrak{c}_4}[\mathcal{S}^{(r-1)}_{\mathfrak{e}_6,2}]\,,
\\
&\text{matching $\mathfrak{f}^{\natural}=\mathfrak{a}_1\oplus \mathfrak{b}_1$}
\qquad
\begin{cases}
\,\,\,k^{(2d)}_{\mathfrak{so}_7}[\mathcal{T}^{(r)}_{\mathfrak{d}_4,2}]=
-\tfrac{3}{2}+k^{(2d)}_{\mathfrak{c}_2}[\mathcal{S}^{(r-1)}_{\mathfrak{d}_4,2}]\,,
\\
2\,k^{(2d)}_{\mathfrak{so}_7}[\mathcal{T}^{(r)}_{\mathfrak{d}_4,2}]=
-4+ k^{(2d)}_{\mathfrak{c}_1}[\mathcal{S}^{(r-1)}_{\mathfrak{d}_4,2}]\,,
\end{cases}
\\
&\text{matching $\mathfrak{f}^{\natural}=\mathfrak{a}_1$}
\qquad
\qquad
k^{(2d)}_{\mathfrak{a}_2}[\mathcal{T}^{(r)}_{\mathfrak{a}_2,2}]=
-1+k^{(2d)}_{\mathfrak{c}_2}[\mathcal{S}^{(r-1)}_{\mathfrak{a}_2,2}]\,,
\\
&\text{matching $\mathfrak{f}^{\natural}=\mathfrak{a}_1$}
\qquad
\qquad
3 k^{(2d)}_{\mathfrak{g}_2}[\mathcal{T}^{(r)}_{\mathfrak{d}_4,3}]=
-5+k^{(2d)}_{\mathfrak{a}_2}[\mathcal{S}^{(r-1)}_{\mathfrak{d}_4,3}]\,,
\end{align}
\end{subequations}
see, e.g., \cite{Beem:2019tfp}. Note also that $k^{(2d)} = -\frac{1}{2} k$, in terms of the four-dimensional flavor central charge $k$. These relations thus provide many consistency conditions on the flavor central charges of the $\SS$ and $\TT$ theories.

  \subsubsection{Inverting the Higgsing $\mathcal{S}^{(r)}\rightarrow \mathcal{T}^{(r)}$}
 We will now describe the Higgsing $\mathcal{S}^{(r)}\rightarrow \mathcal{T}^{(r)}$.
 In the case $r=1$ it reduces to the analysis presented in \cite{WIP_BMMPR}.
In this case analogue of the open patch \eqref{OpenpatchTtoSlowerrank} is given by %
\beq
\label{OpenpatchStoTlowerrank}
\mathcal{M}_{H}[\mathcal{S}^{(r)}_{G,\ell}]
\supset \mathcal{U}\,\simeq\,
\frac{\mathcal{M}_{H}[\mathcal{T}^{(r)}_{G,\ell}]\times 
\mathbb{C}^{2(M-1)}_\xi \times \mathsf{T}^*(\mathbb{C}^*)}{\mathbb{Z}_{\ell}}\,.
\eeq
If we restrict the first factor $\mathcal{M}_{H}[\mathcal{T}^{(r)}_{G,\ell}]$ to its origin,
\eqref{OpenpatchStoTlowerrank} reduces to an approximation of $W=\mathbb{C}^{2M}/\mathbb{Z}_{\ell}$,
where the action of $\mathbb{Z}_{\ell}$ on $\mathbb{C}^{2M}$ was defined below \eqref{ECBSfolds}.
In the following we will describe explicitly only the case $G=E_6$ and $\ell=2$ for concreteness.
In this case the manifest symmetry from the UV to the IR is 
$\mathfrak{c}_{M-1}\oplus \mathfrak{c}_1=\mathfrak{c}_{3}\oplus \mathfrak{c}_1$.
The  relevant decompositions for the IR generators 
\eqref{HBgeneratorsOFTthoeries} 
under this symmetry are
\beq
\mathfrak{f}_4\rightarrow 
 {\color{blue}{\mathfrak{c}_3}}\oplus
  {\color{blue}{\mathfrak{c}_1}}\oplus
    {\color{magenta}{(14',2)}}\,,
    \qquad
    \mathfrak{a}_1\rightarrow  {\color{blue}{\mathfrak{a}_1}}\,,
    \eeq
    \beq
    (26,2)\rightarrow (6,2,2)\oplus (14,1,2)\simeq
   {\color{orange}{ (6,3)}}
    \oplus 
     {\color{red}{(6,1)}}
    \oplus 
  {\color{blue}{(14,2)}}\,.
\eeq 
The flavor symmetry of the UV theory is realized as 
\beq
\mathsf{J}_{\mathfrak{c}_4}^{\text{UV}}=
\begin{cases}
\,\mathsf{e}+\text{desc}\,,\\
\left(
\xi \,\mathsf{e}^{1/2}, (\mathsf{W}^{\text{IR}}_{{\color{red}{(6,1)}}}
+\dots )\mathsf{e}^{-1/2} \right)\,,\\
\, 
\mathsf{J}^{\text{IR}}_ {\color{blue}{\mathfrak{c}_3}}+\xi \xi\,,
\end{cases}
\qquad
\mathsf{J}_{\mathfrak{a}_1}^{\text{UV}}=
\mathsf{J}_{{\color{blue}{\mathfrak{c}_1}}}^{\text{IR}}
+
\mathsf{J}_{{\color{blue}{\mathfrak{a}_1}}}^{\text{IR}}\,.
\eeq
This identification alone is sufficient to determine the levels of the flavor symmetry. 
The generators of the Higgs branch in the list \eqref{generatorsSpart3}, \eqref{higherCMgen} are  given by
\beq
\mathsf{W}_{(42,2)}^{\text{UV}}=
\begin{cases}
\mathsf{J}_{{\color{magenta}{(14',2)}}}^{\text{IR}}\,\mathsf{e}+\text{desc}\,,
\\
\mathsf{W}^{\text{IR}}_{{\color{blue}{(14,2)}}}
+
\mathsf{J}_{{\color{magenta}{(14',2)}}}^{\text{IR}}\xi\big{|}_{(14,2)}\,,
\end{cases}
\qquad
\mathsf{W}_{(36,3)}^{\text{UV}}=
\begin{cases}
\mathsf{J}_{{\color{blue}{\mathfrak{a}_1}}}^{\text{IR}}\,\mathsf{e}+\text{desc}\,,
\\
\left(\mathsf{W}_{{\color{orange} {(6,3)}}}^{\text{IR}}
+\mathsf{J}_{{\color{blue}{\mathfrak{a}_1}}}^{\text{IR}}\,\xi\,\right)\mathsf{e}^{1/2}+\text{desc}\,,
\\
(\dots)_{(21,3)}\,,
\end{cases}
\eeq
%
and similar expressions for 
$\mathsf{W}^{(k)}_{(36,2k-1)}$,
$\mathsf{W}^{(k+\frac{1}{2})}_{(42,2k)}$ 
with
$k=1,\dots,r$.
Let us also provide a few examples for the generators  
\eqref{generatorsSpart4}:
%
\beq
\mathsf{W}_{(8,r+1)}^{\text{UV},(\frac{r+1}{2})}=
\begin{cases}
\mathsf{W}_{(1,r+1)}^{\text{IR},(\frac{r}{2})}\,\mathsf{e}^{1/2}+\text{desc}\,,\\
\mathsf{W}^{\text{IR},(\frac{r+1}{2})}_{{\color{blue}{(6,r+1)}}}+\dots\,,
\end{cases}
\qquad
\mathsf{W}_{(48,r)}^{\text{UV},(\frac{r+2}{2})}=
\begin{cases}
\mathsf{W}_{{\color{red}{(14,r)}}}^{\text{IR},(\frac{r+1}{2})}\,\mathsf{e}^{1/2}+\text{desc}\,,\\
\mathsf{W}^{\text{IR},(\frac{r+2}{2})}_{{\color{blue}{(14',r)}}}+\dots\,, \\
\mathsf{W}^{\text{IR},(\frac{r+2}{2})}_{{\color{blue}{(6,r)}}}+\dots\,, \\
\end{cases}
\eeq
\beq
\mathsf{W}_{(160,r-1)}^{\text{UV},(\frac{r+3}{2})}=
\begin{cases}
\mathsf{W}_{{\color{olive}{(6,r- 1)}}}^{\text{IR},(\frac{r+1}{2})}\,\mathsf{e}+\text{desc}\,,\\
\mathsf{W}^{\text{IR},(\frac{r+2}{2})}_{{\color{red}{(21,r-1)}}}\,\mathsf{e}^{1/2}+\dots\,,\\
\mathsf{W}^{\text{IR},(\frac{r+2}{2})}_{{\color{red}{(14,r-1)}}}\,\mathsf{e}^{1/2}+\dots\,,\\
\dots\\
\end{cases}
\eeq
where we have displayed the IR ingredients 
 \begin{subequations}
\begin{align}
\mathsf{W}^{\text{IR},(\frac{r+1}{2})}_{(26,r)}&
\rightarrow 
\mathsf{W}^{\text{IR},(\frac{r+1}{2})}_{{\color{blue}{(6,r+1)}}}
+
\mathsf{W}^{\text{IR},(\frac{r+1}{2})}_{{\color{olive}{(6,r- 1)}}}
+
\mathsf{W}^{\text{IR},(\frac{r+1}{2})}_{
{\color{red}{(14,r)}}}\,,
\\
\mathsf{W}^{\text{IR},(\frac{r+2}{2})}_{(26,r-1)}&
\rightarrow 
\mathsf{W}^{\text{IR},(\frac{r+2}{2})}_{{\color{blue}{(6,r)}}}
+
\mathsf{W}^{\text{IR},(\frac{r+2}{2})}_{(6,r- 2)}
+
\mathsf{W}^{\text{IR},(\frac{r+2}{2})}_{
{\color{red}{(14,r-1)}}}\,,
\\
\mathsf{W}^{\text{IR},(\frac{r+2}{2})}_{(52,r-1)}&
\rightarrow 
\mathsf{W}^{\text{IR},(\frac{r+2}{2})}_{{\color{red}{(21,r-1)}}}+\dots\,.
\end{align}
\end{subequations}
By using the chain of Higgsing \eqref{chainHiggsings} one can verify the consistency 
of the list of generators proposed in \eqref{allgeneratorsS}
and \eqref{HBgeneratorsOFTthoeries} and add the missing generators in \eqref{generatorsSpart4}, \eqref{generatorsTpart4}.


\section{Construction of \texorpdfstring{$\SS$}{S}-fold theories from six dimensions}\label{6dT2SW} 

In \cite{Ohmori:2018ona} it was observed that the rank-one models $\SS_{G,\ell}^{(1)}$ can be constructed by compactifying certain six-dimensional $\mathcal{N}=(1,0)$ theories on a torus with almost commuting holonomies for the global symmetry along its cycles. Their focus was mainly on models with a dimension six Coulomb branch operator (i.e., $\ell\Delta_7=6$ in our notation), although it was pointed out that a change in the choice of holonomies, effectively implementing a four-dimensional mass deformation, can be used to recover the other rank-one models as well. The purpose of this section is to extend their construction to higher rank $\SS$-fold theories, thereby providing an independent construction of $\SS_{G,\ell}^{(r)}$ models, and to the novel class of theories $\TT_{G,\ell}^{(r)}$. We will again focus on the case $\ell\Delta_7=6$. 

\subsection{\texorpdfstring{$\mathcal{S}^{(r)}_{G,\ell}$}{S} theories from six dimensions}
We start by discussing in detail the construction of $\mathcal{S}^{(r)}_{G,\ell}$ theories (with $\ell\Delta_7=6$) via torus compactifications.

\subsubsection{The six-dimensional theories}\label{6dquivers}

The relevant six-dimensional theories to construct the models $\mathcal{S}^{(r)}_{G,\ell}$ via torus compactification can be realized in M-theory by probing an M9-plane wrapping $\mathbb{R}^6\times\mathbb{C}^2/\mathbb{Z}_{\ell}$ with a stack of $r$ M5-branes wrapping $\mathbb{R}^6$. The resulting six-dimensional SCFTs have been studied in detail in \cite{Mekareeya:2017jgc, Cabrera:2019izd}. In these references it was emphasized that the resulting theory is specified by the choice of holonomy for the $E_8$ symmetry supported on the M9-wall. The global symmetry of the theory is then (at least) $SU(\ell)$ times the subgroup of $E_8$ left unbroken by the holonomy. For our purposes, it is also useful to keep in mind the F-theory realization of these six-dimensional SCFTs in terms of an elliptically fibered Calabi-Yau threefold. By blowing up the singularity in the base, we move to a generic point on the tensor branch of the theory: in the case at hand we get a collection of $r$ curves with self-intersection $-1,-2,\dots,-2$. Notice that by blowing down the $-1$ curve the neighboring $-2$ curve becomes a $-1$ curve and with a sequence of $r$ blow-downs of $-1$ curves we can eliminate the whole configuration of curves. With a further complex structure deformation we can finally remove completely the singularity in the base. Theories with this property have been dubbed very Higgsable in \cite{Ohmori:2015pua}. 

The advantage of going to a generic point on the tensor branch is that there the theory admits a Lagrangian description in terms of a collection of vector, tensor and hypermultiplets and is thus easier to study. The choice of $E_8$ holonomy enters in specifying the gauge theory data and the relevant cases for us are given by linear quivers of $r$ $SU(\ell)$ gauge groups (supported on the $r$ curves in the base), with bifundamental hypermultiplets in between. The quiver ends on one side with $\ell$ fundamental hypermultiplets for the last $SU(\ell)$ gauge group, whereas at the other end (namely the gauge group supported on the $-1$ curve) we have eight fundamental hypermultiplets and a hypermultiplet in the two-index antisymmetric representation of $SU(\ell)$. More concretely, we will consider three classes of SCFTs whose gauge theory phase is described as follows: 
\begin{itemize}
\item For $\ell=2$ we choose the $SO(16)$-preserving holonomy. In the notations of \cite{Mekareeya:2017jgc} this corresponds to $n_2'=1$ and $N_6=r$. The gauge theory is therefore 
\be\label{su2quiver}
\begin{tikzpicture}[thick, scale=0.4]
\node[rectangle, draw, minimum width=.6cm,minimum height=.6cm](L1) at (0,0){8};
\node[](L2) at (3,0){${\color{red} SU(2)}$};
\node[](L3) at (7,0){$SU(2)$};
\node[](L4) at (10.5,0){$\dots$};
\node[](L5) at (14,0){$SU(2)$};
\node[rectangle, draw, minimum width=.6cm,minimum height=.6cm](L6) at (17,0){2};
\node[](L7) at (8.5,2){$r$};

\draw[-] (L1) -- (L2);
\draw[-] (L2) -- (L3);
\draw[-] (L3) -- (L4);
\draw[-] (L4) -- (L5);
\draw[-] (L6) -- (L5);
 \draw[snake=brace]  (2,1) -- (15,1);
\end{tikzpicture}
\ee
where we have colored in red the gauge group supported on the $-1$ curve. As we have already explained, the number $N_6$ of $SU(2)$ gauge groups is $r$. What's more, the two-index antisymmetric hypermultiplet is absent in this case because that representation is trivial for $SU(2)$. The global symmetry of the theory is $SO(16)\times SU(2)\times SU(2)$. The $SU(2)\times SU(2) \cong SO(4)$ symmetry acts on the two fundamental hypermultiplets on the right-hand side of the quiver. Moreover, one of the $SU(2)$-factors simultaneously rotates all bifundamental hypermultiplets as well.
\item For $\ell=3$ we choose the $SU(9)$-preserving holonomy, namely $n_3'=1$ and $N_6=r$. For $SU(3)$ an antisymmetric hypermultiplet is equivalent to a fundamental one, therefore the quiver is 
\be\label{su3quiver}
\begin{tikzpicture}[thick, scale=0.4]
\node[rectangle, draw, minimum width=.6cm,minimum height=.6cm](L1) at (0,0){9};
\node[](L2) at (3,0){${\color{red} SU(3)}$};
\node[](L3) at (7,0){$SU(3)$};
\node[](L4) at (10.5,0){$\dots$};
\node[](L5) at (14,0){$SU(3)$};
\node[rectangle, draw, minimum width=.6cm,minimum height=.6cm](L6) at (17,0){3};
\node[](L7) at (8.5,2){$r$};

\draw[-] (L1) -- (L2);
\draw[-] (L2) -- (L3);
\draw[-] (L3) -- (L4);
\draw[-] (L4) -- (L5);
\draw[-] (L6) -- (L5);
 \draw[snake=brace]  (2,1) -- (15,1);
\end{tikzpicture}
\ee
This model is discussed, for example, in subsection 3.4.5 of \cite{Cabrera:2019izd}. Its global symmetry is $SU(9)\times SU(3)\times U(1)$.
\item For $\ell=4$ we choose the $SU(8)\times SU(2)$-preserving holonomy, namely $n_4'=1$ and $N_6=r$. The quiver is 
\be\label{su4quiver}
\begin{tikzpicture}[thick, scale=0.4]
\node[rectangle, draw, minimum width=.6cm,minimum height=.6cm](L1) at (0,0){8};
\node[](L2) at (3,0){${\color{red} SU(4)}$};
\node[](L3) at (7,0){$SU(4)$};
\node[](L4) at (10.5,0){$\dots$};
\node[](L5) at (14,0){$SU(4)$};
\node[rectangle, draw, minimum width=.6cm,minimum height=.6cm](L6) at (17,0){4};
\node[](L7) at (8.5,2){$r$};
\node[rectangle, draw, minimum width=.6cm,minimum height=.6cm](L8) at (3,-3){1};

\draw[-] (L1) -- (L2);
\draw[-] (L2) -- (L3);
\draw[-] (L3) -- (L4);
\draw[-] (L4) -- (L5);
\draw[-] (L6) -- (L5);
\draw[snake=brace]  (2,1) -- (15,1);
\draw[snake=zigzag,segment aspect=0]  (3,-.5)  -- (3,-2.3);
\end{tikzpicture}
\ee
with global symmetry $SU(8)\times SU(2)\times U(1)\times SU(4)$. The $SU(2)$ flavor symmetry acts on the hypermultiplet in the ${\bf 6}$ (which we denote with a squared 1) of the red $SU(4)$ gauge-group.
\end{itemize}
Notice that for $r=1$, when there's only one gauge group, these models reduce precisely to those considered in \cite{Ohmori:2018ona}. In particular, in this case the fundamental hypermultiplets at the two ends of the quiver are charged under the same gauge group and therefore the global symmetry enhances. This is in perfect agreement with our expectation: the global symmetry of higher rank $\SS$-fold theories is smaller than that of rank-one models. We will now discuss the details of the compactification in the presence of almost commuting holonomies for the flavor symmetry in each of the three cases \eqref{su2quiver}-\eqref{su4quiver}.

\subsubsection{Four-dimensional SCFTs from torus-compactifications}\label{twist4d} 

In this subsection we will show that, starting from the six-dimensional SCFTs described in the previous subsection, with a judicious choice of almost commuting holonomies along the nontrivial cycles of $T^2$, we find four-dimensional SCFTs with the same global symmetry and Coulomb branch spectrum as the  $\SS_{G,\ell}^{(r)}$ theories (with $\ell\Delta_7=6$). We will also show that the dimension of the enhanced Coulomb branch agrees. Moreover, we will be able to read off the transformation properties of the free hypermultiplets constituting the ECB. 

We start by turning on almost commuting holonomies for the symmetries $SO(16)$, $SU(9)$, and $SU(8)$ respectively. In the case $\ell=2$ we consider an $Sp(4)\times SU(2)$ subgroup of $SO(16)$ under which the vector decomposes as $\bf{16}\rightarrow (\bf{8},\bf{2})$ and embed the holonomies in the $SU(2)$ part as follows: 
\be\label{holo2}
P=\begin{pmatrix}J_8 & 0 \\ 0 & -J_8\end{pmatrix}\;, \qquad  Q=\begin{pmatrix}0 & J_8 \\  -J_8 & 0\end{pmatrix}\;,
\ee
where $J_8$ is the $8\times 8$ symplectic form of $Sp(4)$. The matrices $P$ and $Q$ commute up to the $\mathbb{Z}_2$ center of $SO(16)$ and leave just the $Sp(4)$ subgroup unbroken. The eight $SU(2)$ doublets on the left of the quiver (\ref{su2quiver}) are organized into a half hypermultiplet transforming as $(\bf{16},\bf{2})\rightarrow (\bf{8},\bf{2},\bf{2})$, where the second $SU(2)$ is broken by the holonomies (\ref{holo2}). 

The holonomies considered above can be turned on provided there are no operators charged under the $\mathbb{Z}_2$ center. On the other hand, we have just seen that the eight $SU(2)$ fundamental hypermultiplets transform nontrivially. To remedy this, we should accompany the flavor holonomies (\ref{holo2}) with holonomies embedded in the leftmost $SU(2)$ gauge group of the form 
\be\label{holo22}
P=\left(\begin{array}{cc}i & 0 \\ 0 & -i\\\end{array}\right);\quad Q=\left(\begin{array}{cc}0 & i \\  i & 0\\\end{array}\right).
\ee
Now, by turning on these holonomies, the leftmost $SU(2)\times SU(2)$ bifundamental in (\ref{su2quiver}) acquires a nontrivial charge under $\mathbb{Z}_2$ and therefore we should embed the almost commuting holonomies \eqref{holo22} in the second $SU(2)$ gauge group as well. By iterating this argument, we conclude that we should embed the holonomies in all the gauge groups along the quiver (therefore breaking the gauge symmetry completely) and also in one $SU(2)$-factor of the global symmetry acting on the two flavors at the right end of the quiver. In this way we break the global symmetry down to $Sp(4)\times SU(2)$, which is precisely the expected global symmetry for $\SS_{G,\ell}^{(r)}$ theories.\footnote{The $SU(2)$ factor of the global symmetry, as well as the $U(1)$ factors we will find for $\ell=3,4$, are inherited from the global symmetry of the six-dimensional theory. See, e.g., \cite{Apruzzi:2020eqi} for a recent discussion on global symmetries in six dimensions, from which our statements can be easily derived.} In the next subsection, we will recompute the Weyl anomaly coefficients and flavor central charges using the construction of $\SS_{G,\ell}^{(r)}$ as a torus-reduction of six-dimensional SCFTs, and confirm that they also match with our expectations.

This construction also grants easy access to the (quaternionic) dimension of the ECB fiber. At a generic point of the tensor branch of the six-dimensional SCFT the low-energy effective theory consists of $r$ tensor multiplets and the linear quiver (\ref{su2quiver}). Upon torus compactification, we end up at a generic point of the Coulomb branch of the resulting four-dimensional theory, where the low-energy spectrum consists of $r$ vector multiplets coming from the tensor multiplets. The six-dimensional vector multiplets do not contribute because our choice of holonomies has broken the gauge group completely. We also obtain a collection of additional free hypermultiplets: from the half-hypermultiplet transforming in the $ (\bf{8},\bf{2},\bf{2})$ we get four free hypermultiplets, as the holonomy is embedded diagonally in $SU(2)\times SU(2)$ and the $(\bf{2},\bf{2})$ decomposes as $\bf{3}+\bf{1}$. These eight invariant half-hypermultiplets transform as the $\bf{8}$ of $Sp(4)$. The same argument shows that from each bifundamental we get one hypermultiplet, leading to $r$ hypermultiplets all transforming as doublets of $SU(2)$. Overall we find $4+r$ free hypermultiplets at a generic point of the $r$-dimensional Coulomb branch, in perfect agreement with the ECB dimension predicted from F-theory. 

In the case $\ell=3$ we consider an $SU(3)^2$ subgroup of the $SU(9)$ global symmetry such that the fundamental decomposes as $\bf{9}\rightarrow (\bf{3},\bf{3})$. We embed the almost commuting holonomies in one of the $SU(3)$ factors as follows: 
\be\label{holo3}
P=\begin{pmatrix}I_3 & 0 & 0 \\ 0 & \omega^3I_3 & 0\\ 0 & 0 & \omega^6I_3\end{pmatrix}\;,\qquad Q=\begin{pmatrix}0 & I_3 & 0 \\ 0 & 0 & I_3\\ I_3 & 0 & 0\end{pmatrix}\;.
\ee
Here $I_3$ is the $3\times 3$ identity matrix and $\omega=e^{2\pi i/9}$. This choice breaks $SU(9)$ to $SU(3)$, and the two holonomies commute up to a  $\mathbb{Z}_3$. Again, the nine flavors on the left of the quiver (\ref{su3quiver}) transform in the $(\bf{9},\bf{3})\rightarrow (\bf{3},\bf{3},\bf{3})$ and in order to avoid the occurrence of fields charged under the discrete $\mathbb{Z}_3$ group,  we have to embed the holonomies in the $SU(3)$ gauge group as well. The presence of the $SU(3)\times SU(3)$ bifundamentals then forces us to embed the holonomies in all the gauge groups and the global $SU(3)$ rotating the flavors at the right end of the quiver (\ref{su3quiver}). In this way, we break the gauge symmetry completely and we find a four-dimensional theory of rank $r$ with global symmetry $SU(3)\times U(1)$. For $r=1$, this symmetry enhances to $SU(4)$. Again, we can determine the dimension of the ECB by counting the hypermultiplets surviving the projection. We get one from each bifundamental and three from the nine flavors which transform as a triplet of the $SU(3)$ global symmetry. We therefore find a $3+r$-dimensional ECB, in agreement with the expected structure of $\SS_{D_4,3}^{(r)}$ theories. 

Finally, for $\ell=4$ we consider an $SU(4)\times SU(2)$ subgroup of $SU(8)$ such that the fundamental representation decomposes as $\bf{8}\rightarrow (\bf{4},\bf{2})$. We embed the almost commuting holonomies inside $SU(4)$ as
\be\label{holo4}
P=\begin{pmatrix}I_2 & 0 & 0 & 0 \\ 0 & \omega^2I_2 & 0 & 0 \\ 0 & 0 & \omega^4I_2 & 0 \\ 0 & 0 & 0 & \omega^6I_2 \end{pmatrix}\;, \qquad Q=\begin{pmatrix}0 & I_2 & 0 & 0 \\ 0 & 0 & I_2 & 0\\ 0 & 0 & 0 & I_2\\ I_2 & 0 & 0 & 0\end{pmatrix}\;,
\ee
which break $SU(8)$ to $SU(2)$. Here $I_2$ is the $2\times 2$ identity matrix and $\omega=e^{2\pi i/8}$. The two holonomies commute up to an element in $\mathbb{Z}_4$. The eight fundamentals on the left of the quiver (\ref{su4quiver}) transform in the $(\bf{8},\bf{4})\rightarrow (\bf{2},\bf{4},\bf{4})$. Therefore, once again, we embed the holonomies in the $SU(4)$ gauge group as well. As in the previous case the embedding propagates along the quiver and we end up breaking the gauge symmetry completely. We also need to embed the holonomy in the $SU(2)$ carried by the antisymmetric hypermultiplet. The choice (\ref{holo22}) does the job. When the dust settles, we find that the four-dimensional torus-reduced theory has rank $r$ and $SU(2)\times U(1)$ flavor symmetry. What's more, it has $2+r$ free hypermultiplets at a generic point of the Coulomb branch. The two hypermultiplets come from the eight flavors on the left and transform as a doublet of $SU(2)$. (We do not find any massless hypermultiplet coming from the antisymmetric hypermultiplet.) This fits perfectly with the known properties of $\SS_{A_2,\ell}^{(r)}$ theories. 

Finally, for all cases, we can determine the spectrum of the Coulomb branch using the algorithm presented in Appendix B of \cite{Ohmori:2018ona}. As was mentioned before, the six-dimensional theories under investigation here are characterized by a collection of $r$ curves which can be eliminated by repeatedly blowing down the $-1$ curve at one end of the configuration. Combining this with the fact that the chosen holonomies always break the gauge group completely, the algorithm of \cite{Ohmori:2018ona} immediately shows that the Coulomb branch operators have scaling dimension $6,12,\dots, 6r$ as expected.

\subsubsection{Computing central charges}\label{ccsfold}

The goal of this subsection is to compute the central charges of the four-dimensional theories we obtained via torus-compactifications from six dimensions. We apply the method of \cite{Ohmori:2018ona}, which in turn builds on the construction proposed in \cite{Ohmori:2015pua}. The analysis of \cite{Ohmori:2018ona} exploits the fact that in the rank-one case, upon putting the six-dimensional theory on the torus, the Coulomb branch of the resulting four-dimensional theory has three singular points: one at the origin associated to the SCFT we are after and two extra singularities where a hypermultiplet becomes massless and which go off to infinity in the zero area limit of the torus. The key point of the analysis is that the $\mathbb{Z}_{\ell}$ holonomies affect the periodicity of the scalar one gets by integrating the 2-form $B$, which is part of the tensor multiplet, on the torus. This enters in the definition of the gauge invariant coordinate parametrizing the four-dimensional 
Coulomb branch. 

In our case the Coulomb branch is not one-dimensional. However, we can exploit the observation of \cite{Ohmori:2015pua} that for every very Higgsable theory there is a ``distinguished'' one-dimensional submanifold in the Coulomb branch which has exactly the singularity structure described just now. Therefore we can apply the formulae valid for rank-one theories given in \cite{Ohmori:2018ona}. This one-dimensional slice of the Coulomb branch is most easily described in terms of the F-theory description of the six-dimensional theory: we start from the resolved geometry with our collection of $r$ curves $\mathcal{C}_i$ and then we blow down $r-1$ curves with self-intersection $-1$, until we are left with a single $-1$ curve whose volume parametrizes the position in a one-dimensional submanifold of the tensor branch. Upon torus compactification this becomes a one-dimensional submanifold of the Coulomb branch and is precisely the slice we are looking for. At a generic point of this submanifold the low-energy effective theory contains a free vector multiplet coming from the tensor supported on the curve in six dimensions, the free hypermultiplet which is not projected out by the almost commuting holonomies, and a nontrivial SCFT originating from the curves we have blown down. This SCFT is simply $\SS_{G,\ell}^{(r-1)}$. 

The idea is then to argue by induction: we know that for $r=1$ the twisted compactification leads to the central charges of $\SS_{G,\ell}^{(1)}$ theories. This was proven in \cite{Ohmori:2018ona}. We then assume that the central charges for the twisted compactification of rank $r-1$ six-dimensional theories are correctly reproduced,\footnote{Here, by rank we mean the dimension of the tensor branch.} and prove that this implies that also those of the rank-$r$ theories work out. The claim that by blowing down $r-1$ curves we get $\SS_{G,\ell}^{(r-1)}$ theories is part of the inductive step. 

The computation of \cite{Ohmori:2018ona} exploits the topological twist argument of \cite{Shapere:2008zf} which provides the difference between central charges evaluated at the singularity at the origin (which corresponds to the SCFT we are interested in) and those at a generic point of the above-introduced one-dimensional slice of the Coulomb branch. The result for rank-one theories is 
\begin{align}
\label{t2a}(2a-c)_{\text{SCFT}}-(2a-c)_{\text{generic}}&=-\frac{3d}{\ell}-\frac{1}{2}\;,\\
\label{t2c} c_{\text{SCFT}}-c_{\text{generic}}&=\frac{3-3d}{\ell}-1\;,
\end{align}
where $d$ is a coefficient appearing in the anomaly polynomial of the six-dimensional theory. Whenever the effective theory at a generic point on the one-dimensional locus of the tensor branch we are considering is Lagrangian, which is the case for rank-one theories, cancellation of the gauge anomaly requires $d=-h_G^{\vee}$, minus the dual Coxeter number of the gauge group $G$. Since the gauge group supported on any curve is $SU(\ell)$ for all the models we are considering, we have $d=-\ell$ for rank-one theories. 

The equations (\ref{t2a}) and (\ref{t2c}) apply to our case as well, modulo the fact that the quantities $(2a-c)_{\text{generic}}$ and $c_{\text{generic}}$ now include (by induction) the contribution from $\SS_{G,\ell}^{(r-1)}$, which we denote as $(2a-c)_{r-1}$ and $c_{r-1}$ from now on. Also, the value of $d$ is no longer equal to $-h_G^{\vee}$ because the effective theory at a generic point is not Lagrangian anymore. We therefore need to compute $d$ for the models of interest. 

The anomaly polynomial for six-dimensional SCFTs coming from an F-theory compactification includes a Green-Schwarz term of the form \cite{Ohmori:2014kda}
\be\label{gsterm} I_{GS}=\frac{1}{2}\Omega_{ij}I^iI^j\;.\ee 
Here $\Omega_{ij}$ is the inverse of the matrix $\eta^{ij}\equiv -\mathcal{C}_i\cdot\mathcal{C}_j$ encoding the intersection numbers of the curves $\mathcal{C}_i$ in the base and $I^i$ is a 4-form associated with the curve $\mathcal{C}_i$ describing the Green-Schwarz coupling of the corresponding tensor field. In particular each $I^i$ includes the term 
\be I^i=d^ic_2(R)+\dots\ee 
where $c_2(R)$ is the second Chern class of the R-symmetry bundle. We are interested in computing the coefficients $d^i$. Since at a generic point of the tensor branch, before blowing down anything, the low-energy theory is always Lagrangian we can use the gauge anomaly argument mentioned before to conclude that for our models $d^i=-\ell$ for all values of $i$. 
When we blow down the $-1$ curve (which we denote $\mathcal{C}_1$), the $-2$ curve intersecting it (let's call it $\mathcal{C}_2$) becomes a $-1$ curve and its Green-Schwarz term becomes  \cite{Ohmori:2014kda} 
$$I^2\rightarrow I^2+I^1.$$ 
Therefore, when we repeatedly blow down $-1$ curves until we are left with a single $-1$ curve, we find the Green-Schwarz term 
\be I=\sum_i I^i=\sum_id^ic_2(R)+\dots=-r\ell c_2(R)+\dots\;,\ee 
and we are led to the conclusion that we should plug in (\ref{t2a}) and (\ref{t2c}) $d=-r\ell$. 

We therefore find from (\ref{t2a}) 
\be\label{accc}(2a-c)_r=(2a-c)_{r-1}+3r-\frac{1}{4}\;.\ee 
We can now notice that $3r-1/4$ is the contribution of a Coulomb branch operator of dimension $6r$ and therefore we recover (\ref{2a-c}), as expected. From (\ref{t2c}) we find 
\be\label{cccc}c_r=c_{r-1}+\frac{3+3r\ell}{\ell}-\frac{3}{4}\;.\ee 
On the other hand, from (\ref{2a-c}) and (\ref{c-a}) we find 
\be\label{ccr} c_r= \frac{ \ell \Delta_7 r^2  }{4} +\frac{ r (\ell \Delta_7 +2 \Delta_7 -3)}{4 } +   \frac{ \ell (\Delta_7-1)}{12}\;.\ee 
Setting $\ell \Delta_7=6$, we can easily see that (\ref{ccr}) satisfies the recursion (\ref{cccc}). We therefore conclude that the $a$ and $c$ central charges computed from the six-dimensional setup indeed reproduce the expected result.

Next, we can compute the flavor central charge of the nonabelian global symmetry coming from the fundamental hypermultiplets on the left-hand side of the quivers described in subsection \ref{6dquivers} (i.e., the nonabelian groups $H$ in table \ref{global7}). In the rank-one case the arguments of \cite{Ohmori:2018ona} lead to the formula 
\be\label{fcc}k_{\text{SCFT}}-k_{\text{generic}}=\frac{12I}{\ell}\;,\ee 
where $I$ is the embedding index of the four-dimensional global symmetry into the six-di\-men\-sion\-al flavor symmetry. As we have seen, the four-dimensional symmetry group of the higher-rank theories is just a subgroup of the symmetry of the corresponding rank-one theory, but we can notice that the embedding index is always equal to one. Furthermore, in the Green-Schwarz term (\ref{gsterm}) only $I^1$ carries information about the flavor central charge. Combining these considerations, we are led to the conclusion that (\ref{fcc}) holds true for higher-rank theories as well, where again $k_{\text{generic}}$ includes the contribution of $\SS_{G,\ell}^{(r)}$. In this case, the free vector multiplet and the free hypermultiplet do not contribute. The value of $I$ is given in \cite{Ohmori:2018ona}: $I=1$ for $\ell=2$ and $I=\ell$ for $\ell=3,4$. We therefore find for $\ell=2$ 
\be k_r=k_{r-1}+6\;, \qquad k_1=7\;,\ee 
where we have given the value $k_1$ of the flavor central charge for the rank-one theory computed in \cite{Ohmori:2018ona}. We thus obtain $k_r=6r+1$. For $\ell=3,4$ we find instead the formula 
\be k_r=k_{r-1}+12;, \qquad k_1=14\;,\ee 
leading to the result $k_r=12r+2$. These values of the flavor central charge are in perfect agreement with the F-theory computation, see table \ref{flavorcc}. 

Finally, we can compute the flavor central charge of the additional $SU(2)$ flavor symmetry group present for $\ell=2$. Given our constraint $\ell \Delta_7 =6$, we are more specifically looking at $\SS_{E_6,2}^{(r)}$ theories. This case is slightly different because all bifundamental hypermultiplets are charged under this symmetry. Let us start by considering the rank-one case, where we divide the ten flavors into a group of eight and two, to make this case more uniform with higher-rank theories. Clearly only the latter will contribute to the $SU(2)$ flavor central charge. If we denote the $SU(2)$ background curvature with $F$, the Green-Schwarz 4-form $I^1$ will include the term 
$$I^1=\alpha\tr F^2+\dots$$ 
Upon compactification on $T^2$, the quantity $k^{SU(2)}_{\text{SCFT}}-k^{SU(2)}_{\text{generic}}$ will be proportional to $\alpha$,\footnote{Actually $\alpha=1/4$, but for our argument we can leave it generic.} and from (\ref{fcc}) we know that $k^{SU(2)}_{\text{SCFT}}-k^{SU(2)}_{\text{generic}}=6$. 
For generic $r$, we still have $I^1=\alpha\tr F^2+\dots$, but all the 4-forms associated with the $r-1$ curves with self-intersection $-2$ will be $I^i=2\alpha\tr F^2+\dots$ (for $i\neq 1$) since there are bifundamentals both on the left and on the right, and they indeed contribute the same amount. We therefore conclude that after the blow-down we find the Green-Schwarz term 
\be I=\sum_i I^i=(2r-1)\alpha\tr F^2+\dots\ee 
We then deduce the formula\footnote{More precisely, the flavor central charge is encoded in the coefficient of the term $P_1(T)\tr F^2$ of the Green-Schwarz part of the anomaly polynomial $I_{GS}=\frac{1}{2}I^2$. Implicitly we are therefore also using the fact that $I=(2r-1)\alpha\tr F^2+\frac{1}{4}P_1(T)+\dots$. The coefficient of $P_1(T)$ comes entirely from $I^1$, since it vanishes for all other $I^i$'s which are associated with curves of self-intersection $-2$ \cite{Ohmori:2014kda}.} 
\be\label{fcc2} k^{SU(2)}_{\text{SCFT}}-k^{SU(2)}_{\text{generic}}=12r-6\;,\ee 
where $k^{SU(2)}_{\text{generic}}$ includes the contribution of $\SS_{E_6,2}^{(r-1)}$ and the free hypermultiplet which contributes 1 to the flavor central charge. From (\ref{fcc2}), we thus deduce the relation
\be k^{SU(2)}_r=k^{SU(2)}_{r-1}+12r-5\;, \qquad k^{SU(2)}_1=7\;,\ee 
which implies $k^{SU(2)}_r=6r^2+r$, again in perfect agreement with table \ref{flavorcc}. 

\subsubsection{Comments about F/M-theory duality} 

The above discussion makes clear that there are (at least) two descriptions in string theory of the $\SS_{G,\ell}^{(r)}$ theories: one in F-theory involving $r$ D3-branes probing a fourfold singularity obtained by considering a $\mathbb{Z}_{\ell}$ ``orbifold'' of a seven-brane of type $G$ (i.e., the generalized $\mathcal{S}$-fold) and another in M-theory involving $r$ M5-branes probing an M9-plane which is wrapping a $\mathbb{C}^2/\mathbb{Z}_{\ell}$ singularity. The M5-branes wrap a torus with prescribed holonomies around the two cycles. 

This situation is reminiscent of the well-known duality between F-theory and M-theory, usually used as a definition of the F-theory background: F-theory on an elliptically fibered Calabi-Yau times $S^1$ is equivalent to M-theory on the same Calabi-Yau space. Under this duality a D3-brane transverse to the circle is mapped to an M5-brane wrapping the elliptic fiber. This is exactly what happens in our case, since there is a one-to-one correspondence between the D3-probes on the F-theory side and the M5-branes wrapping a $T^2$ in the M-theory setup. However, in our case the standard M/F duality cannot be applied since we do not have a trivially-fibered circle transverse to the D3-branes and accordingly the M-theory background is different from the F-theory $\mathcal{S}$-fold. This suggests that it might be possible to generalize the standard duality beyond the situation in which a trivially-fibered $S^1$ is available in F-theory. This intriguing possibility definitely deserves further investigations.

\subsection{\texorpdfstring{$\mathcal{T}^{(r)}_{G,\ell}$}{T} theories from six dimensions} 

Similarly to the case of $\SS$-fold SCFTs, we can construct the models $\TT_{G,\ell}^{(r)}$ via compactifications of six-dimensional $\mathcal{N}=(1,0)$ theories on a torus.  We again restrict attention to the case $\ell\Delta_7=6$. The relevant six-dimensional SCFTs are realized via the same M-theory setup as before, but we choose the $E_8$ holonomy in such a way that the $-1$ curve does not support any gauge algebra, while on the $r-1$ curves with self-intersection $-2$ we have $SU(\ell)$ gauge groups with bifundamental hypermultiplets in between and $\ell$ flavors at each end of the quiver. We often find it more convenient to blow down the $-1$ curve, resulting in a semi-Lagrangian description consisting of the linear quiver described just now, but where the leftmost $SU(\ell)$ gauge group also couples to the rank-one E-string theory. The global symmetry involves the subgroup of $E_8$ which commutes with the $SU(\ell)$ being gauged: $E_7$ for $\ell=2$, $E_6$ for $\ell=3$ and $SO(10)$ for $\ell=4$. 

Notice that the case $r=1$ simply corresponds to the rank-one E-string theory for all values of $\ell$. Since $E_8$ has trivial center the holonomies we turn on along the cycles of the torus can be rotated into the Cartan subalgebra (see \cite{Ohmori:2018ona}). The reduction thus results in a four-dimensional mass-deformation of the $E_8$ Minahan-Nemeschansky theory. Also, the resulting one-dimensional Coulomb branch does not have the structure described in subsection \ref{ccsfold}, but rather it is an $\ell$-th cover of a $II^{*}$ geometry, with Coulomb branch operator of dimension $\Delta_7 = 6/\ell$. Also, the case $r=2$ is special as we have an enhancement of the six-dimensional global symmetry: in this case there is a single $SU(\ell)$ gauge group with $2\ell$ flavors, which results in a larger flavor symmetry in four dimensions as well.

\subsubsection{The \texorpdfstring{$\ell=2$}{ℓ=2}  case}

In the notation of \cite{Mekareeya:2017jgc} we set $n_2=1$ and $N_6=r$. The quiver is therefore 
\be\label{su2quiver2}
\begin{tikzpicture}[thick, scale=0.4]
\node[rectangle, draw, minimum width=.6cm,minimum height=.6cm](L1) at (-1.5,0){E-string};
\node[](L2) at (3,0){$SU(2)$};
\node[](L3) at (7,0){$SU(2)$};
\node[](L4) at (10.5,0){$\dots$};
\node[](L5) at (14,0){$SU(2)$};
\node[rectangle, draw, minimum width=.6cm,minimum height=.6cm](L8) at (3,-2.5){2};
\node[rectangle, draw, minimum width=.6cm,minimum height=.6cm](L6) at (17,0){2};
\node[](L7) at (8.5,2){$r-1$};

\draw[-] (L1) -- (L2);
\draw[-] (L8) -- (L2);
\draw[-] (L2) -- (L3);
\draw[-] (L3) -- (L4);
\draw[-] (L4) -- (L5);
\draw[-] (L6) -- (L5);
 \draw[snake=brace]  (2,1) -- (15,1);
\end{tikzpicture}
\ee
The full flavor symmetry of the six-dimensional SCFT is $E_7\times SU(2)^3$, with one $SU(2)$ acting simultaneously on all the bifundamental hypermultiplets (and also on the two flavors at each end of the quiver). In the rank-two case the global symmetry enhances to $E_7\times SO(7)$. In order to specify the almost commuting holonomies we consider an $F_4\times SU(2)$ subgroup of $E_7$ and embed the $\mathbb{Z}_2$ holonomies (\ref{holo22}) in the $SU(2)$ part. As explained in subsection \ref{twist4d}, we need to make sure that no operators are charged under $\mathbb{Z}_2$. The only protected E-string operator charged under the $E_8$ global symmetry is the moment map.\footnote{We assume that the full spectrum of unprotected operators is also compatible with our choice of holonomies as determined by considering the moment map operator. It would be important to elucidate this point.} It transforms in the 248-dimensional adjoint representation of $E_8$, which decomposes under $F_4\times SU(2)\times SU(2)$ (where the second $SU(2)$-factor is the gauged $SU(2)$) as 
\be \bf{248}\rightarrow (\bf{52}, \bf{1},\bf{1})+(\bf{1}, \bf{3},\bf{1})+(\bf{26},\bf{3},\bf{1})+(\bf{1},\bf{1},\bf{3})+(\bf{1},\bf{4},\bf{2})+(\bf{26},\bf{2},\bf{2})\;.\ee
Since the last two factors are charged under $\mathbb{Z}_2$, we should embed the holonomy in the gauge group as well and, as in subsection \ref{twist4d}, this propagates along the quiver due to the presence of the bifundamentals, breaking all the gauge groups. At the end of the day, the torus reduction will produce a four-dimensional SCFT of rank $r$ which for $r>2$ has $F_4\times SU(2)$ global symmetry. The ECB fiber has dimension $r$. Indeed, we get one hypermultiplet from each bifundamental. They are charged under $SU(2)$ but are singlets under $F_4$. 

For $r=2$, we have just a single $SU(2)$ gauge group with four flavors which transform in the $\bf{8}$ of the $SO(7)$ global symmetry. The spinor of $SO(7)$ decomposes under $SU(2)^3$ as $\bf{8}\rightarrow (\bf{2},\bf{1},\bf{2})+(\bf{1},\bf{2},\bf{2})$ and by embedding the holonomy in the third $SU(2)$ we get a rank-two SCFT with $F_4\times SU(2)^2$ global symmetry and two-dimensional ECB fiber: at a generic point of the Coulomb branch the low-energy theory includes two free hypermultiplets, each charged under one $SU(2)$ factor. 

\subsubsection{The \texorpdfstring{$\ell=3$}{ℓ=3} case}

For $\ell=3$ we set $n_3=1$ and $N_6=r$. We find the following quiver  
\be\label{su3quiver2}
\begin{tikzpicture}[thick, scale=0.4]
\node[rectangle, draw, minimum width=.6cm,minimum height=.6cm](L1) at (-1.5,0){E-string};
\node[](L2) at (3,0){$SU(3)$};
\node[](L3) at (7,0){$SU(3)$};
\node[](L4) at (10.5,0){$\dots$};
\node[](L5) at (14,0){$SU(3)$};
\node[rectangle, draw, minimum width=.6cm,minimum height=.6cm](L8) at (3,-2.5){3};
\node[rectangle, draw, minimum width=.6cm,minimum height=.6cm](L6) at (17,0){3};
\node[](L7) at (8.5,2){$r-1$};

\draw[-] (L1) -- (L2);
\draw[-] (L8) -- (L2);
\draw[-] (L2) -- (L3);
\draw[-] (L3) -- (L4);
\draw[-] (L4) -- (L5);
\draw[-] (L6) -- (L5);
 \draw[snake=brace]  (2,1) -- (15,1);
\end{tikzpicture}
\ee
whose global symmetry is $E_6\times SU(3)^2\times U(1)$ for generic $r$ and $E_6\times SU(6)$ for $r=2$ \cite{Apruzzi:2020eqi}.
The adjoint of $E_8$ decomposes under $E_6\times SU(3)$ as $\bf{248}\rightarrow (\bf{78},\bf{1})+ (\bf{1},\bf{8})+ (\bf{27},\bar{\bf{3}})+ (\overline{\bf{27}},\bf{3})$. We now consider a $G_2\times SU(3)$ subalgebra of $E_6$ and embed the $\mathbb{Z}_3$ holonomy in the $SU(3)$ part. Again, we should also embed the holonomy in the $SU(3)$ gauge groups. The reason is that under $G_2\times SU(3)\times SU(3)$ (where the second $SU(3)$ is the leftmost gauge group in the quiver (\ref{su3quiver2})) we have the decomposition
\be \bf{248}\rightarrow (\bf{14},\bf{1},\bf{1})+(\bf{1},\bf{8},\bf{1})+(\bf{7},\bf{8},\bf{1})+ (\bf{1},\bf{1},\bf{8})+ (\bf{1},\bar{\bf{6}},\bar{\bf{3}})+(\bf{7},\bf{3},\bar{\bf{3}})+ (\bf{1},\bf{6},\bf{3})+(\bf{7},\bar{\bf{3}},\bf{3})\;,\ee 
of which the last four factors are uncharged under $\mathbb{Z}_3$ only if we embed the holonomy diagonally in the two $SU(3)$ groups. Once again, we end up breaking all the gauge groups in the quiver completely. The torus-reduced theory has rank $r$ and its ECB fiber has dimension $r$. 

The case $r=2$ was already considered in \cite{Ohmori:2018ona}.\footnote{In this reference, it was also identified with the class $\SS$ description of \eqref{T_D4,3^2}.} As was mentioned before, the global symmetry carried by the hypermultiplets enhances to $SU(6)$ and if we consider an $SU(3)\times SU(2)$ subgroup such that the fundamental decomposes as $\bf{6}\rightarrow (\bf{3},\bf{2})$, we can simply embed the holonomy in the $SU(3)$ factor and find a rank-two theory with two-dimensional ECB fiber and global symmetry $G_2\times SU(2)$. The two free hypers at a generic point of the Coulomb branch transform as a doublet of $SU(2)$. Since for $r>2$ the global symmetry of the six-dimensional theory is $E_6\times SU(3)^2\times U(1)$, we expect the resulting four-dimensional SCFT to have global symmetry $G_2\times U(1)$ for $r>2$.\footnote{We thank Fabio Apruzzi for discussions about global symmetries of these models.}

\subsubsection{The \texorpdfstring{$\ell=4$}{ℓ=4}  case}

For $\ell=4$ we choose the $SO(10)\times SU(4)$-preserving holonomy with $n_4=1$ and $N_6=r$. The resulting quiver is 
\be\label{su4quiver2}
\begin{tikzpicture}[thick, scale=0.4]
\node[rectangle, draw, minimum width=.6cm,minimum height=.6cm](L1) at (-1.5,0){E-string};
\node[](L2) at (3,0){$SU(4)$};
\node[](L3) at (7,0){$SU(4)$};
\node[](L4) at (10.5,0){$\dots$};
\node[](L5) at (14,0){$SU(4)$};
\node[rectangle, draw, minimum width=.6cm,minimum height=.6cm](L6) at (17,0){4};
\node[](L7) at (8.5,2){$r-1$};
\node[rectangle, draw, minimum width=.6cm,minimum height=.6cm](L8) at (3,-2.5){4};

\draw[-] (L1) -- (L2);
\draw[-] (L8) -- (L2);
\draw[-] (L2) -- (L3);
\draw[-] (L3) -- (L4);
\draw[-] (L4) -- (L5);
\draw[-] (L6) -- (L5);
\draw[snake=brace]  (2,1) -- (15,1);
\end{tikzpicture}
\ee
This theory has already been discussed in \cite{Mekareeya:2017jgc, Cabrera:2019izd, Frey:2018vpw}. Its global symmetry is $SO(10)\times SU(4)^2\times U(1)$, which enhances to $SO(10)\times SU(8)$ for $r=2$. The adjoint representation of $E_8$ decomposes under $SU(4)\times SO(10)$ as 
\be\label{decso10}\bf{248}\rightarrow (\bf{15},\bf{1})+ (\bf{1},\bf{45})+ (\bf{6},\bf{10})+ (\bar{\bf{4}},\bf{16})+ (\bf{4},\overline{\bf{16}})\;.\ee 
We now consider an $SU(4)\times SU(2)\times SU(2)$ subgroup of $SO(10)$ and embed the $\mathbb{Z}_4$ holonomy in the $SU(4)$ part. Resultingly, the decomposition of the adjoint representation of $E_8$ under $SU(4)_G\times SU(4)\times SU(2)\times SU(2)$, where $SU(4)_G$ denotes the gauged $SU(4)$, contains the adjoint representations of the four factors which are clearly invariant under $\mathbb{Z}_4$ and 
\be\label{dece8} \bf{248}\supset (\bf{4},\bar{\bf{4}},\bf{2},\bf{1})+ (\bar{\bf{4}},\bf{4},\bf{2},\bf{1})+(\bf{6},\bf{6},\bf{1},\bf{1})+(\bf{4},\bf{4},\bf{1},\bf{2})+(\bar{\bf{4}},\bar{\bf{4}},\bf{1},\bf{2})+(\bf{6},\bf{1},\bf{2},\bf{2})+(\bf{1},\bf{6},\bf{2},\bf{2})\;.\ee 
The first three factors indicate that we should embed the holonomies in $SU(4)_G$ as well. On the other hand, the last four factors are invariant only if we also embed the $\mathbb{Z}_2$ holonomy (\ref{holo22}) in the second $SU(2)$ factor. Therefore, only an $SU(2)$ subgroup of $SO(10)$ survives in the four-dimensional reduction. Due to the propagation along the quiver we break completely the gauge symmetry and we end up with a rank $r$ theory with $SU(2)\times U(1)$ global symmetry for $r>2$ and $SU(2)^2$ for $r=2$. The ECB fiber has again dimension $r$.

\subsubsection{Coulomb branch spectrum and central charges from six-dimensions}

The realization of $\TT_{G,\ell}^{(r)}$ as torus-compactifications of six-dimensional SCFTs allows us to (re)derive their Coulomb branch spectrum. We find perfect agreement with the data presented in table \ref{rank-r T-theories}. As mentioned above, the only difference with respect to $\SS_{G,\ell}^{(r)}$ theories is the fact that the $-1$ curve in the fully resolved geometry does not support any gauge algebra. According to the algorithm presented in  \cite{Ohmori:2018ona} we therefore conclude that the Coulomb branch operators have scaling dimensions
\be\label{speccb} 6,12,\dots, 6(r-1), \frac{6r}{\ell}\;,\ee 
in agreement with our expectations. 

Also the $a$ and $c$ central charges can be computed, completely similarly to the discussion in subsection \ref{ccsfold}. The rank-one case is the easiest as we can simply apply equations (\ref{t2a}) and (\ref{t2c}). In this case, we should set $d=-1$, because on the $-1$ curve we do not have any gauge algebra \cite{Ohmori:2014kda}. We find the central charges of instanton-SCFTs (with the free hypermultiplet removed): 
\be\label{rk1cc}4(2a-c)=\frac{12-\ell}{\ell}\;, \qquad c=\frac{6-\ell}{\ell}+\frac{1}{6}\;,\ee
namely the $E_6$ Minahan-Nemeschanksy theory for $\ell=2$, $SU(2)$ SQCD with four flavors for $\ell=3$, and the $(A_1,D_4)$ Argyres-Douglas theory for $\ell=4$. 

For higher-rank cases, it suffices to notice that $d$ receives a contribution equal to $-\ell$ from all the $-2$ curves, since they support an $SU(\ell)$ gauge group, and therefore we should plug $d=-\ell(r-1)-1$ in (\ref{t2a}) and (\ref{t2c}). Also, $(2a-c)_{\text{generic}}$ and $c_{\text{generic}}$ include the contribution from the rank-$(r-1)$ theory, the contribution of a free vector multiplet and a collection of free hypermultiplets: two for $r=2$ and one for higher rank. We therefore discuss the two cases separately. For $r=2$ we find 
\be\label{rk2cc}4(2a-c)=\frac{10\ell+24}{\ell}\;, \qquad c=\frac{3\ell+24}{2\ell}\;.\ee
We proceed by induction for $r>2$ as in subsection \ref{ccsfold}. Substituting $d=-\ell(r-1)-1$ in (\ref{t2a}) and (\ref{t2c}) we find 
\be\label{ccrrank} c_r=c_{r-1}+\frac{12\ell r-15\ell+24}{4\ell}\;, \qquad (2a-c)_r=(2a-c)_{r-1}+\frac{3\ell(r-1)+3}{\ell}-\frac{1}{4}\;,\ee and using the $r=2$ result (\ref{rk2cc}) we get 
\be\label{ccfinal}(2a-c)_r=\frac{6\ell r(r-1)-r\ell+12r}{4\ell}\;, \qquad c_r=\frac{6\ell r(r-1)-3r\ell+24r}{4\ell}\;,\ee 
which is valid for $r\geq2$. Notice that the formula for $2a-c$ is consistent with the Shapere-Tachikawa formula (see (\ref{speccb})).

Finally, we turn attention to the derivation of the flavor central charges of $\TT_{G,\ell}^{(r)}$ using (\ref{fcc}). We start with the global symmetry arising from the E-string sector, namely $F_4$ for $\ell=2$, $G_2$ for $\ell=3$ and $SU(2)$ for $\ell=4$. All these have embedding index one in the six-dimensional global symmetry and the free hypermultiplets at a generic point of the Coulomb branch do not contribute as they are not charged under these groups. We therefore find from (\ref{fcc}) $k_r=k_{r-1}+12/\ell$. Moreover, they also have embedding index one in the larger global symmetry that occurs for $r=1$. Since the flavor central charge in the rank-one case is known to be $12/\ell$, we conclude that
\be\label{ccfe8} k_r^{F_4}=6r\;, \qquad  k_r^{G_2}=4r\;,\qquad  k_r^{SU(2)}=3r\;. \ee 
These values agree for $r=2$ with those read off from the class $\mathcal{S}$ description. See section \ref{sec_classS}. 

The flavor central charges of the various $SU(2)$ factors can be computed as follows. Let us start with the cases $\ell=3,4$ for which we just need to discuss the $r=2$ case. As is clear from the six-dimensional description, the E-string sector does not contribute whereas the two free hypermultiplets on the Coulomb branch form a doublet. Moreover, the embedding index of this $SU(2)$ inside the six-dimensional flavor symmetry is equal to $\ell$. Combining these facts, we easily see that (\ref{fcc}) leads to the relation $k_{SU(2)}=12+2=14$ both for $\ell=3$ and $\ell=4$. In the case $\ell=2$ and $r=2$ we have instead two $SU(2)$ factors, both with embedding index one in the six-dimensional global symmetry. At a generic point of the Coulomb branch there is one hypermultiplet charged under each factor and again the E-string sector does not contribute. Therefore from (\ref{fcc}) we find $k_{SU(2)}=7$ for both factors. 

Finally, for the cases $r>2$ the analysis is a bit more involved and requires us to look at the anomaly polynomial in more detail. Looking at (\ref{su2quiver2}), we see that for any of the $SU(2)$ gauge groups supported on a curve both the bifundamental at its left and at its right contribute to the $SU(2)$ symmetry surviving the compactification.
If we denote the $SU(2)$ background curvature with $F$, the corresponding Green-Schwarz 4-form $I$ will include the term
\begin{equation}
I=\frac{1}{2}\tr F^2+\dots\;,
\end{equation}
and this holds for all the $r-1$ curves supporting a gauge group. We therefore conclude that after the blow-down we find the Green-Schwarz term 
\be I=\sum_i I^i=\frac{r-1}{2}\tr F^2+\dots\ee 
Hence, we we deduce the formula 
\be\label{fcc22} k^{SU(2)}_{\text{SCFT}}-k^{SU(2)}_{\text{generic}}=12r-12\;,\ee 
where $k^{SU(2)}_{\text{generic}}$ includes the contribution of the rank-$(r-1)$ theory and the free hypermultiplet, which contributes one to the flavor central charge. We therefore find from (\ref{fcc2}) the relation
\be\label{rcc2} k^{SU(2)}_r=k^{SU(2)}_{r-1}+12r-11\;, \qquad k^{SU(2)}_2=14\;.\ee 
The central charge in the $r=2$ case is determined exploiting the fact that the E-string does not contribute and that both free hypers are charged under $SU(2)$. Since this is twice the central charge of the two $SU(2)$ factors of the rank-two theory, this tells us that only the diagonal combination of the two survives at higher rank. From the recursion (\ref{rcc2}) we find $k^{SU(2)}_r=6r^2-5r$. 


\section{Class \texorpdfstring{$\mathcal S$}{S} realizations}\label{sec_classS}
A very wide swath of four-dimensional $\mathcal N=2$ superconformal field theories admit a class $\mathcal S$ construction, \ie{}, they can be realized as a topologically twisted compactification of a six-dimensional $\mathcal N=(2,0)$ theory on a Riemann surface, oftentimes in the presence of half-BPS codimension-two defects marking points on the surface \cite{Gaiotto:2009we,Gaiotto:2009hg}. The question we address in this section is whether the newly discovered $S$-fold SCFTs and their cousins $\mathcal T_{G,\ell}^{(r)}$ have a class $\mathcal S$ realization. 

Our strategy to identify class $\mathcal S$ realizations of the higher-rank $S$-fold SCFTs is to perform a systematic scan. Recalling that the exactly marginal couplings of theories of class $\mathcal S$ are encoded in the complex structure moduli of the Riemann surface and taking note of the fact that higher-rank $S$-fold SCFTs are isolated strongly-coupled theories, we can restrict our attention to theories associated with three-punctured spheres.\footnote{Note that it may happen that such theories still possess a frozen exactly marginal coupling, see \cite{Chacaltana:2012ch}.} Theories associated with three-punctured spheres, often called trinion theories, are determined by the choice of simply-laced Lie algebra $\mathfrak j = \mathfrak a_n, \mathfrak d_n,\mathfrak e_{6,7,8}$ specifying the above-lying six-dimensional $(2,0)$ theory and a choice of (twisted) punctures. The punctures of our interest are regular (also called tame). If such puncture is untwisted, it is labeled by an embedding of $\mathfrak{su}(2)$ into $\mathfrak j$. On the other hand, when encircling a twisted puncture, it acts by an element of the outer-automorphism group of $\mathfrak j$. Correspondingly, twisted punctures are specified by an $\mathfrak{su}(2)$-embedding into $\mathfrak g$, the Langlands dual of the subalgebra invariant under that outer-automorphism twist. What's more, because our aim is to find realizations of the entire series of higher-rank $S$-fold theories, our primary focus are untwisted and $\mathbb Z_2$-twisted theories of increasing rank $n$ of the Lie algebras $\mathfrak j = \mathfrak a_n$ or $\mathfrak d_n$. Nevertheless, while (twisted) theories of type $\mathfrak e_{6,7,8}$ or $\mathbb Z_3$-twisted $\mathfrak d_4$ theories cannot accommodate arbitrarily high Coulomb branch scaling dimensions, we will see that they do provide realizations of various low-rank models of our interest.

The necessary technology to perform our scan has been developed in a sequence of papers \cite{Chacaltana:2010ks,Chacaltana:2011ze,Chacaltana:2012zy,Chacaltana:2012ch,Chacaltana:2013oka,Chacaltana:2014jba,Chacaltana:2015bna,Chacaltana:2016shw,Chacaltana:2017boe,Chcaltana:2018zag} for all series of theories except for twisted $A_{2n}$ models. The exploration of the latter has been initiated recently in \cite{Beem:2020pry}. In particular, upon specifying the triple of embeddings labeling the punctures, these papers provide the necessary tools to compute the following data of the theory:
\begin{itemize}
\item the $a$ and $c$ Weyl anomaly coefficients,
\item the Coulomb branch spectrum $\Delta_i, i=1,\ldots, r$,
\item the Schur limit of the superconformal index\footnote{In principle we have also access to the Macdonald limit of the superconformal index, but we'll focus on this simpler limit throughout this paper.}
\begin{equation}
I_{\mathrm{S}}(q, a_j) \colonequals  \tr\, (-1)^F q^{E-R} \prod_{j=1}^{\rank \mathfrak g_F} a_j^{f_j}\;.
\end{equation}
Here the trace runs over the Hilbert space of states of the radially quantized theory. Furthermore, $E$ is the conformal dimension and $R$ the $SU(2)_R$ Cartan generator, while $f_j$ are Cartan generators of the theory's flavor symmetry algebra $\mathfrak g_F$. Writing the index as a plethystic exponential,\footnote{The plethystic exponential is defined as $\mathrm{PE}[g(x_i)] = \exp[\sum_{i=1}^{\infty} \frac{1}{n} g(x_i^n)]$.} the first few terms take the general form
\begin{equation}
I_{\mathrm{S}} = \mathrm{PE}\Big[ \frac{1}{1-q}(\chi_{2h}\, q^{\frac{1}{2}} + \chi_{\mathrm{adj}}\, q + \ldots)   \Big]\;.
\end{equation}
Here $\chi_{2h}$ is the possibly not fully refined character of the fundamental representation of $\mathfrak{sp}(h)$. This term in the exponential indicates the presence of $h$ free hypermultiplets. At order $q$, one encounters $\chi_{\mathrm{adj}}$, the possibly not fully refined character of the adjoint representation of the flavor symmetry algebra of the interacting part of the theory. The Schur limit of the superconformal index is of particular interest because, via the SCFT/VOA correspondence of \cite{Beem:2013sza}, it computes the vacuum character of the associated vertex operator algebra.\footnote{See, for example, \cite{Tachikawa:2013un,Beem:2014rza,Lemos:2014lua,Arakawa:2018egx,Beem:2020pry} for a discussion of the SCFT/VOA correspondence in the context of theories of class $\SS$.} For more details on the computation of (the Schur limit of) the superconformal index, the reader may also wish to consult \cite{Gadde:2011ik,Gadde:2011uv,Gaiotto:2012xa,Mekareeya:2012tn,Lemos:2012ph},
\item the flavor symmetry central charge $k$ for each simple factor in the flavor symmetry group.
\end{itemize}
These data are often sufficient to identify a candidate theory. Further consistency checks can be performed by, for example, investigating if the class $\mathcal S$ realization is consistent with the partial Higgsings of section \ref{sec_modspace}. 

Note that upon increasing the rank $n$ of $\mathfrak j = \mathfrak a_n$ or $\mathfrak d_n$, the scaling dimensions that Coulomb branch chiral ring generators of the trinion theories can have go up as well, while simultaneously the majority of trinion theories have a Coulomb branch of ever larger dimensionality. These features make looking for a particular theory of a particular rank feasible.

Overall, we were successful at identifying various rank-two $S$-fold models, but still higher-rank $S$-fold SCFTs have proved elusive.\footnote{While we consider it unlikely, we do not, however, claim that none of the still higher-rank theories can be realized in class $\mathcal S$, as our scans were necessarily finite and we have not excluded the possibility that these models occur in trinions describing product theories.} We also identify various theories belonging to the collection $\mathcal T_{G,\ell}^{(2)}$.

We organize our class $\mathcal S$ identifications of the models $\mathcal S_{G,\ell}^{(r)}$ and $\mathcal T_{G,\ell}^{(r)}$ according to the choice of group $G$. The same model often has many different class $\mathcal S$ realizations. We do not list all of them, but only a representative one that appears at smallest rank $n$. Exceptions are made if a trinion at higher rank $n$ (or more than one trinion at the same rank $n$) is of relevance to elucidate a relation to another class $\mathcal S$ realization. 

Throughout the rest of this section, we specify $\mathfrak{su}(2)$-embeddings by the dimensions of the $\mathfrak{su}(2)$ representations that appear in the decomposition of the fundamental (or vector) representation of $\mathfrak j$ or $\mathfrak g$. Repeated entries are indicated with a superscript. To easily tell apart $\mathbb Z_2$-twisted punctures, we additionally underline them. For $\mathbb Z_3$-twists, we indicate the outer-automorphism element explicitly as a subscript. Trinions are denoted by a triple of embeddings within parenthesis. A subscript indicates the type $\mathfrak j$ of the theory.

\subsection*{Realizations of $\mathcal S_{E_6,\ell}^{(r)}$ and $\mathcal T_{E_6,\ell}^{(r)}$}
For the models $\mathcal S_{E_6,2}^{(r)}$, we have found the following class $\mathcal S$ realizations:
\begin{align}
\mathcal S_{E_6,2}^{(1)} \otimes \mathrm{HM}^{\otimes 3} ~ &\longleftrightarrow ~ \big([3,2^2,1], [3,2^2,1], [2^2,1^4]\big)_{\mathfrak d_4}\;,\\
\mathcal S_{E_6,2}^{(2)} \otimes \mathrm{HM}^{\otimes 2}~ &\longleftrightarrow ~ \big([4^2,3^2], [5,4^2,1], [5,4^2,1]\big)_{\mathfrak d_7}\;.\label{S_E6,2^2}
\end{align}
Here HM stands for the theory of a single (full) hypermultiplet. As these are our first identifications, let us provide some more details.\footnote{The theory $\mathcal S_{E_6,2}^{(1)}$, often colloquially referred to as the rank-one $C_5$ theory, has been analyzed in detail already in \cite{Chacaltana:2011ze}.} First, we can compute the Coulomb branch spectrum of these models by applying the rules of \cite{Chacaltana:2011ze}. We easily confirm that their Coulomb branch chiral ring is generated by operators of scaling dimensions $\Delta =6$ and $\Delta_1 = 6, \Delta_2=12$, respectively. Next, one can compute the Schur limit of their superconformal indices. For simplicity focusing on the first few orders, one finds for the trinion $\big([3,2^2,1], [3,2^2,1], [2^2,1^4]\big)_{\mathfrak d_4}$
\begin{align}
I_{\mathrm{S}}\big(\big([3,2^2,1], [3,2^2,1], [2^2,1^4]\big)_{\mathfrak d_4}; q,\mathbf a_i\big) = \mathrm{PE}\Big[ \frac{1}{1-q}(\chi_{R_1}(\mathbf a_i)\, q^{\frac{1}{2}} + \chi_{R_2}(\mathbf a_i)\, q + \ldots)   \Big]\;,
\end{align}
where $R_1$ is the direct sum of representations $(\mathbf{1},\mathbf{1};\mathbf{2},\mathbf{1},\mathbf{1})\oplus (\mathbf{1},\mathbf{1};\mathbf{1},\mathbf{2},\mathbf{1})\oplus(\mathbf{1},\mathbf{1};\mathbf{1},\mathbf{1},\mathbf{2})$ of the $SU(2)_7\times SU(2)_7 \times SU(2)_8^3$ flavor symmetries manifested by the punctures of the trinion, and $R_2$ equals $(\mathbf{3},\mathbf{1};\mathbf{1},\mathbf{1},\mathbf{1})\oplus \mathrm{perms}\oplus (\mathbf{2},\mathbf{2};\mathbf{1},\mathbf{1},\mathbf{1})\oplus \mathrm{perms}$. We have also already indicated as subscripts the flavor central charges of the manifest flavor symmetry factors, computed using the results of \cite{Chacaltana:2011ze,Chacaltana:2012zy}. Two important conclusions can be drawn from this expression. First, the trinion contains three full hypermultiplets. Second, the flavor symmetry of the interacting part of the trinion is a 55-dimensional algebra, which decomposes as $R_2$ indicates. We recognize this is as a decomposition of $\mathfrak c_5$. The flavor symmetry manifested by the trinion acting on its interacting part is $SU(2)^5_7$: the free hypermultiplets carried away one unit of flavor central charge from the three $SU(2)$ factors under which they are charged. The decomposition $R_2$ indicates that $SU(2)^5_7$ is embedded inside $C_5$ with embedding index one, so we conclude that the flavor central charge of the enhanced flavor symmetry is indeed $k_{\mathfrak c_5} = 7$. Similarly, 
\begin{align}
I_{\mathrm{S}}\big(\big([4^2,3^2], [5,4^2,1], [5,4^2,1]\big)_{\mathfrak d_7}; q\big) = \mathrm{PE}\Big[ \frac{1}{1-q}(4\, q^{\frac{1}{2}} + 39\, q + \ldots)   \Big]\;.
\end{align}
The trinion contains two hypermultiplets and the flavor symmetry algebra of its interacting subsector is 39-dimensional. Upon refining, one finds that this algebra is $\mathfrak c_4\times \mathfrak a_1$. The flavor central charge of the $\mathfrak c_4$-factor can be confirmed to be $k_{\mathfrak c_4}=13$, while the $\mathfrak a_1$-factor is an enhancement of a $\mathfrak u(1)$ flavor symmetry. Some more details are as follows. The manifest flavor symmetry of the trinion is $SU(2)\times U(1)\times SU(2)\times SU(2)$. One hypermultiplet is a doublet of the first $SU(2)$ factor, while the other one is charged under the $U(1)$ flavor symmetry. The enhanced flavor symmetry of the interacting part of the theory decomposes as $2\times(\mathbf 1,\mathbf 1,\mathbf 1)_{0,\pm 1} \oplus (\mathbf 3,\mathbf 1,\mathbf 1)_{0}\oplus (\mathbf 1,\mathbf 3,\mathbf 1)_{0}\oplus (\mathbf 1,\mathbf 1,\mathbf 3)_{0}\oplus(\mathbf 2,\mathbf 1,\mathbf 1)_{\pm \frac{1}{2}}\oplus(\mathbf 1,\mathbf 2,\mathbf 1)_{\pm \frac{1}{2}}\oplus(\mathbf 1,\mathbf 1,\mathbf 2)_{\pm \frac{1}{2}}\oplus(\mathbf 2,\mathbf 2,\mathbf 1)_{\pm \frac{1}{2}}\oplus(\mathbf 2,\mathbf 1,\mathbf 2)_{\pm \frac{1}{2}}\oplus(\mathbf 1,\mathbf 2,\mathbf 2)_{\pm \frac{1}{2}}$. We recognize the decomposition of $C_4$ into $SU(2)^3\times U(1)$ and an additional $SU(2)$ has emerged. After removing the contribution of the free hypermultiplet, the three $SU(2)$ factors in the $C_4$ decomposition have flavor central charge 13. They appear on equal footing and are embedded with index one. Finally, the conformal anomaly coefficients $a$ and $c$ of the trinions can be most easily computed using the results of \cite{Chacaltana:2012zy}. After subtracting the contribution of the free hypermultiplets, one finds, as expected,
$(a,c)_{\mathcal S_{E_6,2}^{(1)}}=(\frac{41}{12},\frac{49}{12})$ and $(a,c)_{\mathcal S_{E_6,2}^{(2)}}=(\frac{29}{3},\frac{65}{6})$.

In light of the identification of all instanton-SCFTs $\mathcal I_{G}^{(r)}$ as trinions of linearly increasing rank $n$ \cite{Moore:2011ee}, it may be worth pointing out that also in $\mathfrak d_8$ one encounters $\mathcal S_{E_6,2}^{(2)}$. In detail,
\begin{equation}
\mathcal S_{E_6,2}^{(2)} \otimes \mathrm{HM}^{\otimes 7}~ \longleftrightarrow ~ \big([4^4], [5^3,1], [7,4^2,1]\big)_{\mathfrak d_8}\;.
\end{equation}
However, the set of $\mathfrak d_{12}$ theories does not contain $\mathcal S_{E_6,2}^{(3)}$. In fact, we have not encountered $\mathcal S_{E_6,2}^{(3)}$ in any class $\mathcal S$ theory of type $\mathfrak d_{n}$ for $n\leq 14$. Finally, it should be mentioned that $\mathcal S_{E_6,2}^{(2)}$ also occurs as a $\mathbb Z_2$-twisted $\mathfrak e_6$ theory, as was already pointed out in \cite{Apruzzi:2020pmv}.\footnote{It is entry 113 of subsection 3.4 of \cite{Chacaltana:2015bna}.}

Any of the realizations of $\mathcal S_{E_6,2}^{(2)}$ we have presented can be used to compute its Schur index to high order. We obtain for its first several orders
\begin{align}
I_{\mathrm{S}}\big(\mathcal S_{E_6,2}^{(2)};q,\mathbf a, \mathbf b \big) =  \mathrm{PE}\Big[ \frac{1}{1-q}\Big\{&\big(\chi^{\mathfrak c_4}_{\mathrm{adj}}(\mathbf a) + \chi^{\mathfrak a_1}_{\mathrm{adj}}(\mathbf b) \big) q + \big(\chi^{\mathfrak c_4}_{\mathbf{8}}(\mathbf a) \chi^{\mathfrak a_1}_{\mathrm{adj}}(\mathbf b) + \chi^{\mathfrak c_4}_{\mathbf{42}}(\mathbf a) \chi^{\mathfrak a_1}_{\mathbf{2}}(\mathbf b) \big)q^{\frac{3}{2}} \nn \\
& + \big(1 + \chi^{\mathfrak a_1}_{\mathbf{5}}(\mathbf b) + \chi^{\mathfrak c_4}_{\mathrm{adj}}(\mathbf a) \chi^{\mathfrak a_1}_{\mathrm{adj}}(\mathbf b)  + \chi^{\mathfrak c_4}_{\mathbf{48}}(\mathbf a) \chi^{\mathfrak a_1}_{\mathbf{2}}(\mathbf b)\big)q^2 \nn \\
& + \big( \chi^{\mathfrak c_4}_{\mathbf{160}}(\mathbf a) +   \chi^{\mathfrak c_4}_{\mathbf{42}}(\mathbf a) \chi^{\mathfrak a_1}_{\mathbf{4}}(\mathbf b) \big)q^{\frac{5}{2}} + \ldots\Big\}   \Big]\;.
\end{align}
Here $\mathbf a, \mathbf b$ denote the $\mathfrak c_4$ and $\mathfrak a_1$ flavor symmetry fugacities respectively.

For the models $\mathcal T_{E_6,2}^{(r)}$, we have found the following simple class $\mathcal S$ realizations:\footnote{The $D_4$ realization has already appeared explicitly in \cite{Chacaltana:2011ze}.}\textsuperscript{,}\footnote{As above, among others, there is also a realization in $\mathfrak d_8$, to wit, $\big([4^4], [5^3,1], [7,5,3,1]\big)_{\mathfrak d_8}$, and in twisted $\mathfrak e_6$ (see entry 30 of subsection 3.5 of \cite{Chacaltana:2015bna}).}
\begin{align}
\mathcal T_{E_6,2}^{(2)} ~ &\longleftrightarrow ~ \big([3,2^2,1], [3,2^2,1], [1^8]\big)_{\mathfrak d_4}\;,\\
\mathcal T_{E_6,2}^{(2)} \otimes \mathrm{HM}^{\otimes 5}~ &\longleftrightarrow ~ \big([4^2,3^2], [5,4^2,1], [5,5,3,1]\big)_{\mathfrak d_7}\;.\label{T_E6,2^2}
\end{align}
The interacting part of this trinion correctly reproduces the conformal anomaly coefficients $(a,c)_{\mathcal T_{E_6,2}^{(2)}}=(\frac{13}{2},\frac{15}{2})$, the Coulomb branch spectrum $\Delta_1=\Delta_2 = 6$, and the flavor symmetry $(F_4)_{12} \times SU(2)_7 \times SU(2)_7$ (the flavor central charges of the $SU(2)$ factors cannot be determined directly from the second realization). The class $\mathcal S$ realization allows for a straightforward evaluation of the Schur index of $\mathcal T_{E_6,2}^{(2)}$:
\begin{align}
I_{\mathrm{S}}\big(\mathcal T_{E_6,2}^{(2)};q,\mathbf a, \mathbf b_1, \mathbf b_2 \big) =  \mathrm{PE}\Big[ \frac{1}{1-q}\Big\{&\big(\chi^{\mathfrak f_4}_{\mathrm{adj}}(\mathbf a) + \chi^{\mathfrak a_1}_{\mathrm{adj}}(\mathbf b_1) + \chi^{\mathfrak a_1}_{\mathrm{adj}}(\mathbf b_2) \big) q \nn \\
&+ \big(\chi^{\mathfrak f_4}_{\mathbf{26}}(\mathbf a)\chi^{\mathfrak a_1}_{\mathbf{2}}(\mathbf b_1) + \chi^{\mathfrak f_4}_{\mathbf{26}}(\mathbf a)\chi^{\mathfrak a_1}_{\mathbf{2}}(\mathbf b_2)\big)q^{\frac{3}{2}} \nn \\
& + \big(\chi^{\mathfrak f_4}_{\mathrm{adj}}(\mathbf a)\chi^{\mathfrak a_1}_{\mathbf{2}}(\mathbf b_1)\chi^{\mathfrak a_1}_{\mathbf{2}}(\mathbf b_2) + \chi^{\mathfrak f_4}_{\mathbf{26}}(\mathbf a)\big)q^2 + \ldots\Big\}   \Big]\;.
\end{align}

Notice that the Higgsing of $\mathcal S_{E_6,2}^{(2)}$ to $\mathcal T_{E_6,2}^{(2)}$, triggered by giving a vacuum expectation value to the $\mathfrak c_4$ moment map, is manifest in the $D_7$ class $\mathcal S$ realization. Indeed the realizations \eqref{S_E6,2^2} and \eqref{T_E6,2^2} differ in the specification of one puncture. If we depict a puncture by a Young diagram whose columns have heights equal to the integers specifying the $\mathfrak{su}(2)$-embedding, our Higgsing corresponds to moving one box in the Young diagram that captures a factor of the $\mathfrak c_4$ flavor symmetry. The operation of Higgsing by moving one box at a time was studied in detail in \cite{Gaiotto:2012uq}. The realizations in $D_4$ also make manifest the Higgsing of $\mathcal T_{E_6,2}^{(2)}$ to the rank-two instanton SCFT $\mathcal I_{E_6}^{(2)}$: The former theory is described by the trinion  $\big([3,2^2,1], [3,2^2,1], [1^8]\big)_{\mathfrak d_4}$ and the latter by the trinion $\big([3^2,1^2], [3,2^2,1], [1^8]\big)_{\mathfrak d_4}$.

\subsection*{Realizations of $\mathcal S_{D_4,\ell}^{(r)}$ and $\mathcal T_{D_4,\ell}^{(r)}$}
Looking for class $\mathcal S$ realizations of $\mathcal S_{D_4,\ell}^{(r)}$ and  $\mathcal T_{D_4,\ell}^{(r)}$ proceeds similarly. Recall that now nontrivial $S$-fold theories correspond to two possible values of $\ell$, namely $\ell = 2,3$. We first remind the reader of one of the realizations of $\mathcal S_{D_4,2}^{(1)}$ as a twisted $A_3$ theory \cite{Chacaltana:2012ch}:
\begin{equation}
\mathcal S_{D_4,2}^{(1)} \otimes \mathrm{HM}^{\otimes 1}~ \longleftrightarrow ~ \big([2,1^2], \underline{[2^2,1]}, \underline{[2^2,1]}\big)_{\mathfrak a_3}\;.\label{S_D4,2^1}
\end{equation}
Within this same family of twisted $A_{2n+1}$ theories, we have scanned up to (and including) $A_{25}$, and additionally found the rank-two $S$-fold theory
\begin{equation}
\mathcal S_{D_4,2}^{(2)} \otimes \mathrm{HM}^{\otimes 1}~ \longleftrightarrow ~ \big([4^2], \underline{[3^3]}, \underline{[4^2,1]}\big)_{\mathfrak a_7}\;.\label{S_D4,2^2}
\end{equation}
The usual checks can be performed. Its Coulomb branch spectrum works out to be $\Delta_1 = 4, \Delta_2=8$. The Weyl anomaly coefficients of the interacting part of the theory match the expectation of table \ref{rank-r S-folds}: $(a,c)_{\mathcal S_{D_4,2}^{(2)}}=(\frac{73}{12},\frac{20}{3})$. Its interacting flavor symmetry is $Sp(2)_{9}\times SU(2)_{18}\times SU(2)_{16}$ -- all levels can be easily confirmed from this particular class $\mathcal S$ description --, and finally its Schur index can be computed straightforwardly:
\begin{align}
&I_{\mathrm{S}}\big(\mathcal S_{D_4,2}^{(2)};q,\mathbf a, \mathbf b_1,\mathbf b_2 \big) =  \nn\\
& \mathrm{PE}\Big[ \frac{1}{1-q}\Big\{\big(\chi^{\mathfrak c_2}_{\mathrm{adj}}(\mathbf a) + \chi^{\mathfrak a_1}_{\mathrm{adj}}(\mathbf b_1) + \chi^{\mathfrak a_1}_{\mathrm{adj}}(\mathbf b_2) \big) q \nn\\
&\qquad+ \big(\chi^{\mathfrak c_2}_{\mathbf{4}}(\mathbf a)\chi^{\mathfrak a_1}_{\mathrm{adj}}(\mathbf b_2) + \chi^{\mathfrak c_2}_{\mathbf{5}}(\mathbf a)\chi^{\mathfrak a_1}_{\mathrm{adj}}(\mathbf b_1)\chi^{\mathfrak a_1}_{\mathbf{2}}(\mathbf b_2) \big)
 q^{\frac{3}{2}} \nn \\
&\qquad + \big(1+\chi^{\mathfrak a_1}_{\mathbf{5}}(\mathbf b_2) +\chi^{\mathfrak c_2}_{\mathrm{adj}}(\mathbf a)\chi^{\mathfrak a_1}_{\mathrm{adj}}(\mathbf b_2) +\chi^{\mathfrak a_1}_{\mathrm{adj}}(\mathbf b_1)\chi^{\mathfrak a_1}_{\mathrm{adj}}(\mathbf b_2)  
 + \chi^{\mathfrak c_2}_{\mathbf{4}}(\mathbf a)\chi^{\mathfrak a_1}_{\mathrm{adj}}(\mathbf b_1)\chi^{\mathfrak a_1}_{\mathbf{2}}(\mathbf b_2)\big) q^2 \nn \\
 &\qquad + \big( \chi^{\mathfrak c_2}_{\mathbf{5}}(\mathbf a)\chi^{\mathfrak a_1}_{\mathrm{adj}}(\mathbf b_1)\chi^{\mathfrak a_1}_{\mathbf{4}}(\mathbf b_2)+  \chi^{\mathfrak c_2}_{\mathbf{16}}(\mathbf a)
- \chi^{\mathfrak c_2}_{\mathbf{4}}(\mathbf a)
 \chi^{\mathfrak a_1}_{\mathrm{adj}}(\mathbf b_1)
  \big)
   q^{\frac{5}{2}} + \ldots\Big\}   \Big]\;.
\end{align}

Among the theories $\mathcal T_{D_4,2}^{(r)}$, we find a realization of the rank-two model in twisted $A_3$
\begin{equation}
\mathcal T_{D_4,2}^{(2)} ~ \longleftrightarrow ~ \big([1^4], \underline{[2^2,1]}, \underline{[2^2,1]}\big)_{\mathfrak a_3}\;.\label{T_D4,2^2}
\end{equation}
This theory has all the correct properties -- $\Delta_1=\Delta_2=4$, $(a,c)_{\mathcal T_{D_4,2}^{(2)}}=(4,\frac{9}{2})$, and flavor symmetry $SO(7)_8\times SU(2)^2_5$ --, see \cite{Chacaltana:2012ch}. This theory can be easily seen to admit Higgsings to the rank-two $\mathfrak d_4$ instanton SCFT, which can be realized as $\big([1^4], \underline{[2^2,1]}, \underline{[3,1,1]}\big)_{\mathfrak a_3}$, and to $\mathcal S_{D_4,2}^{(1)}$ presented above. It may also be useful to observe that in the set of twisted $A_7$ trinions, one also encounters this model:
\begin{equation}
\mathcal T_{D_4,2}^{(2)} \otimes \mathrm{HM}^{\otimes 2}~ \longleftrightarrow ~ \big([4^2], \underline{[3^3]}, \underline{[5,3,1]}\big)_{\mathfrak a_7}\;.\label{T_D4,2^2bis}
\end{equation}
This description clearly shows that $\mathcal T_{D_4,2}^{(2)} $ can be obtained from $\mathcal S_{D_4,2}^{(2)}$ by performing a partial Higgsing triggered by a vacuum expectation value of the $Sp(2)$ moment map. The Schur index of $\mathcal T_{D_4,2}^{(2)} $ reads
\begin{align}
I_{\mathrm{S}}\big(\mathcal T_{D_4,2}^{(2)} ;q,\mathbf a, \mathbf b_1,\mathbf b_2 \big) = \mathrm{PE}\Big[ \frac{1}{1-q}\Big\{&\big(\chi^{\mathfrak b_3}_{\mathrm{adj}}(\mathbf a) + \chi^{\mathfrak a_1}_{\mathrm{adj}}(\mathbf b_1) + \chi^{\mathfrak a_1}_{\mathrm{adj}}(\mathbf b_2) \big) q \nn \\
&+ \big(\chi^{\mathfrak b_3}_{\mathbf{7}}(\mathbf a)\chi^{\mathfrak a_1}_{\mathbf{2}}(\mathbf b_1) + \chi^{\mathfrak b_3}_{\mathbf{7}}(\mathbf a)\chi^{\mathfrak a_1}_{\mathbf{2}}(\mathbf b_2)\big)q^{\frac{3}{2}} \nn \\
& + \chi^{\mathfrak b_3}_{\mathrm{adj}}(\mathbf a)\chi^{\mathfrak a_1}_{\mathbf{2}}(\mathbf b_1)\chi^{\mathfrak a_1}_{\mathbf{2}}(\mathbf b_2)\, q^2 + \ldots\Big\}   \Big]\;.
\end{align}

Turning attention to $\ell=3$, we note that in \cite{Chacaltana:2016shw} the theory $\mathcal S_{D_4,3}^{(1)}$ has been identified as a particular $\mathbb Z_3$ twisted $D_4$ theory:
\begin{equation}
\mathcal S_{D_4,3}^{(1)} ~ \longleftrightarrow ~ \big([5,3], [A_1]_{\omega}, [A_1]_{\omega^2}\big)_{\mathfrak d_4}\;,\label{S_D4,3^1}
\end{equation}
where the twist by the $\mathbb Z_3$ element $\omega$ is indicated explicitly with a subscript. Moreover, $\mathfrak{su}(2)$-embeddings into $\mathfrak g_2$ are specified by their Bala-Carter labels.\footnote{The $\mathfrak{su}(2)$-embedding labeled by $A_1$ has commutant $\mathfrak{su}(2)$ inside $\mathfrak g_2$. It is determined by the decomposition $\mathbf 7 \rightarrow (\mathbf 1,\mathbf 3) + (\mathbf 2, \mathbf 2)$ under $\mathfrak{su}(2)\times \mathfrak{su}(2)$. For later purposes, let us also remark that the $\mathfrak{su}(2)$-embedding labeled by $\widetilde{A_1}$ also has centralizer $\mathfrak{su}(2)$ and is is determined by the decomposition $\mathbf 7 \rightarrow (\mathbf 3,\mathbf 1) + (\mathbf 2, \mathbf 2)$.} We have not been able to find another realization of this model or its higher rank cousins. However, the class of $\mathbb Z_3$ twisted $D_4$ theories does contain $\mathcal T_{D_4,3}^{(2)}$ as well:
\begin{equation}
\mathcal T_{D_4,3}^{(2)} ~ \longleftrightarrow ~ \big([5,3], [0]_{\omega}, [A_1]_{\omega^2}\big)_{\mathfrak d_4}\;,\label{T_D4,3^2}
\end{equation}
where $0$ denotes the trivial embedding into $\mathfrak g_2$. This theory exhibits all expected properties, see table \ref{rank-r T-theories}. Its Schur index is given by 
\begin{align}
I_{\mathrm{S}}\big(\mathcal T_{D_4,3}^{(2)} ;q,\mathbf a, \mathbf b\big) = \mathrm{PE}\Big[ \frac{1}{1-q}\Big\{&\big(\chi^{\mathfrak g_2}_{\mathrm{adj}}(\mathbf a) + \chi^{\mathfrak a_1}_{\mathrm{adj}}(\mathbf b)\big) q + \big(\chi^{\mathfrak a_1}_{\mathbf{4}}(\mathbf b) + 2\chi^{\mathfrak g_2}_{\mathbf{7}}(\mathbf a)\chi^{\mathfrak a_1}_{\mathbf{2}}(\mathbf b)\big)q^{\frac{3}{2}} \nn \\
& + \big(3 + \chi^{\mathfrak g_2}_{\mathrm{adj}}(\mathbf a) + 2 \chi^{\mathfrak g_2}_{\mathbf{7}}(\mathbf a)\chi^{\mathfrak a_1}_{\mathbf{3}}(\mathbf b) \big) q^2 \nn \\
& + \big( \chi^{\mathfrak g_2}_{\mathrm{adj}}(\mathbf a)\chi^{\mathfrak a_1}_{\mathbf{4}}(\mathbf b) + \chi^{\mathfrak g_2}_{\mathbf{7}}(\mathbf a)\chi^{\mathfrak a_1}_{\mathbf{2}}(\mathbf b)  \big) q^{\frac{5}{2}} + \ldots\Big\}   \Big]\;.
\end{align}
The Higgsing from $\mathcal T_{D_4,3}^{(2)}$ to $\mathcal S_{D_4,3}^{(1)}$ is obvious in class $\mathcal S$, and so is the Higgsing to the rank-two $\mathfrak d_4$ instanton SCFT $\mathcal S_{D_4,1}^{(2)}$, which admits a realization as $\big([5,3], [0]_{\omega}, [\widetilde{A_1}]_{\omega^2}\big)_{\mathfrak d_4} $.

\subsection*{Realizations of $\mathcal S_{A_2,\ell}^{(r)}$ and $\mathcal T_{A_2,\ell}^{(r)}$}
Let us next turn attention to $\mathcal S_{A_2,\ell}^{(r)}$ and $\mathcal T_{A_2,\ell}^{(r)}$, which are defined for $\ell = 2,4$. While for $\ell = 4$, not even $\mathcal S_{A_2,4}^{(1)}$ has a known class $\mathcal S$ description, for $\ell=2$, it was found in \cite{Chacaltana:2012ch,Chacaltana:2014nya} that
\begin{equation}
\mathcal S_{A_2,2}^{(1)} ~ \longleftrightarrow ~ \big([2,1],\underline{[1,1]},\underline{[1,1]}\big)_{\mathfrak a_2}\;.\label{S_A2,2^1}
\end{equation}
We may expect to find some of the other models of our interest in the collection of twisted $A_{\mathrm{even}}$ theories. A scan through twisted $A_{2n}$ trinions is however inconvenienced by our current lack of understanding of their Coulomb branch spectra. Nevertheless, a systematic study of such theories has recently been initiated in \cite{Beem:2020pry}. In that work, $\mathcal T_{A_2,2}^{(2)}$ was also identified. It was called $\widetilde T_3$ and is given by 
\begin{equation}
\mathcal T_{A_2,2}^{(2)} ~ \longleftrightarrow ~ \big([1,1,1],\underline{[1,1]},\underline{[1,1]}\big)_{\mathfrak a_2}\;.\label{T_A2,2^2}
\end{equation}
Its index was already computed in \cite{Beem:2020pry} and reads
\begin{align}
I_{\mathrm{S}}\big(\mathcal T_{A_2,2}^{(2)} ;q,\mathbf a, \mathbf b_1, \mathbf b_2\big)=\mathrm{PE}\bigg[\frac{1}{1-q}\Big\{&\big(\chi_{\text{adj}}^{\mathfrak{c}_1}(\mathbf b_1)+\chi_{\text{adj}}^{\mathfrak{c}_1}(\mathbf b_2) + \chi_{\text{adj}}^{\mathfrak{a}_2}(\mathbf a)\big)q \nn \\
&+ \big(\chi_{\mathbf{2}}^{\mathfrak{c}_1}(\mathbf b_1)\chi_{\mathbf{2}}^{\mathfrak{c}_1}(\mathbf b_2)\chi_{\text{adj}}^{\mathfrak{a}_2}(\mathbf a) -1 \big)q^2 + \ldots \bigg]\;.
\end{align}
We have not been able to identify $\mathcal S_{A_2,2}^{(2)}$.\footnote{It is noteworthy that among twisted $A_4$ theories one encounters a trinion with the same Weyl anomaly coefficients and the same global symmetry as $\mathcal S_{A_2,2}^{(2)}$, but incorrect flavor central charges. The same model reappears in twisted $A_{2n}$ of higher rank $2n$.} Finally, we note that the family of twisted $A_2$ theories makes manifest a variety of partial Higgsings, in particular the Higgsing of $\mathcal T_{A_2,2}^{(2)}$ to $\mathcal S_{A_2,2}^{(1)}$ and to $\mathcal I_{A_2}^{(2)} \longleftrightarrow \big([1,1,1],\underline{[1,1]},\underline{[2]}\big)_{\mathfrak a_2}$.

\subsection*{Realizations of $\mathcal S_{A_1,\ell}^{(r)}$ and $\mathcal T_{A_1,\ell}^{(r)}$}
We have not encountered any class $\mathcal S$ realizations of $\mathcal S_{A_1,\ell}^{(r)}$ or $\mathcal T_{A_1,\ell}^{(r)}$.


\acknowledgments

The authors would like to thank Fabio Apruzzi, Christopher Beem, Antoine Bourget, Sergio Cecotti, Jacques Distler, Julius Grimminger, Amihay Hanany, Mario Martone, Leonardo Rastelli, Sakura Sch\"afer-Nameki and Yuji Tachikawa for useful discussions. The work of S.G. is supported by the ERC Consolidator Grant 682608 ``Higgs bundles: Supersymmetric Gauge Theories and Geometry (HIGGSBNDL).'' C.M. and W.P. are partially supported by grant \#494786 from the Simons Foundation.

\bibliographystyle{./auxiliary/JHEP}
\bibliography{./auxiliary/refs}
\end{document}